\documentclass[
reprint,
superscriptaddress,
 amsmath,amssymb,
 aps,
]{revtex4-2}

\usepackage{xparse}
\usepackage{graphicx}\usepackage{dcolumn}\usepackage{bm}\usepackage{physics}
\usepackage[shortlabels]{enumitem}
\usepackage{xr-hyper}
\usepackage{hyperref}
\def\P{{\operatorname{P}}}
\newcommand{\Var}[1]{\mathrm{Var}\left[ #1 \right]}
\newcommand{\qqc}{,\qquad}
\DeclareMathOperator*{\argmin}{argmin}
\DeclareMathOperator*{\argmax}{argmax}

\begin{document}

\preprint{APS/123-QED}

\title{
Phase transition to chaos in complex ecosystems with non-reciprocal species-resource interactions
}

\author{Emmy Blumenthal}
\email{emmyb320@bu.edu}
\author{Jason W. Rocks}\email{jrocks@bu.edu}
\author{Pankaj Mehta}\email{pankajm@bu.edu}
\affiliation{Department of Physics, Boston University, Boston, MA 02215, USA}
\affiliation{Faculty of Computing and Data Science, Boston University, Boston, MA 02215, USA}

\date{\today}             
\begin{abstract}

Non-reciprocal interactions between microscopic constituents can profoundly shape the large-scale properties of complex systems.
Here, we investigate the effects of non-reciprocity in the context of theoretical ecology  by analyzing a generalization of MacArthur's consumer-resource model with asymmetric interactions between species and resources. 
Using a mixture of analytic cavity calculations and numerical simulations, we show that such ecosystems generically undergo a phase transition to chaotic dynamics as the amount of non-reciprocity is increased. We analytically construct the phase diagram for this model and show that the emergence of chaos is controlled by a single quantity: the ratio of surviving species to surviving resources. 
We also numerically calculate the Lyapunov exponents in the chaotic phase and carefully analyze finite-size effects. 
Our findings show how non-reciprocal interactions can give rise to complex and unpredictable dynamical behaviors even in the simplest ecological consumer-resource models.

\end{abstract}

                              \maketitle

Many complex systems operate out of equilibrium where components generically interact non-reciprocally. Significant current research aims to untangle the implications of non-reciprocal interactions for self-organization and pattern formation. While much progress has been made towards understanding non-reciprocity in systems composed of a few types of species or fields, the consequences of non-reciprocity in more complex systems composed of many interacting components are less clear and presents interesting questions in studies of ecosystems, pattern formation, active matter, mechanical networks, and neural networks~\cite{fruchart2021nonrecip,ivlev2015brokenthird,sompolinsky1986temporal,sompolinsky1988chaos,gomex2011sheep}.

Large, diverse ecosystems with many types of species and resources provide a natural setting for exploring this open problem. Over the last decade, researchers have adapted methods from the statistical physics of disordered systems (e.g., replicas, the cavity method, random matrix theory) to analyze such ecosystems~\cite{may1972stability,yoshino2007ecostability,diederich1989replicators,mahadevan2023spatiotemporal,fisher2014nicheneutral,dickens2016analytically,marsland2020microbial,advani2018MCRMCavity,cui2021diverse,roy2021glassyLV}. Much of this work has focused on systems with reciprocal interactions in which dynamics are often implicitly governed by an optimization function and reach a fixed point~\cite{marsland2020mepp,gatto1990general,case1980minimization,chesson1990macarthur}.

One notable exception are recent studies of the random Generalized Lotka--Volterra (GLV) model in which species interact non-reciprocally~\cite{pearce2020stabilization,roy2020persistent, hu2022emergent, ros2023rGLVtypical,ros2023rGLVquenched,pirey2023aging}.
These systems can exhibit novel behaviors such as dynamic fluctuations and chaos, including unpredictable ``boom-and-bust'' dynamics where low-abundance species suddenly bloom to high abundance~\cite{depirey2023manyspecies}. These observations suggest that non-reciprocal interactions can qualitatively change ecological dynamics in species-only models. However, the generalization of these observations to more complex ecosystems with multiple trophic layers or environmentally-mediated interactions remains unexplored.

Here, we introduce a generalization of the classic MacArthur Consumer Resource Model (MCRM) that includes non-reciprocal interactions between species and resources. Consumer-resource models, first introduced by MacArthur and Levins~\cite{macarthurlevins1967, Macarthur1970, chesson1990macarthur}, have played a foundational role in modern theoretical ecology and undergird many powerful theoretical frameworks for understanding ecological competition, including contemporary niche theory and Tilman's R* principle~\cite{tilman1982resource,chase2009ecological}.

\begin{figure}[t]
	\centering
	\includegraphics[width=0.9\linewidth]{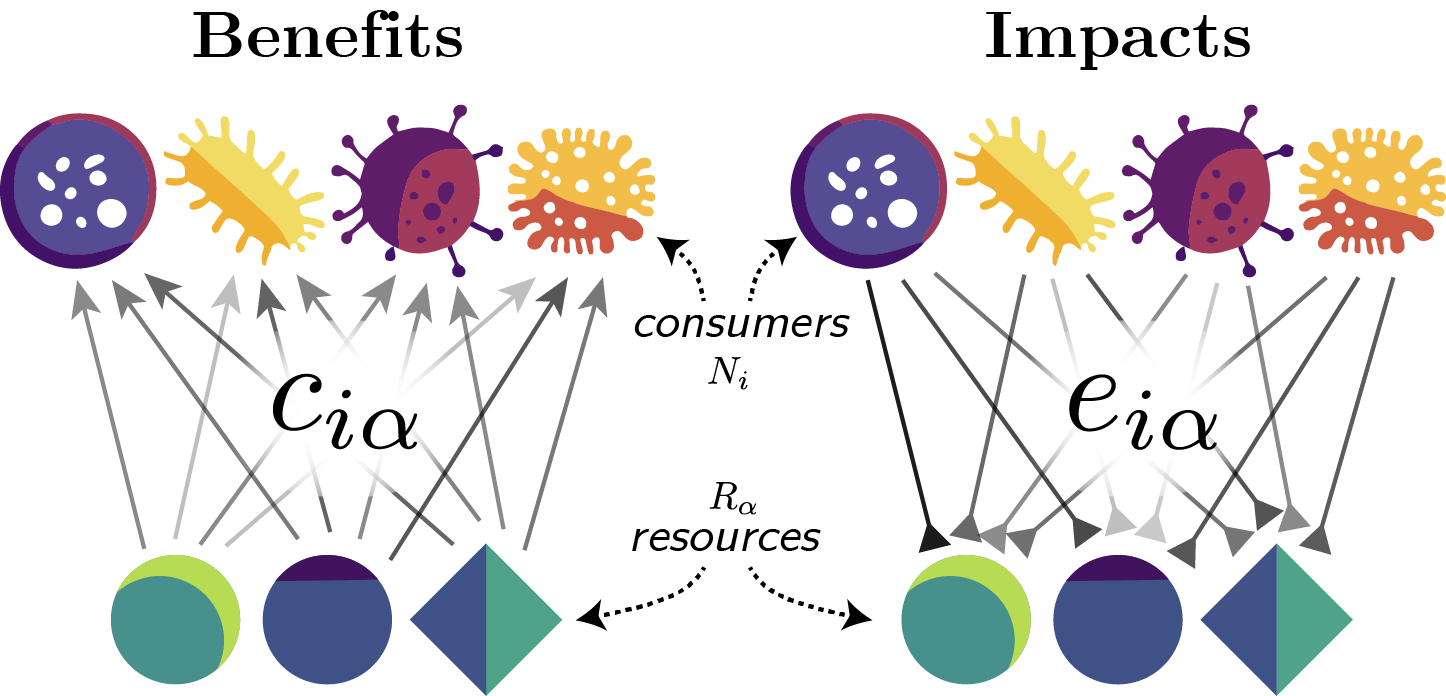}
	\caption{
		Schematic of the asymmetric MacArthur Consumer Resource Model (aMCRM).
		Species $i$ benefits with relative weight $c_{i\alpha}$ from consuming resource $\alpha$ and impacts the abundance of the resource with relative weight $e_{i\alpha}$.
			\label{fig:aMCRM-schematic}
	}
				  \end{figure}

\noindent \textbf{Theoretical setup}. We consider an ecosystem with ${i=1,\ldots, S}$ species which may consume ${\alpha=1,\ldots, M}$ distinct self-replenishing resources with dynamics governed by the equations,
\begin{align}
	\dv{N_i}{t}
	&=
	N_i
	\qty(
	  \sum_{\alpha = 1}^M c_{i\alpha} R_\alpha - m_i
	),
	\\
	\dv{R_\alpha}{t}
	&=
	R_\alpha
	\qty(
	  K_\alpha - R_\alpha
	  ) 
	- \sum_{i=1}^S N_i e_{i\alpha} R_\alpha, \label{Eq:Rdynamics}
\end{align}
where $N_i$ is the population size of species $i$, $R_\alpha$ is the abundance of resource $\alpha$, $c_{i\alpha}$ is the relative consumption preference of species $i$ for resource $\alpha$,  $e_{i\alpha}$ describes the impact of species $i$ on resource $\alpha$, $m_i$ is the natural mortality rate of species $i$, and  $K_\alpha$ is the carrying capacity of resource $\alpha$ in the absence of consumption.  We call this model the asymmetric MacArthur Consumer Resource Model (aMCRM) with a schematic provided in Fig.~\ref{fig:aMCRM-schematic}. When $e_{i \alpha}=c_{i \alpha}$ the species-resource interactions become reciprocal, or symmetric, and the aMCRM reduces to the classical MacArthur Consumer Resource Model (MCRM).

To develop intuition for the role of non-reciprocity in the aMCRM, we consider the limit where the resource dynamics are fast and the resource abundances become entrained to species dynamics. In this case, we take the RHS of Eq.~\eqref{Eq:Rdynamics} to be zero and solve to find ${R_\alpha=\max \qty{0, K_\alpha - \sum_i N_i e_{i\alpha} R_\alpha}}$. Substituting this result into the equation for species dynamics yields an effective Generalized Lotka--Volterra (GLV) equation,
\begin{equation}
\begin{aligned}
	\dv{N_i}{t}
	=&
	N_i
	\qty(
	  \kappa_i - \sum_{j=1}^S A_{ij} N_j
	)
	,\\
	\kappa_i = &\sum_{\alpha=1}^M c_{i \alpha} K_\alpha - m_i,\\
	A_{ij} =& \sum_{\alpha=1}^M c_{i \alpha} e_{j \alpha}\Theta\qty(R_\alpha),
	\end{aligned}
\end{equation}
where $\kappa_i$ is the effective carrying capacity for species $i$ and $A_{ij}$ is the effective species-species interaction matrix, encoding how species $j$ impacts species $i$ ($\Theta$ is the Heaviside function). Although typically not quantitatively accurate, this approximation provides useful qualitative insight into the nature of the non-reciprocal interactions.

In MacArthur's original consumer-resource model, impacts and benefits are identical, ${e_{i \alpha} = c_{i \alpha}}$. In this case, $A_{ij}$ is symmetric, all interactions are reciprocal, the ecosystem has a unique fixed point, and the resulting steady state can be derived using an optimization principle~\cite{marsland2020mepp}. Such behavior is expected because choosing ${c_{i \alpha}=e_{i \alpha}}$ implicitly assumes that each species consumes resources proportional to the marginal utility conferred to that species (in the context of game theory and microeconomics, this is a ``rational strategy''). When the resource-species interactions are non-reciprocal, ${e_{i \alpha} \neq c_{i \alpha}}$, $A_{ij}$ is no longer symmetric, the resulting dynamics can no longer be described using an optimization principle, and there is no guarantee that the dynamics will reach a stable fixed point.

Numerical integration of the aMCRM is performed with a small immigration rate to numerically regularize simulations and ensure that when a steady state is reached, it is uninvadable (see SI section~\ref{appendix:immigration-and-cutoffs} for details).

\noindent \textbf{Thermodynamic limit.} To investigate the aMCRM, we work in the thermodynamic limit where the numbers of species $S$ and resources $M$ become very large while their ratio $M/S$ is held fixed. We assume that parameters are drawn randomly from a fixed distribution analogous to quenched disorder. To ensure a proper thermodynamic limit, parameters are drawn as follows: 
\begin{equation}
\begin{gathered}
K_\alpha = K + \sigma_K \delta K_\alpha, \quad m_i = m + \sigma_m \delta m_i,\\
c_{i\alpha} = \frac{\mu_c}{M} + \frac{\sigma_c}{\sqrt{M}} d_{i\alpha},\\
e_{i\alpha} = \frac{\mu_e}{M} + \frac{\sigma_e}{\sqrt{M}} \qty(
	\rho d_{i\alpha} + \sqrt{1-\rho^2} x_{i\alpha}
)
\end{gathered}
\end{equation}
where $\delta K_\alpha, \delta m_i, d_{i\alpha}, x_{i\alpha}$ are independent standard random variables (i.e., zero mean and unit variance) and $\abs{\rho} \leq 1$ is the interaction reciprocity parameter.
For simplicity, we take $\mu_c = \mu_e \equiv \mu$ and $\sigma_c = \sigma_e \equiv \sigma$ in all figures and simulations.
The central limit theorem ensures that, in the thermodynamic limit, our results are agnostic to the exact form of the underlying distributions and depend only on first and second moments.
Therefore, we sample all parameters from normal distributions unless otherwise specified.

With this parameterization, $\rho$ controls the level of reciprocity of species-resource interactions through the correlation of consumption benefits and impacts:
\begin{equation}
	\operatorname{corr}(c_{i\alpha}, e_{j\beta}) = \rho \; \delta_{ij} \delta_{\alpha\beta}.
\end{equation}
When $\rho = 1$, the aMCRM reduces to the fully symmetric MCRM; when $\rho = 0$, the aMCRM models completely non-reciprocal species-resource interactions. By tuning $\rho$, we can systematically explore the effects of non-reciprocity.

\noindent \textbf{Cavity method.}
Just as in the original MCRM, we can analytically calculate the thermodynamic-limit behavior using the cavity method~\cite{advani2018MCRMCavity,cui2021diverse,Bunin2017,nishimori2001}. Unlike replicas, the cavity method does not require the existence of an energy function and therefore can be extended to the aMCRM. We assume dynamics are self-averaging and described by a replica-symmetric ansatz. Using this ansatz, we derive self-consistent mean-field equations for the fraction of surviving species, the fraction of non-depleted resources, the first and second moments of the steady-state species and resource abundances, and the average linear-order responses of a resource's abundance to a small change in its own carrying capacity and of a species' population to a small change in its own natural mortality rate (see SI section~\ref{appendix:cavity-calc} for detailed derivations). As seen in Figs.~\ref{fig:cavity-slice} and \ref{fig:cavity-heatmaps}, numerical simulations and analytical predictions agree remarkably well for moderate non-reciprocity.

  \begin{figure}[t]
	\centering
	  \includegraphics[width=0.91\linewidth]{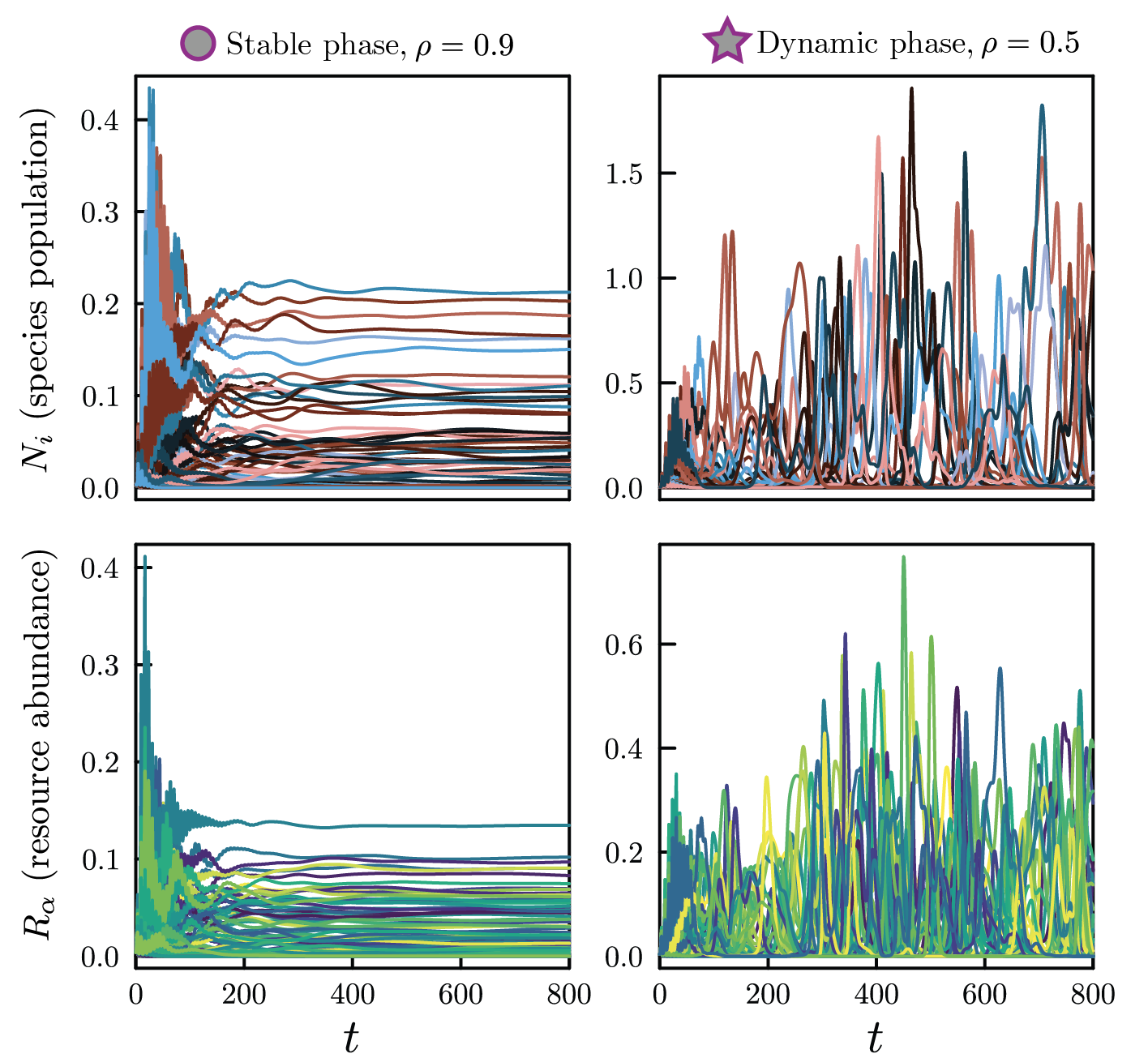}
    	\caption{\label{fig:stable-dynamic-dyn}
		  Example dynamics of the aMCRM in a community of $S=M=256$ species and resources. 
		  Left: dynamics in the stable phase; species-resource interactions are nearly reciprocal.
		  Right: dynamics in the dynamic phase; species-resource interactions are less reciprocal.
		  The parameter values for the stable-phase and dynamic-phase simulations are respectively marked with a circle and star in Fig.~\ref{fig:phase-diagram}(a).
			  }
																						  \end{figure}

\begin{figure}[ht]
	\centering
	\includegraphics[width=\linewidth]{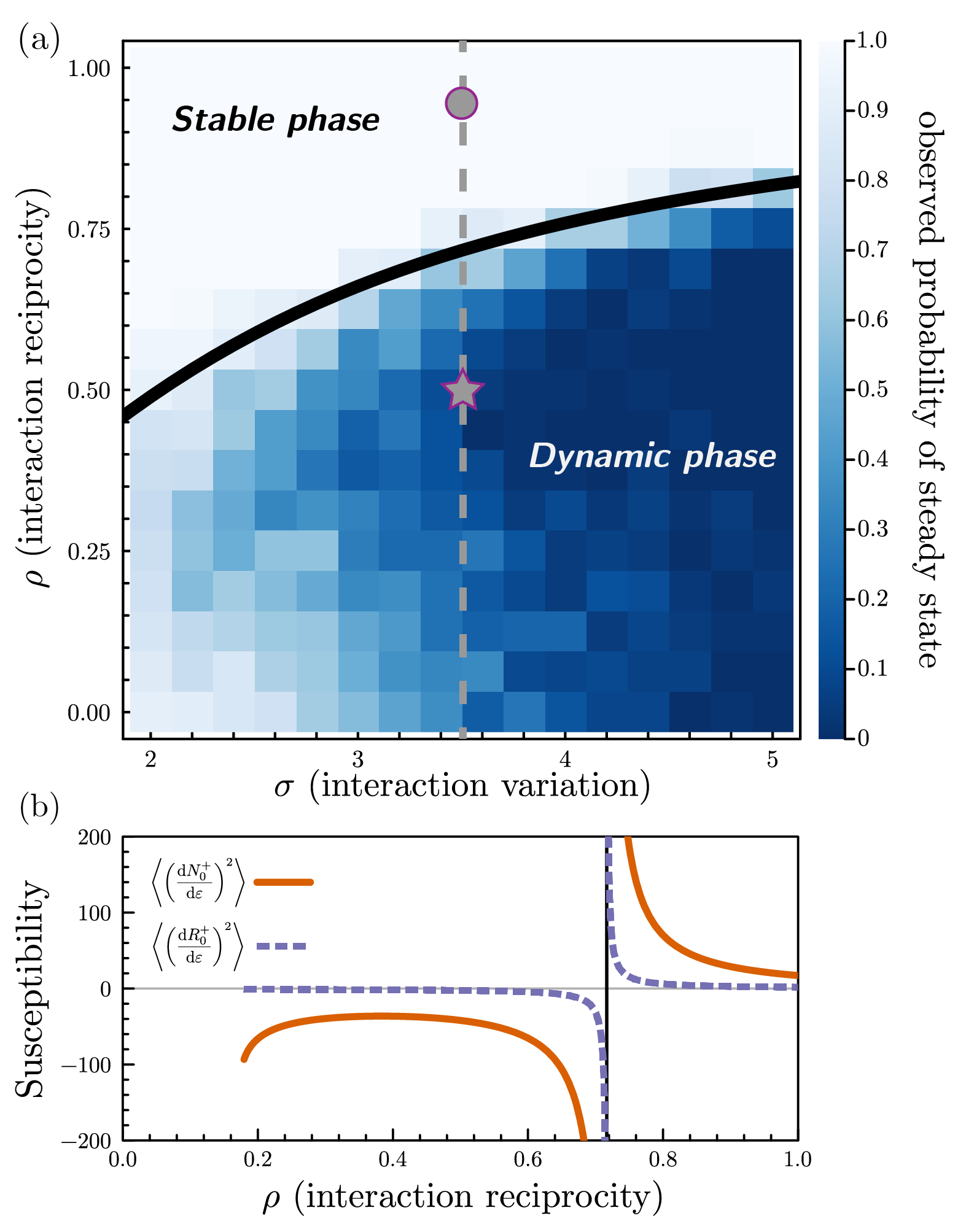}
				\caption{\label{fig:phase-diagram}
		Phase diagram of the aMCRM and diverging susceptibility.
		(a) Heatmap of the fraction of simulations which reached steady state in finite simulation time for various values of $\rho$, the level of reciprocity of species-resource interactions, and $\sigma$, the magnitude of fluctuations in species-resource interactions.
		Overlain is the cavity method-calculated phase boundary.
		(b)
		Variances of susceptibilities of mean-field species and resources as a function of $\rho$, with $\sigma$ fixed at the value indicated by the dashed line in (a).
																							}
																								  \end{figure}

\noindent {\textbf{Transition to dynamic phase.}}
Without reciprocal interactions, the aMCRM has no guarantee of reaching a steady state. We find that when the interaction reciprocity $\rho$ is less than a critical $\rho^\star$, the aMCRM exhibits a phase transition from a unique self-averaging steady state to a chaotic dynamic phase. Fig.~\ref{fig:stable-dynamic-dyn} shows numerical simulations of typical resource and species dynamics observed in each phase (see SI section~\ref{appendix:simulation-and-numerical-methods} for simulation details ~\cite{rackauckas2017differentialequations, julia2017,tsitouras2011runge, VCABM1993,JuliaBBO, optimizationRackauckas2023}).

Using the cavity method, we can analytically compute the phase boundary between the stable and dynamic phases~\cite{Bunin2017}. We perturb the nonzero steady-state species and resource abundances, ${N_i \to N_i + \varepsilon \eta_i^{(N)}}$ and $R_\alpha \to R_\alpha + \varepsilon \eta_\alpha^{(R)}$, where $\varepsilon$ is a small parameter and $\eta_i^{(N)}, \eta_\alpha^{(R)}$ are independent standard random variables, and calculate the susceptibilities $\dv*{N_i}{\varepsilon}$, $\dv*{R_\alpha}{\varepsilon}$. Because of the disordered nature of the perturbation, the expectations of the first moments of the susceptibilities are zero, but the second moments, $\expval*{(\dv*{N_i}{\varepsilon})^2}$, $\expval*{(\dv*{R_\alpha}{\varepsilon})^2}$, are nonzero (see SI section~\ref{appendix:stability-phase-transition} for details). 

The phase transition to the dynamic phase is signaled by the divergence of the these susceptibilities' second moments (see Fig.~\ref{fig:phase-diagram}). Surprisingly, we find that $\rho^\star$, the critical value marking the phase transition to chaos, depends on model parameters only through the species-packing fraction, the ratio of surviving species to non-depleted resources, via the expression (see SI section~\ref{appendix:stability-phase-transition}):

\begin{equation}
	\rho^\star
	=
	\sqrt{
	\frac{
	  \text{\# of surviving species}
	  }
	  {
		\text{\# of non-depleted resources}
	  }}.
	  \label{eq:instability-condition}
\end{equation}
When $\rho<\rho^\star$ the ecosystem undergoes a phase transition to chaos. As the number of surviving species and non-depleted resources are fixed by model parameters, the above equation defines a co-dimension-one phase boundary in the parameter space. Beyond this boundary in the dynamic phase, the second moments of the susceptibilities become negative, indicating that the replica-symmetric ansatz no longer holds, and its results are unstable to any perturbation.

Fig.~\ref{fig:phase-diagram}(a) shows a phase diagram overlain on a heatmap of the fraction of simulations that reach steady state within a chosen finite runtime.
We highlight the locations of the simulations in the stable and dynamic phases in Fig.~\ref{fig:stable-dynamic-dyn} with a circle and a star, respectively.
In Fig.~\ref{fig:phase-diagram}(b), we plot the second moments of the susceptibilities as a function of $\rho$ with fixed $\sigma$ along the slice of phase space indicated by the dashed line in Fig.~\ref{fig:phase-diagram}(a).
The susceptibilities' variances diverge at the phase transition and become invalidly negative in the dynamic phase.
As the phase transition is approached, the fraction of simulations that reach steady state in a finite simulation time sharply decreases. 
An alternative phase diagram with parameters drawn from uniform distributions is shown in Fig.~\ref{fig:uniform-heatmap}.

Finally, we note that for certain choices of parameters, the replica-symmetric self-consistent equations do not have a solution. This transition to infeasibility has an interesting interpretation but is not physically realized because it occurs within the dynamic phase where the replica-symmetric solution is unstable (see SI section~\ref{appendix:cavity-infeas}).
When $\rho = 1$, the instability transition and transition to cavity infeasibility coincide, and neither transition is ever achieved because in the MCRM, the competitive exclusion principle applies, keeping the right-hand side of Eq.~\eqref{eq:instability-condition} less than or equal to one.
When $\rho = 0$, the system is beyond the instability transition, and the cavity infeasibility transition is achieved, meaning no replica-symmetric solution exists.
Mathematically, for $\rho = 0$, the only solution to the mean-field equations is the trivial solution where all species are extinct.

\begin{figure}[t]
	\centering
			\includegraphics[width=0.95\linewidth]{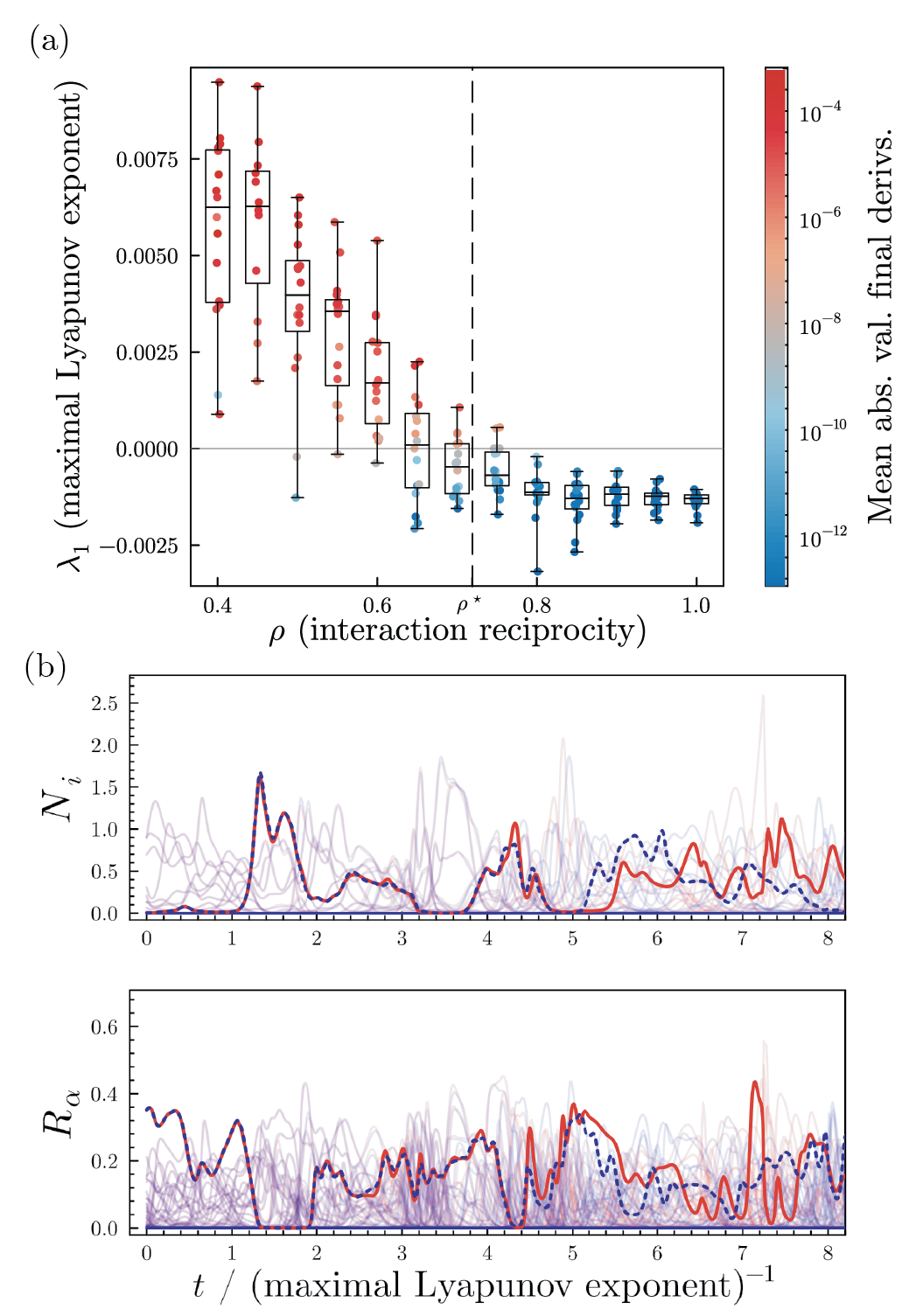}
	\caption{
		\label{fig:diverging-trajecs-maxlyap_dotplot}
				Chaos in the dynamic phase of the aMCRM.
				(a)
				Dot and box-and-whisker plot of $\lambda_1$s, maximal Lyapunov exponents, for simulations at various values of $\rho$, colored by the mean of absolute values of derivatives of all species and resources at the end of the simulation which is an indicator of whether the simulation has reached steady state.
						(b)
		Two trajectories (red and blue) with slightly different initial conditions in the dynamic phase of the aMCRM.
		A species and a resource are highlighted to emphasize the chaotic dynamics; all other species and resources are shown at low opacity for clarity.
		The units of time are given by the inverse of the maximal Lyapunov exponent, $\lambda_1^{-1} =  190$.
																	}
												\end{figure}

\noindent \textbf{Chaos.} 
In order to better understand the transition to chaos, we numerically computed the maximal Lyapunov exponent $\lambda_1$ of the aMCRM in the dynamic and stable phases using the ``H2'' method of Geist~\cite{Benettin1980,Geist1990,Datseris2022,Datseris2018}. The maximal Lyapunov exponent characterizes how quickly trajectories from nearby initial conditions diverge (positive exponent) or converge (negative exponent). 
As seen in Fig.~\ref{fig:diverging-trajecs-maxlyap_dotplot}(a), typically, in the dynamic phase, $\lambda_1 > 0 $, while in the stable phase, $\lambda_1 < 0$.
For the parameters used in Fig.~\ref{fig:stable-dynamic-dyn}, $|\lambda_1| \approx 5\times 10^{-3}$, indicating that the divergence or convergence of nearby trajectories occurs on a timescale of $\lambda_1^{-1}\approx 2\times 10^2$ time units. We further confirmed the existence of chaos by analyzing the generalized alignment index (GALI) which measures how a volume element formed by tangent vectors to a trajectory changes over time~\cite{galiSkokos2007,Datseris2018,Datseris2022} (see Fig.~\ref{fig:gali}).
Additionally, we estimated and analyzed the Kaplan--Yorke dimension and found that it is less than the number of surviving species and resources~\cite{kaplan1979dimension}.
Further details are given in SI section~\ref{appendix:chaos}.

A direct signature of chaotic dynamics is high sensitivity to initial conditions as observed in Fig.~\ref{fig:diverging-trajecs-maxlyap_dotplot}(b). The red and blue lines show the simulated trajectory of a single species (top) and resource (bottom) started from initial conditions with slight differences. Initially, the trajectories are almost identical before diverging from each other significantly after a few Lyapunov times.

\noindent \textbf{Finite-size effects.}
Like most phase transitions, the transition between the stable and dynamic phases is a thermodynamic-limit phenomenon.
In small ecosystems, the aMCRM may approach steady state even when in the dynamic phase due to finite-size effects.
As a result, it is not clear in Fig.~\ref{fig:phase-diagram} what the true probability of steady state is in the thermodynamic limit.
In SI section~\ref{appendix:finite-size-scaling}, we quantify these effects by performing a numerical analysis to extrapolate the steady-state probabilities to infinite system size for each of the two points highlighted in Fig.~\ref{fig:phase-diagram}.
For both sets of parameters, we measure the distribution of steady-state times for many simulations for a variety of system sizes.
Using a custom method based on maximum-likelihood estimation, we then perform a finite-size scaling collapse on these distributions, allowing us to approximately determine the steady-state probabilities as a function of system size.
Our scaling collapses provide strong evidence that the probability of reaching steady state in the thermodynamic limit approaches exactly zero in the dynamic phase and one in the stable phase.

\noindent \textbf{Discussion.}
In this letter, we analyzed the effects of non-reciprocal species-resource interactions on the stability of ecosystems. We introduced the asymmetric MacArthur Consumer Resource Model (aMCRM), a generalization of the MacArthur Consumer Resource Model (MCRM). Using the cavity method, we identified a phase transition between a stable phase in which a unique, uninvadable, self-averaging steady state exists and a dynamic phase with chaotic fluctuations. Remarkably, the phase boundary depends on model parameters only through the species-packing ratio---the ratio of surviving species to non-depleted resources.

Tilman analyzed stability in a two-species, two-resource system where the yields of the species on resources differs from their growth and determined ``the equilibrium point will be stable if each species consumes proportionately more of the resource that more limits its own growth'' \cite{tilman1982resource}.
The divergence between $c_{i\alpha}$ and $e_{i\alpha}$ in the aMCRM is analogous to the divergence between the yield and growth rate in Tilman's analysis.
Our results suggest that this principle generalizes to ecosystems with many species and resources; however, our analysis takes a statistical approach and uses a different model of dynamics.

We found that the chaotic regime is generic and occurs robustly and shares features with GLV models with asymmetric interactions where chaos can also be found \cite{pearce2020stabilization,mahadevan2023spatiotemporal}.
In consumer-resource models, chaotic dynamics generically occurs when the systems are well below the competitive exclusion bound, while the dynamics in GLV systems can violate the competitive exclusion principle.
Unlike previous work on dynamical fluctuations in consumer-resource models \cite{dalmedigos2020dynamical,ladeira2019chaotic, din2021discrete, williams2004stabilization, beninca2015species}, the aMCRM does not require the introduction of explicit species-species interactions to exhibit chaotic dynamics, chaos occurs below the competitive exclusion bound, the resource carrying capacities are static, the dynamics are continuous and not discrete, and the onset of chaos requires no fine-tuning and is analyzed in high dimensions.
Additionally, our analysis works explicitly with the consumer-resource model and not the effective GLV model.

Collectively, these works suggest that non-reciprocal interactions can lead to complex, chaotic dynamics in systems with many types of species/fields. In particular, like GLV models, we also find that species and resources often jump rapidly between low and high abundances. In the future, it will be interesting to see if the methods developed in Ref.~\onlinecite{depirey2023manyspecies} in the context of GLV systems generalize to explain boom-and-bust dynamics in consumer-resource models and derive relevant correlation functions and dynamical susceptibilities.
Preliminary results suggest that other consumer resource models with non-reciprocal species-resource interactions, such as that with externally supplied resources~\cite{cui2020effect}, also exhibit chaotic dynamics; we hope to explore this in future work.
Finally, further investigations may seek to understand these phenomena in the context of ecological processes such as immigration, alternative resource dynamics \cite{cui2020effect,marsland2020mepp}, the addition of network and metabolic structure into interactions \cite{marsland2020community, dal2021resource, marcus2023local}, the inclusion of additional trophic structure~\cite{feng2023multitrophic}, and spatial and temporal structure \cite{ebrahimi2022particle}.

\textbf{Acknowledgements.}
We would like to thank Zhijie (Sarah) Feng, Claudio Chamon, and Chris Laumann for useful discussions.
Additionally, we thank the Boston University Research Computing Services for managing computational resources.
This work was funded by NIH NIGMS R35GM119461 to P.M. and the Boston University Undergraduate Research Opportunities Program to E.B.
\footnote{See Supplementary Material [url] for complete derivations, further numerical methods and results, and additional figures, which includes Refs.~[51-52].}

\nocite{Bunin2017,Datseris2018,Datseris2022,Benettin1980,Geist1990,galiSkokos2007,Datseris2018,Datseris2022,kaplan1979dimension,frederickson1983dimension,datseris2023FractalDimensions,rackauckas2017differentialequations,julia2017,tsitouras2011runge,VCABM1993,rackauckas2017differentialequations,JuliaBBO,optimizationRackauckas2023,Benettin1980,Geist1990,Datseris2022,Datseris2018,marsland2020mepp}
\bibliography{aMCRM}

\begin{thebibliography}{58}%
\makeatletter
\providecommand \@ifxundefined [1]{%
 \@ifx{#1\undefined}
}%
\providecommand \@ifnum [1]{%
 \ifnum #1\expandafter \@firstoftwo
 \else \expandafter \@secondoftwo
 \fi
}%
\providecommand \@ifx [1]{%
 \ifx #1\expandafter \@firstoftwo
 \else \expandafter \@secondoftwo
 \fi
}%
\providecommand \natexlab [1]{#1}%
\providecommand \enquote  [1]{``#1''}%
\providecommand \bibnamefont  [1]{#1}%
\providecommand \bibfnamefont [1]{#1}%
\providecommand \citenamefont [1]{#1}%
\providecommand \href@noop [0]{\@secondoftwo}%
\providecommand \href [0]{\begingroup \@sanitize@url \@href}%
\providecommand \@href[1]{\@@startlink{#1}\@@href}%
\providecommand \@@href[1]{\endgroup#1\@@endlink}%
\providecommand \@sanitize@url [0]{\catcode `\\12\catcode `\$12\catcode
  `\&12\catcode `\#12\catcode `\^12\catcode `\_12\catcode `\%12\relax}%
\providecommand \@@startlink[1]{}%
\providecommand \@@endlink[0]{}%
\providecommand \url  [0]{\begingroup\@sanitize@url \@url }%
\providecommand \@url [1]{\endgroup\@href {#1}{\urlprefix }}%
\providecommand \urlprefix  [0]{URL }%
\providecommand \Eprint [0]{\href }%
\providecommand \doibase [0]{https://doi.org/}%
\providecommand \selectlanguage [0]{\@gobble}%
\providecommand \bibinfo  [0]{\@secondoftwo}%
\providecommand \bibfield  [0]{\@secondoftwo}%
\providecommand \translation [1]{[#1]}%
\providecommand \BibitemOpen [0]{}%
\providecommand \bibitemStop [0]{}%
\providecommand \bibitemNoStop [0]{.\EOS\space}%
\providecommand \EOS [0]{\spacefactor3000\relax}%
\providecommand \BibitemShut  [1]{\csname bibitem#1\endcsname}%
\let\auto@bib@innerbib\@empty
\bibitem [{\citenamefont {Fruchart}\ \emph {et~al.}(2021)\citenamefont
  {Fruchart}, \citenamefont {Hanai}, \citenamefont {Littlewood},\ and\
  \citenamefont {Vitelli}}]{fruchart2021nonrecip}%
  \BibitemOpen
  \bibfield  {author} {\bibinfo {author} {\bibfnamefont {M.}~\bibnamefont
  {Fruchart}}, \bibinfo {author} {\bibfnamefont {R.}~\bibnamefont {Hanai}},
  \bibinfo {author} {\bibfnamefont {P.~B.}\ \bibnamefont {Littlewood}},\ and\
  \bibinfo {author} {\bibfnamefont {V.}~\bibnamefont {Vitelli}},\ }\bibfield
  {title} {\bibinfo {title} {Non-reciprocal phase transitions},\ }\href
  {https://doi.org/10.1038/s41586-021-03375-9} {\bibfield  {journal} {\bibinfo
  {journal} {Nature}\ }\textbf {\bibinfo {volume} {592}},\ \bibinfo {pages}
  {363} (\bibinfo {year} {2021})}\BibitemShut {NoStop}%
\bibitem [{\citenamefont {Ivlev}\ \emph {et~al.}(2015)\citenamefont {Ivlev},
  \citenamefont {Bartnick}, \citenamefont {Heinen}, \citenamefont {Du},
  \citenamefont {Nosenko},\ and\ \citenamefont
  {L\"owen}}]{ivlev2015brokenthird}%
  \BibitemOpen
  \bibfield  {author} {\bibinfo {author} {\bibfnamefont {A.~V.}\ \bibnamefont
  {Ivlev}}, \bibinfo {author} {\bibfnamefont {J.}~\bibnamefont {Bartnick}},
  \bibinfo {author} {\bibfnamefont {M.}~\bibnamefont {Heinen}}, \bibinfo
  {author} {\bibfnamefont {C.-R.}\ \bibnamefont {Du}}, \bibinfo {author}
  {\bibfnamefont {V.}~\bibnamefont {Nosenko}},\ and\ \bibinfo {author}
  {\bibfnamefont {H.}~\bibnamefont {L\"owen}},\ }\bibfield  {title} {\bibinfo
  {title} {Statistical mechanics where newton's third law is broken},\ }\href
  {https://doi.org/10.1103/PhysRevX.5.011035} {\bibfield  {journal} {\bibinfo
  {journal} {Phys. Rev. X}\ }\textbf {\bibinfo {volume} {5}},\ \bibinfo {pages}
  {011035} (\bibinfo {year} {2015})}\BibitemShut {NoStop}%
\bibitem [{\citenamefont {Sompolinsky}\ and\ \citenamefont
  {Kanter}(1986)}]{sompolinsky1986temporal}%
  \BibitemOpen
  \bibfield  {author} {\bibinfo {author} {\bibfnamefont {H.}~\bibnamefont
  {Sompolinsky}}\ and\ \bibinfo {author} {\bibfnamefont {I.}~\bibnamefont
  {Kanter}},\ }\bibfield  {title} {\bibinfo {title} {Temporal association in
  asymmetric neural networks},\ }\href
  {https://doi.org/10.1103/PhysRevLett.57.2861} {\bibfield  {journal} {\bibinfo
   {journal} {Phys. Rev. Lett.}\ }\textbf {\bibinfo {volume} {57}},\ \bibinfo
  {pages} {2861} (\bibinfo {year} {1986})}\BibitemShut {NoStop}%
\bibitem [{\citenamefont {Sompolinsky}\ \emph {et~al.}(1988)\citenamefont
  {Sompolinsky}, \citenamefont {Crisanti},\ and\ \citenamefont
  {Sommers}}]{sompolinsky1988chaos}%
  \BibitemOpen
  \bibfield  {author} {\bibinfo {author} {\bibfnamefont {H.}~\bibnamefont
  {Sompolinsky}}, \bibinfo {author} {\bibfnamefont {A.}~\bibnamefont
  {Crisanti}},\ and\ \bibinfo {author} {\bibfnamefont {H.~J.}\ \bibnamefont
  {Sommers}},\ }\bibfield  {title} {\bibinfo {title} {Chaos in random neural
  networks},\ }\href {https://doi.org/10.1103/PhysRevLett.61.259} {\bibfield
  {journal} {\bibinfo  {journal} {Phys. Rev. Lett.}\ }\textbf {\bibinfo
  {volume} {61}},\ \bibinfo {pages} {259} (\bibinfo {year} {1988})}\BibitemShut
  {NoStop}%
\bibitem [{\citenamefont {G{\'o}mez-Nava}\ \emph {et~al.}(2022)\citenamefont
  {G{\'o}mez-Nava}, \citenamefont {Bon},\ and\ \citenamefont
  {Peruani}}]{gomex2011sheep}%
  \BibitemOpen
  \bibfield  {author} {\bibinfo {author} {\bibfnamefont {L.}~\bibnamefont
  {G{\'o}mez-Nava}}, \bibinfo {author} {\bibfnamefont {R.}~\bibnamefont
  {Bon}},\ and\ \bibinfo {author} {\bibfnamefont {F.}~\bibnamefont {Peruani}},\
  }\bibfield  {title} {\bibinfo {title} {Intermittent collective motion in
  sheep results from alternating the role of leader and follower},\ }\href
  {https://doi.org/10.1038/s41567-022-01769-8} {\bibfield  {journal} {\bibinfo
  {journal} {Nature Physics}\ }\textbf {\bibinfo {volume} {18}},\ \bibinfo
  {pages} {1494} (\bibinfo {year} {2022})}\BibitemShut {NoStop}%
\bibitem [{\citenamefont {MAY}(1972)}]{may1972stability}%
  \BibitemOpen
  \bibfield  {author} {\bibinfo {author} {\bibfnamefont {R.~M.}\ \bibnamefont
  {MAY}},\ }\bibfield  {title} {\bibinfo {title} {Will a large complex system
  be stable?},\ }\href {https://doi.org/10.1038/238413a0} {\bibfield  {journal}
  {\bibinfo  {journal} {Nature}\ }\textbf {\bibinfo {volume} {238}},\ \bibinfo
  {pages} {413} (\bibinfo {year} {1972})}\BibitemShut {NoStop}%
\bibitem [{\citenamefont {Yoshino}\ \emph {et~al.}(2007)\citenamefont
  {Yoshino}, \citenamefont {Galla},\ and\ \citenamefont
  {Tokita}}]{yoshino2007ecostability}%
  \BibitemOpen
  \bibfield  {author} {\bibinfo {author} {\bibfnamefont {Y.}~\bibnamefont
  {Yoshino}}, \bibinfo {author} {\bibfnamefont {T.}~\bibnamefont {Galla}},\
  and\ \bibinfo {author} {\bibfnamefont {K.}~\bibnamefont {Tokita}},\
  }\bibfield  {title} {\bibinfo {title} {Statistical mechanics and stability of
  a model eco-system},\ }\href
  {https://doi.org/10.1088/1742-5468/2007/09/P09003} {\bibfield  {journal}
  {\bibinfo  {journal} {Journal of Statistical Mechanics: Theory and
  Experiment}\ }\textbf {\bibinfo {volume} {2007}},\ \bibinfo {pages} {P09003}
  (\bibinfo {year} {2007})}\BibitemShut {NoStop}%
\bibitem [{\citenamefont {Diederich}\ and\ \citenamefont
  {Opper}(1989)}]{diederich1989replicators}%
  \BibitemOpen
  \bibfield  {author} {\bibinfo {author} {\bibfnamefont {S.}~\bibnamefont
  {Diederich}}\ and\ \bibinfo {author} {\bibfnamefont {M.}~\bibnamefont
  {Opper}},\ }\bibfield  {title} {\bibinfo {title} {Replicators with random
  interactions: A solvable model},\ }\href
  {https://doi.org/10.1103/PhysRevA.39.4333} {\bibfield  {journal} {\bibinfo
  {journal} {Phys. Rev. A}\ }\textbf {\bibinfo {volume} {39}},\ \bibinfo
  {pages} {4333} (\bibinfo {year} {1989})}\BibitemShut {NoStop}%
\bibitem [{\citenamefont {Mahadevan}\ \emph {et~al.}(2023)\citenamefont
  {Mahadevan}, \citenamefont {Pearce},\ and\ \citenamefont
  {Fisher}}]{mahadevan2023spatiotemporal}%
  \BibitemOpen
  \bibfield  {author} {\bibinfo {author} {\bibfnamefont {A.}~\bibnamefont
  {Mahadevan}}, \bibinfo {author} {\bibfnamefont {M.~T.}\ \bibnamefont
  {Pearce}},\ and\ \bibinfo {author} {\bibfnamefont {D.~S.}\ \bibnamefont
  {Fisher}},\ }\bibfield  {title} {\bibinfo {title} {Spatiotemporal ecological
  chaos enables gradual evolutionary diversification without niches or
  tradeoffs},\ }\href {https://doi.org/10.7554/eLife.82734} {\bibfield
  {journal} {\bibinfo  {journal} {eLife}\ }\textbf {\bibinfo {volume} {12}},\
  \bibinfo {pages} {e82734} (\bibinfo {year} {2023})}\BibitemShut {NoStop}%
\bibitem [{\citenamefont {Fisher}\ and\ \citenamefont
  {Mehta}(2014)}]{fisher2014nicheneutral}%
  \BibitemOpen
  \bibfield  {author} {\bibinfo {author} {\bibfnamefont {C.~K.}\ \bibnamefont
  {Fisher}}\ and\ \bibinfo {author} {\bibfnamefont {P.}~\bibnamefont {Mehta}},\
  }\bibfield  {title} {\bibinfo {title} {The transition between the niche and
  neutral regimes in ecology},\ }\href
  {https://doi.org/10.1073/pnas.1405637111} {\bibfield  {journal} {\bibinfo
  {journal} {Proceedings of the National Academy of Sciences}\ }\textbf
  {\bibinfo {volume} {111}},\ \bibinfo {pages} {13111} (\bibinfo {year}
  {2014})},\ \Eprint
  {https://arxiv.org/abs/https://www.pnas.org/doi/pdf/10.1073/pnas.1405637111}
  {https://www.pnas.org/doi/pdf/10.1073/pnas.1405637111} \BibitemShut {NoStop}%
\bibitem [{\citenamefont {Dickens}\ \emph {et~al.}(2016)\citenamefont
  {Dickens}, \citenamefont {Fisher},\ and\ \citenamefont
  {Mehta}}]{dickens2016analytically}%
  \BibitemOpen
  \bibfield  {author} {\bibinfo {author} {\bibfnamefont {B.}~\bibnamefont
  {Dickens}}, \bibinfo {author} {\bibfnamefont {C.~K.}\ \bibnamefont
  {Fisher}},\ and\ \bibinfo {author} {\bibfnamefont {P.}~\bibnamefont
  {Mehta}},\ }\bibfield  {title} {\bibinfo {title} {Analytically tractable
  model for community ecology with many species},\ }\href
  {https://doi.org/10.1103/PhysRevE.94.022423} {\bibfield  {journal} {\bibinfo
  {journal} {Phys. Rev. E}\ }\textbf {\bibinfo {volume} {94}},\ \bibinfo
  {pages} {022423} (\bibinfo {year} {2016})}\BibitemShut {NoStop}%
\bibitem [{\citenamefont {Marsland}\ \emph
  {et~al.}(2020{\natexlab{a}})\citenamefont {Marsland}, \citenamefont {Cui},\
  and\ \citenamefont {Mehta}}]{marsland2020microbial}%
  \BibitemOpen
  \bibfield  {author} {\bibinfo {author} {\bibfnamefont {R.}~\bibnamefont
  {Marsland}}, \bibinfo {author} {\bibfnamefont {W.}~\bibnamefont {Cui}},\ and\
  \bibinfo {author} {\bibfnamefont {P.}~\bibnamefont {Mehta}},\ }\bibfield
  {title} {\bibinfo {title} {A minimal model for microbial biodiversity can
  reproduce experimentally observed ecological patterns},\ }\href
  {https://doi.org/10.1038/s41598-020-60130-2} {\bibfield  {journal} {\bibinfo
  {journal} {Scientific Reports}\ }\textbf {\bibinfo {volume} {10}},\ \bibinfo
  {pages} {3308} (\bibinfo {year} {2020}{\natexlab{a}})}\BibitemShut {NoStop}%
\bibitem [{\citenamefont {Advani}\ \emph {et~al.}(2018)\citenamefont {Advani},
  \citenamefont {Bunin},\ and\ \citenamefont {Mehta}}]{advani2018MCRMCavity}%
  \BibitemOpen
  \bibfield  {author} {\bibinfo {author} {\bibfnamefont {M.}~\bibnamefont
  {Advani}}, \bibinfo {author} {\bibfnamefont {G.}~\bibnamefont {Bunin}},\ and\
  \bibinfo {author} {\bibfnamefont {P.}~\bibnamefont {Mehta}},\ }\bibfield
  {title} {\bibinfo {title} {Statistical physics of community ecology: a cavity
  solution to macarthur's consumer resource model},\ }\href
  {https://doi.org/10.1088/1742-5468/aab04e} {\bibfield  {journal} {\bibinfo
  {journal} {Journal of Statistical Mechanics: Theory and Experiment}\ ,\
  \bibinfo {pages} {033406}} (\bibinfo {year} {2018})}\BibitemShut {NoStop}%
\bibitem [{\citenamefont {Cui}\ \emph {et~al.}(2021)\citenamefont {Cui},
  \citenamefont {Marsland~III},\ and\ \citenamefont {Mehta}}]{cui2021diverse}%
  \BibitemOpen
  \bibfield  {author} {\bibinfo {author} {\bibfnamefont {W.}~\bibnamefont
  {Cui}}, \bibinfo {author} {\bibfnamefont {R.}~\bibnamefont {Marsland~III}},\
  and\ \bibinfo {author} {\bibfnamefont {P.}~\bibnamefont {Mehta}},\ }\bibfield
   {title} {\bibinfo {title} {Diverse communities behave like typical random
  ecosystems},\ }\href@noop {} {\bibfield  {journal} {\bibinfo  {journal}
  {Physical Review E}\ }\textbf {\bibinfo {volume} {104}},\ \bibinfo {pages}
  {034416} (\bibinfo {year} {2021})}\BibitemShut {NoStop}%
\bibitem [{\citenamefont {Altieri}\ \emph {et~al.}(2021)\citenamefont
  {Altieri}, \citenamefont {Roy}, \citenamefont {Cammarota},\ and\
  \citenamefont {Biroli}}]{roy2021glassyLV}%
  \BibitemOpen
  \bibfield  {author} {\bibinfo {author} {\bibfnamefont {A.}~\bibnamefont
  {Altieri}}, \bibinfo {author} {\bibfnamefont {F.}~\bibnamefont {Roy}},
  \bibinfo {author} {\bibfnamefont {C.}~\bibnamefont {Cammarota}},\ and\
  \bibinfo {author} {\bibfnamefont {G.}~\bibnamefont {Biroli}},\ }\bibfield
  {title} {\bibinfo {title} {Properties of equilibria and glassy phases of the
  random lotka-volterra model with demographic noise},\ }\href
  {https://doi.org/10.1103/PhysRevLett.126.258301} {\bibfield  {journal}
  {\bibinfo  {journal} {Phys. Rev. Lett.}\ }\textbf {\bibinfo {volume} {126}},\
  \bibinfo {pages} {258301} (\bibinfo {year} {2021})}\BibitemShut {NoStop}%
\bibitem [{\citenamefont {Marsland}\ \emph
  {et~al.}(2020{\natexlab{b}})\citenamefont {Marsland}, \citenamefont {Cui},\
  and\ \citenamefont {Mehta}}]{marsland2020mepp}%
  \BibitemOpen
  \bibfield  {author} {\bibinfo {author} {\bibfnamefont {R.}~\bibnamefont
  {Marsland}}, \bibinfo {author} {\bibfnamefont {W.}~\bibnamefont {Cui}},\ and\
  \bibinfo {author} {\bibfnamefont {P.}~\bibnamefont {Mehta}},\ }\bibfield
  {title} {\bibinfo {title} {The minimum environmental perturbation principle:
  A new perspective on niche theory},\ }\href {https://doi.org/10.1086/710093}
  {\bibfield  {journal} {\bibinfo  {journal} {The American Naturalist}\
  }\textbf {\bibinfo {volume} {196}},\ \bibinfo {pages} {291} (\bibinfo {year}
  {2020}{\natexlab{b}})}\BibitemShut {NoStop}%
\bibitem [{\citenamefont {Gatto}(1990)}]{gatto1990general}%
  \BibitemOpen
  \bibfield  {author} {\bibinfo {author} {\bibfnamefont {M.}~\bibnamefont
  {Gatto}},\ }\bibfield  {title} {\bibinfo {title} {A general minimum principle
  for competing populations: Some ecological and evolutionary consequences},\
  }\href {https://doi.org/https://doi.org/10.1016/0040-5809(90)90044-V}
  {\bibfield  {journal} {\bibinfo  {journal} {Theoretical Population Biology}\
  }\textbf {\bibinfo {volume} {37}},\ \bibinfo {pages} {369} (\bibinfo {year}
  {1990})}\BibitemShut {NoStop}%
\bibitem [{\citenamefont {Case}(1980)}]{case1980minimization}%
  \BibitemOpen
  \bibfield  {author} {\bibinfo {author} {\bibfnamefont {T.~J.}\ \bibnamefont
  {Case}},\ }\bibfield  {title} {\bibinfo {title} {Macarthur's minimization
  principle: A footnote},\ }\href {https://doi.org/10.1086/283550} {\bibfield
  {journal} {\bibinfo  {journal} {The American Naturalist}\ }\textbf {\bibinfo
  {volume} {115}},\ \bibinfo {pages} {133} (\bibinfo {year} {1980})},\ \Eprint
  {https://arxiv.org/abs/https://doi.org/10.1086/283550}
  {https://doi.org/10.1086/283550} \BibitemShut {NoStop}%
\bibitem [{\citenamefont {Chesson}(1990)}]{chesson1990macarthur}%
  \BibitemOpen
  \bibfield  {author} {\bibinfo {author} {\bibfnamefont {P.}~\bibnamefont
  {Chesson}},\ }\bibfield  {title} {\bibinfo {title} {Macarthur's
  consumer-resource model},\ }\href@noop {} {\bibfield  {journal} {\bibinfo
  {journal} {Theoretical Population Biology}\ }\textbf {\bibinfo {volume}
  {37}},\ \bibinfo {pages} {26} (\bibinfo {year} {1990})}\BibitemShut {NoStop}%
\bibitem [{\citenamefont {Pearce}\ \emph {et~al.}(2020)\citenamefont {Pearce},
  \citenamefont {Agarwala},\ and\ \citenamefont
  {Fisher}}]{pearce2020stabilization}%
  \BibitemOpen
  \bibfield  {author} {\bibinfo {author} {\bibfnamefont {M.~T.}\ \bibnamefont
  {Pearce}}, \bibinfo {author} {\bibfnamefont {A.}~\bibnamefont {Agarwala}},\
  and\ \bibinfo {author} {\bibfnamefont {D.~S.}\ \bibnamefont {Fisher}},\
  }\bibfield  {title} {\bibinfo {title} {Stabilization of extensive fine-scale
  diversity by ecologically driven spatiotemporal chaos},\ }\href
  {https://doi.org/10.1073/pnas.1915313117} {\bibfield  {journal} {\bibinfo
  {journal} {Proceedings of the National Academy of Sciences}\ }\textbf
  {\bibinfo {volume} {117}},\ \bibinfo {pages} {14572} (\bibinfo {year}
  {2020})},\ \Eprint
  {https://arxiv.org/abs/https://www.pnas.org/doi/pdf/10.1073/pnas.1915313117}
  {https://www.pnas.org/doi/pdf/10.1073/pnas.1915313117} \BibitemShut {NoStop}%
\bibitem [{\citenamefont {Roy}\ \emph {et~al.}(2020)\citenamefont {Roy},
  \citenamefont {Barbier}, \citenamefont {Biroli},\ and\ \citenamefont
  {Bunin}}]{roy2020persistent}%
  \BibitemOpen
  \bibfield  {author} {\bibinfo {author} {\bibfnamefont {F.}~\bibnamefont
  {Roy}}, \bibinfo {author} {\bibfnamefont {M.}~\bibnamefont {Barbier}},
  \bibinfo {author} {\bibfnamefont {G.}~\bibnamefont {Biroli}},\ and\ \bibinfo
  {author} {\bibfnamefont {G.}~\bibnamefont {Bunin}},\ }\bibfield  {title}
  {\bibinfo {title} {Complex interactions can create persistent fluctuations in
  high-diversity ecosystems},\ }\href
  {https://doi.org/10.1371/journal.pcbi.1007827} {\bibfield  {journal}
  {\bibinfo  {journal} {PLOS Computational Biology}\ }\textbf {\bibinfo
  {volume} {16}},\ \bibinfo {pages} {1} (\bibinfo {year} {2020})}\BibitemShut
  {NoStop}%
\bibitem [{\citenamefont {Hu}\ \emph {et~al.}(2022)\citenamefont {Hu},
  \citenamefont {Amor}, \citenamefont {Barbier}, \citenamefont {Bunin},\ and\
  \citenamefont {Gore}}]{hu2022emergent}%
  \BibitemOpen
  \bibfield  {author} {\bibinfo {author} {\bibfnamefont {J.}~\bibnamefont
  {Hu}}, \bibinfo {author} {\bibfnamefont {D.~R.}\ \bibnamefont {Amor}},
  \bibinfo {author} {\bibfnamefont {M.}~\bibnamefont {Barbier}}, \bibinfo
  {author} {\bibfnamefont {G.}~\bibnamefont {Bunin}},\ and\ \bibinfo {author}
  {\bibfnamefont {J.}~\bibnamefont {Gore}},\ }\bibfield  {title} {\bibinfo
  {title} {Emergent phases of ecological diversity and dynamics mapped in
  microcosms},\ }\href@noop {} {\bibfield  {journal} {\bibinfo  {journal}
  {Science}\ }\textbf {\bibinfo {volume} {378}},\ \bibinfo {pages} {85}
  (\bibinfo {year} {2022})}\BibitemShut {NoStop}%
\bibitem [{\citenamefont {Ros}\ \emph {et~al.}(2023{\natexlab{a}})\citenamefont
  {Ros}, \citenamefont {Roy}, \citenamefont {Biroli}, \citenamefont {Bunin},\
  and\ \citenamefont {Turner}}]{ros2023rGLVtypical}%
  \BibitemOpen
  \bibfield  {author} {\bibinfo {author} {\bibfnamefont {V.}~\bibnamefont
  {Ros}}, \bibinfo {author} {\bibfnamefont {F.}~\bibnamefont {Roy}}, \bibinfo
  {author} {\bibfnamefont {G.}~\bibnamefont {Biroli}}, \bibinfo {author}
  {\bibfnamefont {G.}~\bibnamefont {Bunin}},\ and\ \bibinfo {author}
  {\bibfnamefont {A.~M.}\ \bibnamefont {Turner}},\ }\bibfield  {title}
  {\bibinfo {title} {Generalized lotka-volterra equations with random,
  nonreciprocal interactions: The typical number of equilibria},\ }\href
  {https://doi.org/10.1103/PhysRevLett.130.257401} {\bibfield  {journal}
  {\bibinfo  {journal} {Phys. Rev. Lett.}\ }\textbf {\bibinfo {volume} {130}},\
  \bibinfo {pages} {257401} (\bibinfo {year} {2023}{\natexlab{a}})}\BibitemShut
  {NoStop}%
\bibitem [{\citenamefont {Ros}\ \emph {et~al.}(2023{\natexlab{b}})\citenamefont
  {Ros}, \citenamefont {Roy}, \citenamefont {Biroli},\ and\ \citenamefont
  {Bunin}}]{ros2023rGLVquenched}%
  \BibitemOpen
  \bibfield  {author} {\bibinfo {author} {\bibfnamefont {V.}~\bibnamefont
  {Ros}}, \bibinfo {author} {\bibfnamefont {F.}~\bibnamefont {Roy}}, \bibinfo
  {author} {\bibfnamefont {G.}~\bibnamefont {Biroli}},\ and\ \bibinfo {author}
  {\bibfnamefont {G.}~\bibnamefont {Bunin}},\ }\bibfield  {title} {\bibinfo
  {title} {Quenched complexity of equilibria for asymmetric generalized
  lotka–volterra equations},\ }\href
  {https://doi.org/10.1088/1751-8121/ace00f} {\bibfield  {journal} {\bibinfo
  {journal} {Journal of Physics A: Mathematical and Theoretical}\ }\textbf
  {\bibinfo {volume} {56}},\ \bibinfo {pages} {305003} (\bibinfo {year}
  {2023}{\natexlab{b}})}\BibitemShut {NoStop}%
\bibitem [{\citenamefont {Arnoulx~de Pirey}\ and\ \citenamefont
  {Bunin}(2023)}]{pirey2023aging}%
  \BibitemOpen
  \bibfield  {author} {\bibinfo {author} {\bibfnamefont {T.}~\bibnamefont
  {Arnoulx~de Pirey}}\ and\ \bibinfo {author} {\bibfnamefont {G.}~\bibnamefont
  {Bunin}},\ }\bibfield  {title} {\bibinfo {title} {Aging by near-extinctions
  in many-variable interacting populations},\ }\href
  {https://doi.org/10.1103/PhysRevLett.130.098401} {\bibfield  {journal}
  {\bibinfo  {journal} {Phys. Rev. Lett.}\ }\textbf {\bibinfo {volume} {130}},\
  \bibinfo {pages} {098401} (\bibinfo {year} {2023})}\BibitemShut {NoStop}%
\bibitem [{\citenamefont {de~Pirey}\ and\ \citenamefont
  {Bunin}(2023)}]{depirey2023manyspecies}%
  \BibitemOpen
  \bibfield  {author} {\bibinfo {author} {\bibfnamefont {T.~A.}\ \bibnamefont
  {de~Pirey}}\ and\ \bibinfo {author} {\bibfnamefont {G.}~\bibnamefont
  {Bunin}},\ }\href@noop {} {\bibinfo {title} {Many-species ecological
  fluctuations as a jump process from the brink of extinction}} (\bibinfo
  {year} {2023}),\ \Eprint {https://arxiv.org/abs/2306.13634} {arXiv:2306.13634
  [q-bio.PE]} \BibitemShut {NoStop}%
\bibitem [{\citenamefont {MacArthur}\ and\ \citenamefont
  {Levins}(1967)}]{macarthurlevins1967}%
  \BibitemOpen
  \bibfield  {author} {\bibinfo {author} {\bibfnamefont {R.}~\bibnamefont
  {MacArthur}}\ and\ \bibinfo {author} {\bibfnamefont {R.}~\bibnamefont
  {Levins}},\ }\bibfield  {title} {\bibinfo {title} {The limiting similarity,
  convergence, and divergence of coexisting species},\ }\href@noop {}
  {\bibfield  {journal} {\bibinfo  {journal} {The american naturalist}\
  }\textbf {\bibinfo {volume} {101}},\ \bibinfo {pages} {377} (\bibinfo {year}
  {1967})}\BibitemShut {NoStop}%
\bibitem [{\citenamefont {Macarthur}(1970)}]{Macarthur1970}%
  \BibitemOpen
  \bibfield  {author} {\bibinfo {author} {\bibfnamefont {R.}~\bibnamefont
  {Macarthur}},\ }\bibfield  {title} {\bibinfo {title} {Species packing and
  competitive equilibrium for many species*},\ }\href@noop {} {\bibfield
  {journal} {\bibinfo  {journal} {POPULATION BIOLOGY}\ }\textbf {\bibinfo
  {volume} {1}},\ \bibinfo {pages} {11} (\bibinfo {year} {1970})}\BibitemShut
  {NoStop}%
\bibitem [{\citenamefont {Tilman}(1982)}]{tilman1982resource}%
  \BibitemOpen
  \bibfield  {author} {\bibinfo {author} {\bibfnamefont {D.}~\bibnamefont
  {Tilman}},\ }\href@noop {} {\emph {\bibinfo {title} {Resource competition and
  community structure}}}\ (\bibinfo  {publisher} {Princeton university press},\
  \bibinfo {year} {1982})\BibitemShut {NoStop}%
\bibitem [{\citenamefont {Chase}\ and\ \citenamefont
  {Leibold}(2009)}]{chase2009ecological}%
  \BibitemOpen
  \bibfield  {author} {\bibinfo {author} {\bibfnamefont {J.~M.}\ \bibnamefont
  {Chase}}\ and\ \bibinfo {author} {\bibfnamefont {M.~A.}\ \bibnamefont
  {Leibold}},\ }\href@noop {} {\emph {\bibinfo {title} {Ecological niches:
  linking classical and contemporary approaches}}}\ (\bibinfo  {publisher}
  {University of Chicago Press},\ \bibinfo {year} {2009})\BibitemShut {NoStop}%
\bibitem [{\citenamefont {Bunin}(2017)}]{Bunin2017}%
  \BibitemOpen
  \bibfield  {author} {\bibinfo {author} {\bibfnamefont {G.}~\bibnamefont
  {Bunin}},\ }\bibfield  {title} {\bibinfo {title} {Ecological communities with
  lotka-volterra dynamics},\ }\href
  {https://doi.org/10.1103/PhysRevE.95.042414} {\bibfield  {journal} {\bibinfo
  {journal} {Phys. Rev. E}\ }\textbf {\bibinfo {volume} {95}},\ \bibinfo
  {pages} {042414} (\bibinfo {year} {2017})}\BibitemShut {NoStop}%
\bibitem [{\citenamefont {Nishimori}(2001)}]{nishimori2001}%
  \BibitemOpen
  \bibfield  {author} {\bibinfo {author} {\bibfnamefont {H.}~\bibnamefont
  {Nishimori}},\ }\href {https://books.google.com/books?id=nO0T1VzfhZcC} {\emph
  {\bibinfo {title} {Statistical Physics of Spin Glasses and Information
  Processing: An Introduction}}},\ International series of monographs on
  physics\ (\bibinfo  {publisher} {Oxford University Press},\ \bibinfo {year}
  {2001})\BibitemShut {NoStop}%
\bibitem [{\citenamefont {Rackauckas}\ and\ \citenamefont
  {Nie}(2017)}]{rackauckas2017differentialequations}%
  \BibitemOpen
  \bibfield  {author} {\bibinfo {author} {\bibfnamefont {C.}~\bibnamefont
  {Rackauckas}}\ and\ \bibinfo {author} {\bibfnamefont {Q.}~\bibnamefont
  {Nie}},\ }\bibfield  {title} {\bibinfo {title} {Differential{E}quations.jl--a
  performant and feature-rich ecosystem for solving differential equations in
  {J}ulia},\ }\href@noop {} {\bibfield  {journal} {\bibinfo  {journal} {Journal
  of Open Research Software}\ }\textbf {\bibinfo {volume} {5}} (\bibinfo {year}
  {2017})}\BibitemShut {NoStop}%
\bibitem [{\citenamefont {Bezanson}\ \emph {et~al.}(2017)\citenamefont
  {Bezanson}, \citenamefont {Edelman}, \citenamefont {Karpinski},\ and\
  \citenamefont {Shah}}]{julia2017}%
  \BibitemOpen
  \bibfield  {author} {\bibinfo {author} {\bibfnamefont {J.}~\bibnamefont
  {Bezanson}}, \bibinfo {author} {\bibfnamefont {A.}~\bibnamefont {Edelman}},
  \bibinfo {author} {\bibfnamefont {S.}~\bibnamefont {Karpinski}},\ and\
  \bibinfo {author} {\bibfnamefont {V.~B.}\ \bibnamefont {Shah}},\ }\href
  {https://doi.org/10.1137/141000671} {\bibinfo {title} {Julia: A fresh
  approach to numerical computing}} (\bibinfo {year} {2017})\BibitemShut
  {NoStop}%
\bibitem [{\citenamefont {Tsitouras}(2011)}]{tsitouras2011runge}%
  \BibitemOpen
  \bibfield  {author} {\bibinfo {author} {\bibfnamefont {C.}~\bibnamefont
  {Tsitouras}},\ }\bibfield  {title} {\bibinfo {title} {Runge–kutta pairs of
  order 5 (4) satisfying only the first column simplifying assumption},\
  }\href@noop {} {\bibfield  {journal} {\bibinfo  {journal} {Computers \&
  Mathematics with Applications}\ }\textbf {\bibinfo {volume} {62}},\ \bibinfo
  {pages} {770–775} (\bibinfo {year} {2011})}\BibitemShut {NoStop}%
\bibitem [{\citenamefont {Ernst~Hairer}(1993)}]{VCABM1993}%
  \BibitemOpen
  \bibfield  {author} {\bibinfo {author} {\bibfnamefont {S.~P.~N.}\
  \bibnamefont {Ernst~Hairer}, \bibfnamefont {Gerhard~Wanner}},\ }\href
  {https://doi.org/10.1007/978-3-540-78862-1} {\emph {\bibinfo {title} {Solving
  Ordinary Differential Equations I}}}\ (\bibinfo  {publisher} {Springer Berlin
  Heidelberg},\ \bibinfo {year} {1993})\BibitemShut {NoStop}%
\bibitem [{\citenamefont {SciML}(2022)}]{JuliaBBO}%
  \BibitemOpen
  \bibfield  {author} {\bibinfo {author} {\bibfnamefont {J.}~\bibnamefont
  {SciML}},\ }\href {https://github.com/robertfeldt/BlackBoxOptim.jl} {\bibinfo
  {title} {Blackboxoptim.jl}},\ \bibinfo {howpublished} {GitHub} (\bibinfo
  {year} {2022})\BibitemShut {NoStop}%
\bibitem [{\citenamefont {Dixit}\ and\ \citenamefont
  {Rackauckas}(2023)}]{optimizationRackauckas2023}%
  \BibitemOpen
  \bibfield  {author} {\bibinfo {author} {\bibfnamefont {V.~K.}\ \bibnamefont
  {Dixit}}\ and\ \bibinfo {author} {\bibfnamefont {C.}~\bibnamefont
  {Rackauckas}},\ }\href {https://doi.org/10.5281/zenodo.7738525} {\bibinfo
  {title} {Optimization.jl: A unified optimization package}} (\bibinfo {year}
  {2023})\BibitemShut {NoStop}%
\bibitem [{\citenamefont {Benettin}\ \emph {et~al.}(1980)\citenamefont
  {Benettin}, \citenamefont {Galgani}, \citenamefont {Giorgilli},\ and\
  \citenamefont {Strelcyn}}]{Benettin1980}%
  \BibitemOpen
  \bibfield  {author} {\bibinfo {author} {\bibfnamefont {G.}~\bibnamefont
  {Benettin}}, \bibinfo {author} {\bibfnamefont {L.}~\bibnamefont {Galgani}},
  \bibinfo {author} {\bibfnamefont {A.}~\bibnamefont {Giorgilli}},\ and\
  \bibinfo {author} {\bibfnamefont {J.-M.}\ \bibnamefont {Strelcyn}},\
  }\bibfield  {title} {\bibinfo {title} {Lyapunov characteristic exponents for
  smooth dynamical systems and for hamiltonian systems; a method for computing
  all of them. part 1: Theory},\ }\href {https://doi.org/10.1007/BF02128236}
  {\bibfield  {journal} {\bibinfo  {journal} {Meccanica}\ }\textbf {\bibinfo
  {volume} {15}},\ \bibinfo {pages} {9} (\bibinfo {year} {1980})}\BibitemShut
  {NoStop}%
\bibitem [{\citenamefont {Geist}\ \emph {et~al.}(1990)\citenamefont {Geist},
  \citenamefont {Parlitz},\ and\ \citenamefont {Lauterborn}}]{Geist1990}%
  \BibitemOpen
  \bibfield  {author} {\bibinfo {author} {\bibfnamefont {K.}~\bibnamefont
  {Geist}}, \bibinfo {author} {\bibfnamefont {U.}~\bibnamefont {Parlitz}},\
  and\ \bibinfo {author} {\bibfnamefont {W.}~\bibnamefont {Lauterborn}},\
  }\bibfield  {title} {\bibinfo {title} {{Comparison of Different Methods for
  Computing Lyapunov Exponents}},\ }\href {https://doi.org/10.1143/PTP.83.875}
  {\bibfield  {journal} {\bibinfo  {journal} {Progress of Theoretical Physics}\
  }\textbf {\bibinfo {volume} {83}},\ \bibinfo {pages} {875} (\bibinfo {year}
  {1990})},\ \Eprint
  {https://arxiv.org/abs/https://academic.oup.com/ptp/article-pdf/83/5/875/5302061/83-5-875.pdf}
  {https://academic.oup.com/ptp/article-pdf/83/5/875/5302061/83-5-875.pdf}
  \BibitemShut {NoStop}%
\bibitem [{\citenamefont {Datseris}\ and\ \citenamefont
  {Parlitz}(2022)}]{Datseris2022}%
  \BibitemOpen
  \bibfield  {author} {\bibinfo {author} {\bibfnamefont {G.}~\bibnamefont
  {Datseris}}\ and\ \bibinfo {author} {\bibfnamefont {U.}~\bibnamefont
  {Parlitz}},\ }\href {https://doi.org/10.1007/978-3-030-91032-7} {\emph
  {\bibinfo {title} {Nonlinear Dynamics}}}\ (\bibinfo  {publisher} {Springer
  International Publishing},\ \bibinfo {year} {2022})\BibitemShut {NoStop}%
\bibitem [{\citenamefont {Datseris}(2018)}]{Datseris2018}%
  \BibitemOpen
  \bibfield  {author} {\bibinfo {author} {\bibfnamefont {G.}~\bibnamefont
  {Datseris}},\ }\bibfield  {title} {\bibinfo {title} {Dynamicalsystems.jl: A
  julia software library for chaos and nonlinear dynamics},\ }\href
  {https://doi.org/10.21105/joss.00598} {\bibfield  {journal} {\bibinfo
  {journal} {Journal of Open Source Software}\ }\textbf {\bibinfo {volume}
  {3}},\ \bibinfo {pages} {598} (\bibinfo {year} {2018})}\BibitemShut {NoStop}%
\bibitem [{\citenamefont {Skokos}\ \emph {et~al.}(2007)\citenamefont {Skokos},
  \citenamefont {Bountis},\ and\ \citenamefont
  {Antonopoulos}}]{galiSkokos2007}%
  \BibitemOpen
  \bibfield  {author} {\bibinfo {author} {\bibfnamefont {C.}~\bibnamefont
  {Skokos}}, \bibinfo {author} {\bibfnamefont {T.}~\bibnamefont {Bountis}},\
  and\ \bibinfo {author} {\bibfnamefont {C.}~\bibnamefont {Antonopoulos}},\
  }\bibfield  {title} {\bibinfo {title} {Geometrical properties of local
  dynamics in hamiltonian systems: The generalized alignment index (gali)
  method},\ }\href
  {https://doi.org/https://doi.org/10.1016/j.physd.2007.04.004} {\bibfield
  {journal} {\bibinfo  {journal} {Physica D: Nonlinear Phenomena}\ }\textbf
  {\bibinfo {volume} {231}},\ \bibinfo {pages} {30} (\bibinfo {year}
  {2007})}\BibitemShut {NoStop}%
\bibitem [{\citenamefont {Kaplan}\ and\ \citenamefont
  {Yorke}(1979)}]{kaplan1979dimension}%
  \BibitemOpen
  \bibfield  {author} {\bibinfo {author} {\bibfnamefont {J.~L.}\ \bibnamefont
  {Kaplan}}\ and\ \bibinfo {author} {\bibfnamefont {J.~A.}\ \bibnamefont
  {Yorke}},\ }\bibfield  {title} {\bibinfo {title} {Chaotic behavior of
  multidimensional difference equations},\ }in\ \href@noop {} {\emph {\bibinfo
  {booktitle} {Functional differential equations and approximations of fixed
  points. Lecture notes in mathematics Vol. 730}}},\ \bibinfo {editor} {edited
  by\ \bibinfo {editor} {\bibfnamefont {H.~O.}\ \bibnamefont {Peitgen}}\ and\
  \bibinfo {editor} {\bibfnamefont {H.~O.}\ \bibnamefont {Walter}}}\ (\bibinfo
  {publisher} {Springer},\ \bibinfo {address} {Berlin},\ \bibinfo {year}
  {1979})\BibitemShut {NoStop}%
\bibitem [{\citenamefont {Dalmedigos}\ and\ \citenamefont
  {Bunin}(2020)}]{dalmedigos2020dynamical}%
  \BibitemOpen
  \bibfield  {author} {\bibinfo {author} {\bibfnamefont {I.}~\bibnamefont
  {Dalmedigos}}\ and\ \bibinfo {author} {\bibfnamefont {G.}~\bibnamefont
  {Bunin}},\ }\bibfield  {title} {\bibinfo {title} {Dynamical persistence in
  high-diversity resource-consumer communities},\ }\href
  {https://doi.org/10.1371/journal.pcbi.1008189} {\bibfield  {journal}
  {\bibinfo  {journal} {PLOS Computational Biology}\ }\textbf {\bibinfo
  {volume} {16}},\ \bibinfo {pages} {1} (\bibinfo {year} {2020})}\BibitemShut
  {NoStop}%
\bibitem [{\citenamefont {LADEIRA}\ and\ \citenamefont
  {de~OLIVEIRA}(2019)}]{ladeira2019chaotic}%
  \BibitemOpen
  \bibfield  {author} {\bibinfo {author} {\bibfnamefont {D.~G.}\ \bibnamefont
  {LADEIRA}}\ and\ \bibinfo {author} {\bibfnamefont {M.~M.}\ \bibnamefont
  {de~OLIVEIRA}},\ }\bibfield  {title} {\bibinfo {title} {Chaotic coexistence
  in a resource–consumer model},\ }\href
  {https://doi.org/10.1142/S0218339019500086} {\bibfield  {journal} {\bibinfo
  {journal} {Journal of Biological Systems}\ }\textbf {\bibinfo {volume}
  {27}},\ \bibinfo {pages} {167} (\bibinfo {year} {2019})}\BibitemShut
  {NoStop}%
\bibitem [{\citenamefont {Din}\ and\ \citenamefont
  {Khan}(2021)}]{din2021discrete}%
  \BibitemOpen
  \bibfield  {author} {\bibinfo {author} {\bibfnamefont {Q.}~\bibnamefont
  {Din}}\ and\ \bibinfo {author} {\bibfnamefont {M.~I.}\ \bibnamefont {Khan}},\
  }\bibfield  {title} {\bibinfo {title} {A discrete-time model for
  consumer--resource interaction with stability, bifurcation and chaos
  control},\ }\href {https://doi.org/10.1007/s12346-021-00488-4} {\bibfield
  {journal} {\bibinfo  {journal} {Qualitative Theory of Dynamical Systems}\
  }\textbf {\bibinfo {volume} {20}},\ \bibinfo {pages} {56} (\bibinfo {year}
  {2021})}\BibitemShut {NoStop}%
\bibitem [{\citenamefont {Williams}\ and\ \citenamefont
  {Martinez}(2004)}]{williams2004stabilization}%
  \BibitemOpen
  \bibfield  {author} {\bibinfo {author} {\bibfnamefont {R.~J.}\ \bibnamefont
  {Williams}}\ and\ \bibinfo {author} {\bibfnamefont {N.~D.}\ \bibnamefont
  {Martinez}},\ }\bibfield  {title} {\bibinfo {title} {Stabilization of chaotic
  and non-permanent food-web dynamics},\ }\href@noop {} {\bibfield  {journal}
  {\bibinfo  {journal} {The European Physical Journal B}\ }\textbf {\bibinfo
  {volume} {38}},\ \bibinfo {pages} {297} (\bibinfo {year} {2004})}\BibitemShut
  {NoStop}%
\bibitem [{\citenamefont {Beninc{\`a}}\ \emph {et~al.}(2015)\citenamefont
  {Beninc{\`a}}, \citenamefont {Ballantine}, \citenamefont {Ellner},\ and\
  \citenamefont {Huisman}}]{beninca2015species}%
  \BibitemOpen
  \bibfield  {author} {\bibinfo {author} {\bibfnamefont {E.}~\bibnamefont
  {Beninc{\`a}}}, \bibinfo {author} {\bibfnamefont {B.}~\bibnamefont
  {Ballantine}}, \bibinfo {author} {\bibfnamefont {S.~P.}\ \bibnamefont
  {Ellner}},\ and\ \bibinfo {author} {\bibfnamefont {J.}~\bibnamefont
  {Huisman}},\ }\bibfield  {title} {\bibinfo {title} {Species fluctuations
  sustained by a cyclic succession at the edge of chaos},\ }\href@noop {}
  {\bibfield  {journal} {\bibinfo  {journal} {Proceedings of the National
  Academy of Sciences}\ }\textbf {\bibinfo {volume} {112}},\ \bibinfo {pages}
  {6389} (\bibinfo {year} {2015})}\BibitemShut {NoStop}%
\bibitem [{\citenamefont {Cui}\ \emph {et~al.}(2020)\citenamefont {Cui},
  \citenamefont {Marsland~III},\ and\ \citenamefont {Mehta}}]{cui2020effect}%
  \BibitemOpen
  \bibfield  {author} {\bibinfo {author} {\bibfnamefont {W.}~\bibnamefont
  {Cui}}, \bibinfo {author} {\bibfnamefont {R.}~\bibnamefont {Marsland~III}},\
  and\ \bibinfo {author} {\bibfnamefont {P.}~\bibnamefont {Mehta}},\ }\bibfield
   {title} {\bibinfo {title} {Effect of resource dynamics on species packing in
  diverse ecosystems},\ }\href@noop {} {\bibfield  {journal} {\bibinfo
  {journal} {Physical Review Letters}\ }\textbf {\bibinfo {volume} {125}},\
  \bibinfo {pages} {048101} (\bibinfo {year} {2020})}\BibitemShut {NoStop}%
\bibitem [{\citenamefont {Marsland}\ \emph
  {et~al.}(2020{\natexlab{c}})\citenamefont {Marsland}, \citenamefont {Cui},
  \citenamefont {Goldford},\ and\ \citenamefont
  {Mehta}}]{marsland2020community}%
  \BibitemOpen
  \bibfield  {author} {\bibinfo {author} {\bibfnamefont {R.}~\bibnamefont
  {Marsland}}, \bibinfo {author} {\bibfnamefont {W.}~\bibnamefont {Cui}},
  \bibinfo {author} {\bibfnamefont {J.}~\bibnamefont {Goldford}},\ and\
  \bibinfo {author} {\bibfnamefont {P.}~\bibnamefont {Mehta}},\ }\bibfield
  {title} {\bibinfo {title} {The community simulator: A python package for
  microbial ecology},\ }\href@noop {} {\bibfield  {journal} {\bibinfo
  {journal} {Plos one}\ }\textbf {\bibinfo {volume} {15}},\ \bibinfo {pages}
  {e0230430} (\bibinfo {year} {2020}{\natexlab{c}})}\BibitemShut {NoStop}%
\bibitem [{\citenamefont {Dal~Bello}\ \emph {et~al.}(2021)\citenamefont
  {Dal~Bello}, \citenamefont {Lee}, \citenamefont {Goyal},\ and\ \citenamefont
  {Gore}}]{dal2021resource}%
  \BibitemOpen
  \bibfield  {author} {\bibinfo {author} {\bibfnamefont {M.}~\bibnamefont
  {Dal~Bello}}, \bibinfo {author} {\bibfnamefont {H.}~\bibnamefont {Lee}},
  \bibinfo {author} {\bibfnamefont {A.}~\bibnamefont {Goyal}},\ and\ \bibinfo
  {author} {\bibfnamefont {J.}~\bibnamefont {Gore}},\ }\bibfield  {title}
  {\bibinfo {title} {Resource--diversity relationships in bacterial communities
  reflect the network structure of microbial metabolism},\ }\href@noop {}
  {\bibfield  {journal} {\bibinfo  {journal} {Nature Ecology \& Evolution}\
  }\textbf {\bibinfo {volume} {5}},\ \bibinfo {pages} {1424} (\bibinfo {year}
  {2021})}\BibitemShut {NoStop}%
\bibitem [{\citenamefont {Marcus}\ \emph {et~al.}(2023)\citenamefont {Marcus},
  \citenamefont {Turner},\ and\ \citenamefont {Bunin}}]{marcus2023local}%
  \BibitemOpen
  \bibfield  {author} {\bibinfo {author} {\bibfnamefont {S.}~\bibnamefont
  {Marcus}}, \bibinfo {author} {\bibfnamefont {A.~M.}\ \bibnamefont {Turner}},\
  and\ \bibinfo {author} {\bibfnamefont {G.}~\bibnamefont {Bunin}},\ }\bibfield
   {title} {\bibinfo {title} {Local and extensive fluctuations in
  sparsely-interacting ecological communities},\ }\href@noop {} {\bibfield
  {journal} {\bibinfo  {journal} {arXiv preprint arXiv:2308.01828}\ } (\bibinfo
  {year} {2023})}\BibitemShut {NoStop}%
\bibitem [{\citenamefont {Feng}\ \emph {et~al.}(2023)\citenamefont {Feng},
  \citenamefont {au2}, \citenamefont {Rocks},\ and\ \citenamefont
  {Mehta}}]{feng2023multitrophic}%
  \BibitemOpen
  \bibfield  {author} {\bibinfo {author} {\bibfnamefont {Z.}~\bibnamefont
  {Feng}}, \bibinfo {author} {\bibfnamefont {R.~M.~I.}\ \bibnamefont {au2}},
  \bibinfo {author} {\bibfnamefont {J.~W.}\ \bibnamefont {Rocks}},\ and\
  \bibinfo {author} {\bibfnamefont {P.}~\bibnamefont {Mehta}},\ }\href@noop {}
  {\bibinfo {title} {Emergent competition shapes the ecological properties of
  multi-trophic ecosystems}} (\bibinfo {year} {2023}),\ \Eprint
  {https://arxiv.org/abs/2303.02983} {arXiv:2303.02983 [q-bio.PE]} \BibitemShut
  {NoStop}%
\bibitem [{\citenamefont {Ebrahimi}\ \emph {et~al.}(2022)\citenamefont
  {Ebrahimi}, \citenamefont {Goyal},\ and\ \citenamefont
  {Cordero}}]{ebrahimi2022particle}%
  \BibitemOpen
  \bibfield  {author} {\bibinfo {author} {\bibfnamefont {A.}~\bibnamefont
  {Ebrahimi}}, \bibinfo {author} {\bibfnamefont {A.}~\bibnamefont {Goyal}},\
  and\ \bibinfo {author} {\bibfnamefont {O.~X.}\ \bibnamefont {Cordero}},\
  }\bibfield  {title} {\bibinfo {title} {Particle foraging strategies promote
  microbial diversity in marine environments},\ }\href@noop {} {\bibfield
  {journal} {\bibinfo  {journal} {Elife}\ }\textbf {\bibinfo {volume} {11}},\
  \bibinfo {pages} {e73948} (\bibinfo {year} {2022})}\BibitemShut {NoStop}%
\bibitem [{Note1()}]{Note1}%
  \BibitemOpen
  \bibinfo {note} {See Supplementary Material [url] for complete derivations,
  further numerical methods and results, and additional figures, which includes
  Refs.~[51-52].}\BibitemShut {Stop}%
\bibitem [{\citenamefont {Frederickson}\ \emph {et~al.}(1983)\citenamefont
  {Frederickson}, \citenamefont {Kaplan}, \citenamefont {Yorke},\ and\
  \citenamefont {Yorke}}]{frederickson1983dimension}%
  \BibitemOpen
  \bibfield  {author} {\bibinfo {author} {\bibfnamefont {P.}~\bibnamefont
  {Frederickson}}, \bibinfo {author} {\bibfnamefont {J.~L.}\ \bibnamefont
  {Kaplan}}, \bibinfo {author} {\bibfnamefont {E.~D.}\ \bibnamefont {Yorke}},\
  and\ \bibinfo {author} {\bibfnamefont {J.~A.}\ \bibnamefont {Yorke}},\
  }\bibfield  {title} {\bibinfo {title} {The liapunov dimension of strange
  attractors},\ }\href
  {https://doi.org/https://doi.org/10.1016/0022-0396(83)90011-6} {\bibfield
  {journal} {\bibinfo  {journal} {Journal of Differential Equations}\ }\textbf
  {\bibinfo {volume} {49}},\ \bibinfo {pages} {185} (\bibinfo {year}
  {1983})}\BibitemShut {NoStop}%
\bibitem [{\citenamefont {Datseris}\ \emph {et~al.}(2023)\citenamefont
  {Datseris}, \citenamefont {Kottlarz}, \citenamefont {Braun},\ and\
  \citenamefont {Parlitz}}]{datseris2023FractalDimensions}%
  \BibitemOpen
  \bibfield  {author} {\bibinfo {author} {\bibfnamefont {G.}~\bibnamefont
  {Datseris}}, \bibinfo {author} {\bibfnamefont {I.}~\bibnamefont {Kottlarz}},
  \bibinfo {author} {\bibfnamefont {A.~P.}\ \bibnamefont {Braun}},\ and\
  \bibinfo {author} {\bibfnamefont {U.}~\bibnamefont {Parlitz}},\ }\bibfield
  {title} {\bibinfo {title} {Estimating fractal dimensions: A comparative
  review and open source implementations},\ }\bibfield  {journal} {\bibinfo
  {journal} {Chaos: An Interdisciplinary Journal of Nonlinear Science}\
  }\textbf {\bibinfo {volume} {33}},\ \href {https://doi.org/10.1063/5.0160394}
  {10.1063/5.0160394} (\bibinfo {year} {2023})\BibitemShut {NoStop}%
\end{thebibliography}%

\clearpage

\begin{widetext}

\begin{widetext}
\tableofcontents

\appendix

\renewcommand\thefigure{\thesection\arabic{figure}}

\clearpage
\section{Cavity calculation\label{appendix:cavity-calc}}

The objective of the cavity calculation is to find the steady-state behavior of the aMCRM.
In particular, we will find the distribution of the steady-state abundances of the species and resources.
The cavity method takes advantage of the system's self-averaging behavior.
A system is said to have self-averaging behavior if the distribution of properties of constituents are independent of the exact realization of quenched disorder.
This means that constituents' properties can be treated as random variates drawn from a distribution that is common to all systems with the same distribution of quenched disorder.
A consequence of this is that an average of some observable taken over constituents given a fixed realization of quenched disorder is equal to the average of the observable for one constituent taken over all realizations of quenched disorder.
The emergence of this self-averaging behavior is a consequence of the central limit theorem and quenched disorder, as we will see.
In Fig.~\ref{fig:cavity-eCDFs}, we compare the distribution of steady-state resource and species abundance for a single realization of quenched disorder to the distribution of abundances predicted by the cavity method.

\begin{figure}[ht]
	\centering
	\includegraphics[width=0.5\linewidth]{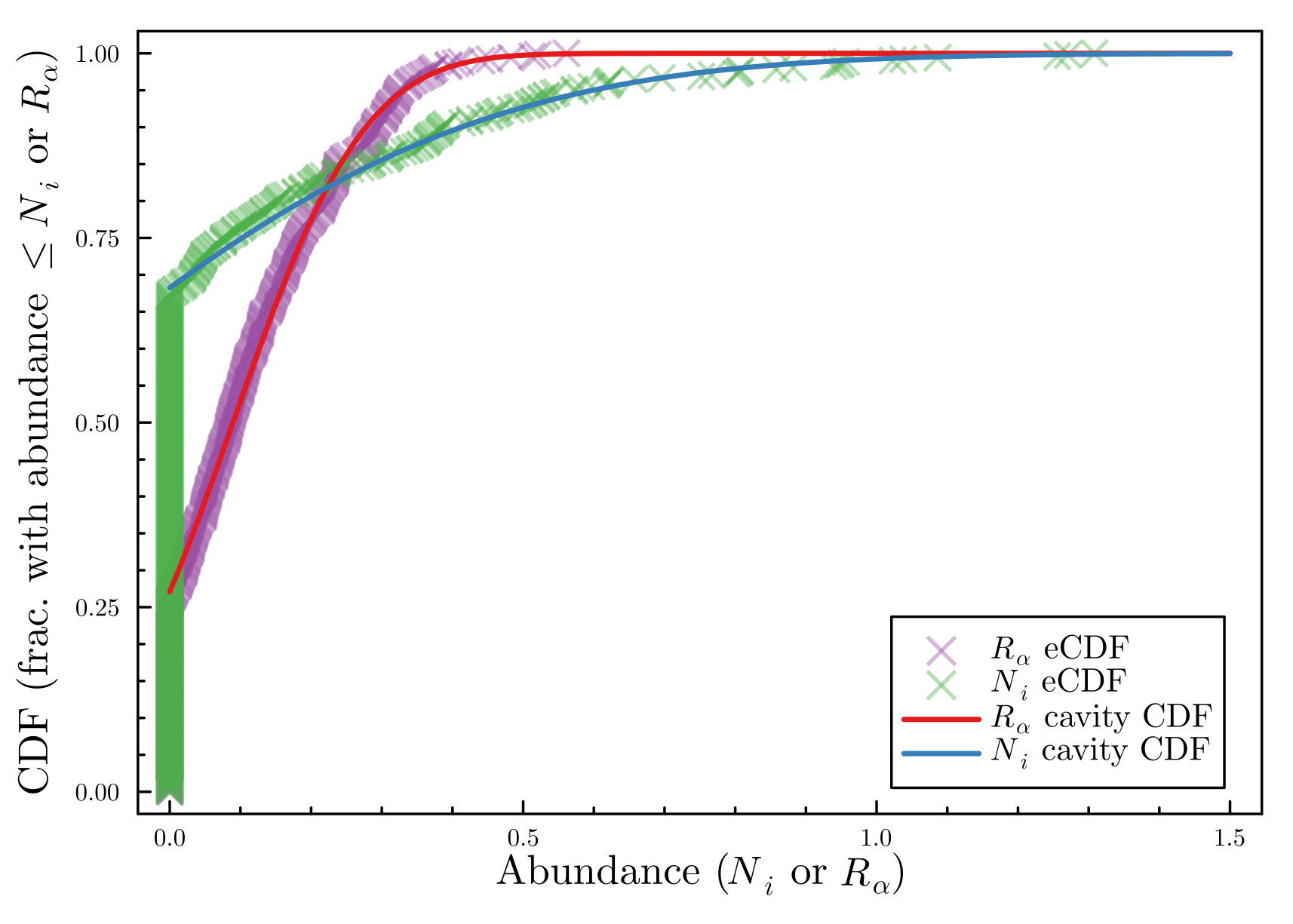}
	\caption{\label{fig:cavity-eCDFs}
	Comparison of the empirical cumulative distribution functions (eCDFs) of the steady-state abundances of species ($S=512$) and resources ($M=512$) for a single realization of quenched disorder and the distribution predicted by the cavity calculation; simulations are performed in the stable phase.
	See appendix \ref{appendix:fig-details} for simulation parameters.
	}
\end{figure}

\subsection{Setup}


We begin by introducing the constituent averages,
\begin{equation}
	\expval{ R}
	\equiv
	\frac{1}{M}
	\sum_{\alpha=1}^M {\overline R_\alpha}\qqc
	\expval{ N}
	\equiv
	\frac{1}{S}
	\sum_{i=1}^S {\overline N_i},
\end{equation}
where we have introduced the notation $\overline X$ to denote the steady-state value of a quantity $X$.
With these quantities defined, we can write the steady-state aMCRM as,
\begin{align}
	0
	=
	\dv{N_i}{t}
	&=
	\overline N_i
	\qty[
	g
	+
	\frac{\sigma_c}{\sqrt{M}}
	\sum_{\alpha=1}^M 
	d_{i\alpha}
	\overline R_\alpha
	-
	\sigma_m
	\delta m_i
	],
	\\
	0
	=
	\dv{R_\alpha}{t}
	&=
	\overline R_\alpha
	\qty[		\kappa
		-
		\overline R_\alpha
		-
		\frac{\sigma_e}{\sqrt{M}}
		\sum_{i=1}^{S}
		N_i
		\left(
			\rho d_{i\alpha} + \sqrt{1-\rho^2} x_{i\alpha}
		\right)
		+ 
		\sigma_K \delta K_\alpha
	],
\end{align}
where
\begin{gather}
	g 
	\equiv
	\mu_c \expval{R}
	-
	m\qqc
	\kappa
	\equiv
	K 
	-
	\mu_e \gamma^{-1} \expval{N}\qqc
	\gamma \equiv \frac{M}{S}.
\end{gather}


\subsection{Cavity solution}

The cavity method begins by introducing a new species $i = 0$ and a new resource $\alpha = 0$. We will later treat the new terms introduced by these variables as small perturbations.
With the new species and resource, we can write the steady-state aMCRM for the existing species $i = 1,\dots,S$ and resources $\alpha = 1,\dots,M$ as,
\begin{align}
	0
	=
	\dv{N_i}{t}
	&=
	\overline
	N_i
	\qty[
		\mu_c \expval{R} -
		\qty(m - \frac{\sigma_c}{\sqrt{M}} d_{i0}R_0)
		+
		\frac{\sigma_c }{\sqrt{M}} \sum_{\alpha=1}^M d_{i\alpha} \overline R_\alpha
		-
		\sigma_m
		\delta m_i
	],
	\label{eq:aMCRM-N-pert}
	\\
	0
	=
	\dv{R_\alpha}{t}
	&=
	\overline R_\alpha
	\Bigg[
		\qty(
			K 
			-
			\frac{\sigma_e}{ \sqrt{M}} N_0 \qty(\rho d_{0\alpha} + \sqrt{1-\rho^2} x_{0\alpha})
		)
		-
		\mu_e \gamma^{-1}\expval{N}
		-
		\overline R_\alpha
	\nonumber
		\\
		&
		\qquad
		\qquad
		-
		\frac{\sigma_e }{\sqrt{M}}
		\sum_{i=1}^{S}
		\overline N_i
		\qty(
			\rho d_{i\alpha}
			+
			\sqrt{1-\rho^2} x_{i\alpha}
		)
		+
		\sigma_K \delta K_\alpha
	\Bigg].
	\label{eq:aMCRM-R-pert}
\end{align}
For the new species and resource, the steady-state aMCRM says,
\begin{align}
	0
	=
	\dv{N_0}{t}
	&=
	\overline N_0
	\qty[
		g
		+
		\frac{\sigma_c }{\sqrt{M}}
		\sum_{\alpha=1}^M
		d_{0\alpha} \overline R_\alpha
		+
		M^{-1/2}\sigma_c
		d_{00}\overline R_0
		-
		\sigma_m 
		\delta m_0
	],
	\label{eq:aMCRM-N-pert-new}
	\\
	0
	=
	\dv{R_0}{t}
	&=
	\overline R_0
	\Bigg[
		\kappa
		-
		R_0
		-
		\frac{\sigma_e}{\sqrt{M}}
		\sum_{i=1}^{S}
		\overline N_j \left(
			\rho d_{i0}
			+
			\sqrt{1-\rho^2}x_{i0}
		\right)
	\nonumber
		\\
		&\qquad\qquad
		-
		\frac{\sigma_e}{\sqrt{M}}
		\overline N_0
		\left(
			\rho d_{00}
			+
			\sqrt{1-\rho^2}x_{00}
		\right)
		+
		\sigma_K
		\delta K_\alpha
	\Bigg].
	\label{eq:aMCRM-R-pert-new}
\end{align}
Next, we will analyze the perturbed system relative to the unperturbed system.
A quantity with $\setminus 0$ represents the value before adding the new variables to the system.
Looking to Eqs.~\ref{eq:aMCRM-N-pert} and \ref{eq:aMCRM-R-pert}, we see that the presence of the new species and resource effectively perturbs the model parameters as,
\begin{align}
	m_i \to m_i - \frac{\sigma_c}{\sqrt{M}} d_{i0} \overline R_0,
	\qquad
	K_\alpha \to K_\alpha - \frac{\sigma_e}{\sqrt{M}} \overline N_0 \left(
		\rho d_{0\alpha} + \sqrt{1-\rho^2} x_{0\alpha}
		\right).
\end{align}
In the thermodynamic limit where $M$ and $S$ are large, we model the perturbation using linear response:
\begin{align}
	\overline N_i
	&=
	\overline N_{i\setminus 0}
	-
	\frac{\sigma_e}{\sqrt{M}}
	\sum_{\beta = 1}^M
	\chi_{i\beta}^{(N)}
	 \left(
		\rho d_{0\beta} + \sqrt{1-\rho^2} x_{0\beta}
	\right)\overline N_0
	-
	\frac{\sigma_c}{\sqrt{M}}
	\sum_{j=1}^S
	\nu_{i j}^{(N)}
	d_{j0}
	\overline R_0
	,
	\label{eq:linear-response-N}
	\\
	\overline R_\alpha
	&=
	\overline R_{\alpha \setminus 0}
	-
	\frac{\sigma_e}{\sqrt{M}}
	\sum_{\beta=1}^{M}
	\chi_{\alpha\beta}^{(R)}
	\left(
		\rho d_{0\beta}
		+
		\sqrt{1-\rho^2} x_{0\beta}
	\right)
	\overline N_0
	-
	\frac{\sigma_c}{\sqrt{M}}
	\sum_{j=1}^S
	\nu_{\alpha j}^{(R)} d_{j0} \overline R_0
	\label{eq:linear-response-R}
	,
\end{align}
where we have defined the susceptibility matrices:
\begin{align}
	&
	\label{Eq:Ksuscept}
	\chi_{i\beta}^{(N)}
	\equiv
	\frac{\partial \overline N_i}{\partial K_\beta},
	&&
	\chi_{\alpha\beta}^{(R)}
	\equiv
	\frac{\partial \overline R_\alpha}{\partial K_\beta},
	\\
	\label{Eq:msuscept}
	&
	\nu_{ij}^{(N)}
	\equiv
	\frac{\partial\overline N_i}{\partial m_j},
	&&
	\nu_{\alpha j}^{(R)}
	\equiv
	\frac{\partial \overline R_\alpha}{\partial m_j}.
\end{align}

\subsubsection{Self-consistency equations for species populations}

With the perturbation introduced and susceptibilities defined, we will exploit the linear response approximation to derive the self-consistency equations for the species populations.
Substituting Eqs.~\ref{eq:linear-response-N}, \ref{eq:linear-response-R} into the aMCRM equation for the additional species [Eq.~\eqref{eq:aMCRM-N-pert-new}], we obtain,
\begin{align}
	0
	&=
	\overline N_0
	\left[
		g
		+
			\frac{\sigma_c }{\sqrt{M}}
			\sum_{\alpha=1}^M
			d_{0\alpha} 
			\overline R_{\alpha \setminus 0}
			-
			\frac{\sigma_c \sigma_e}{M}
			\sum_{\alpha,\beta=1}^M
			\chi_{\alpha\beta}^{(R)}
			d_{0\alpha} 
			\left(
				\rho d_{0\beta}
				+
				\sqrt{1-\rho^2} x_{0\beta}
			\right)
			\overline N_0
			\right.
			\nonumber
			\\
			&\qquad\qquad
			\left.
			-
			\frac{\sigma_c^2}{M}
			\sum_{\alpha=1}^M
			\sum_{j=1}^S
			\nu_{\alpha j}^{(R)}
			d_{0\alpha} 
			d_{j0}
			\overline N_0
		+
		\frac{\sigma_c}{\sqrt{M}}
		d_{00}
		\overline R_0
		-
		\sigma_m 
		\delta m_0
	\right].
	\label{substituteSusceptN0}
\end{align}
Taking the mean of the third term with respect to the new matrix elements $d_{0\alpha}$ and $x_{0\alpha}$, we obtain
\begin{align}
	&\expval{
		\frac{\sigma_c\sigma_e}{M}
		\sum_{\alpha,\beta=1}^M
			\chi_{\alpha\beta}^{(R)}
			d_{0\alpha} 
			\left(
				\rho d_{0\beta}
				+
				\sqrt{1-\rho^2} x_{0\beta}
			\right)
			\overline N_0
	}
	\nonumber
	\\
	&\qquad=
	\overline N_0
	\frac{\sigma_c\sigma_e}{M}
	\sum_{\alpha,\beta=1}^M
		{\chi_{\alpha\beta}^{(R)}}
		\left(
			\rho 
			\expval{
			d_{0\alpha} 
			d_{0\beta}
			}
			+
			\expval{
			d_{0\alpha} 
			}
			\expval{
			x_{0\beta}
			}
			\sqrt{1-\rho^2}
		\right)
		\nonumber
		\\
	&\qquad=
	\overline N_0
	\frac{\sigma_c\sigma_e}{M}
	\sum_{\alpha,\beta=1}^M
		{\chi_{\alpha\beta}^{(R)}}
		\left(
			\rho 
			\delta_{\alpha\beta}
			+
			0
			\times 
			0
			\sqrt{1-\rho^2}
		\right)
	=
	\overline N_0 \sigma_c\sigma_e \rho \chi,
\end{align}
where we have used that $d_{0\alpha}$ and $x_{0\beta}$ are independent and have defined,
\begin{align}
	\chi \equiv \frac{1}{M} \sum_{\alpha = 1}^M {\chi_{\alpha\alpha}^{(R)}},
	\label{eq:chi-def}
\end{align}
to be the trace of the $\overline R_\alpha \leftarrow K_\beta$ susceptibility matrix divided by the number of resources.
The variance of this term is $O(M^{-1})$, which can be verified by expanding its second moment.
The mean of the fourth term is zero because $d_{0\alpha}$ and $d_{j0}$ are uncorrelated when $\alpha = 1,\dots,M,\; j =1,\dots, S$ and its variance is $O(M^{-1})$.
Discarding terms of order $O(M^{-1/2})$ and higher, we obtain,
\begin{align}
	0 = \overline N_0
	\left(
		g 
		-
		\sigma_c \sigma_e \rho \chi \overline N_0
		+
		\frac{\sigma_c}{\sqrt{M}}
		\sum_{\alpha = 1}^M d_{0\alpha} \overline R_{\alpha \setminus 0}
		-
		\sigma_m \delta m_0
	\right)
	+
	O(M^{-1/2}).
	\label{eq:N0beforeconsist}
\end{align}
The last two terms above are a sum of many independent random variables, so, by the central limit theorem, we can model these terms as a sum of normal random variables.
The mean of these terms is,
\begin{align}
	\expval{
		\frac{\sigma_c}{\sqrt{M}}\sum_{\alpha=1}^M
		d_{0\alpha} \overline R_{\alpha \setminus 0}
		-
		\sigma_m \delta m_0
	}
	\nonumber
	&
	=
		\frac{\sigma_c}{\sqrt{M}}\sum_{\alpha=1}^M
		\expval{d_{0\alpha}}  \overline R_{\alpha \setminus 0}
		-
		\sigma_m \expval{\delta m_0}
	\\
&
		=
		\frac{\sigma_c}{\sqrt{M}}\sum_{\alpha=1}^M
		0\times{ \overline R_{\alpha \setminus 0}}
		-
		\sigma_m \times 0
		=0.
\end{align}
The variance of these terms is,
\begin{align}
	\sigma_g^2
	&\equiv
	\Var{
		\frac{\sigma_c}{\sqrt{M}}
		\sum_{\alpha=1}^M d_{0\alpha} \overline R_{\alpha\setminus 0}
		-
		\sigma_m\delta m_0
	}
	=
	\frac{\sigma_c^2}{M}
	\sum_{\alpha=1}^M 
		{\overline R^2_{\alpha\setminus 0}}
		\Var{d_{0\alpha}}
	+
	\sigma_m^2
	\Var{\delta m_0}
	\nonumber
	\\
	&=
	\frac{\sigma_c^2}{M}
	\sum_{\alpha=1}^{M}
	{\overline R_{\alpha\setminus 0}^2}
	+
	\sigma_m^2
	=
	\sigma_c^2 q_R + \sigma_m^2,
\end{align}
where we have defined
  $
	q_R 
	\equiv 
	M^{-1}
	\sum_{\alpha = 1}^M
	{\overline R_{\alpha \setminus 0}^2}.
$
Due to the self-averaging nature of the system and that the size of the perturbation is of order $O(M^{-1/2})$, we have $q_R \approx \expval*{\overline R_0^2}$.
Let $Z_N\sim \mathcal N(0,1)$ be a unit normal random variable. 
The large-$M$ limit approximate steady-state condition for the perturbing species becomes,
\begin{align}
	0 = 
	\overline N_0
	\left(
		g 
		-
		\sigma_c\sigma_e\rho\chi\overline N_0
		+
		\sigma_g Z_N
	\right).
	\label{eq:N0selfconsist}
\end{align}
Solving for $\overline N_0$ and discarding invadable solutions, we obtain,
\begin{align}
	\boxed{\overline N_0
	=
	\max\left\{
		0,
		\frac{g + \sigma_g Z_N}{\sigma_c \sigma_e \rho\chi}
	\right\}}
	.
	\label{eq:N0selfconsistsoln}
\end{align}
Additionally, observe that if $\chi \to 0$, the cavity solution diverges; this will be discussed in further detail in SI section~\ref{appendix:cavity-infeas}.

\subsubsection{Self-consistency equations for resource abundances}
Now, we repeat this process to find a self-consistency equation for the resources.
We substitute the linear response approximation for species into the aMCRM steady-state equation for the additional resource:
\begin{align}
	0
	&=
	\overline R_0
	\Bigg[
		\kappa
		-
		\overline R_0
		-
		\frac{\sigma_e}{\sqrt{M}}
		\sum_{i=1}^{S}
		\overline N_{i\setminus 0}
		\left(
			\rho d_{i0}
			+
			\sqrt{1-\rho^2}x_{i0}
		\right)
	\label{eq:substituteSusceptR0}
	\\
	&\qquad
		+
		\frac{\sigma_e^2}{M}
		\sum_{i=1}^{S}
		\sum_{\beta = 1}^M
		\chi_{i\beta}^{(N)}
		\left(
			\rho d_{i0}
			+
			\sqrt{1-\rho^2}x_{i0}
		\right)
		\left(
			\rho d_{0\beta} + \sqrt{1-\rho^2} x_{0\beta}
		\right)\overline N_0
	\nonumber
		\\
		&\qquad
		+
		\frac{\sigma_e\sigma_c}{M}
		\sum_{i,j=1}^{S}
		\nu_{i j}^{(N)}
		\left(
			\rho d_{i0}
			+
			\sqrt{1-\rho^2}x_{i0}
		\right)
		d_{j0}
		\overline R_0
		-
		\frac{\sigma_e}{\sqrt{M}}
		\overline N_0
		\left(
			\rho d_{00}
			+
			\sqrt{1-\rho^2}x_{00}
		\right)
		+
		\sigma_K
		\delta K_\alpha
	\Bigg].
	\nonumber
\end{align}
Observe that the fourth term (involving $\chi_{i\beta}^{(N)}$) has zero mean and variance of order $O(M^{-1})$; this can be seen by recalling that $d_{i0},d_{0\beta}, x_{i0},x_{0\beta}$ are all independent for $i,\beta \geq 1$.
Similarly, we see that the variance of the fifth term (involving $\nu_{ij}^{(N)}$) is of order $O(M^{-1})$.
We will ignore fluctuations of order $O(M^{-{1/2}})$.
The mean of the fifth term is,
\begin{align}
	&\nonumber{
		\frac{\sigma_e \sigma_c}{M}
		\sum_{i,j = 1}^S
		\nu_{ij}^{(N)}
		\left(
			\rho d_{i0} + \sqrt{1-\rho^2} x_{i0}
		\right)
		d_{j0} \overline R_0
	}
  =
	\overline R_0
	\frac{\sigma_e \sigma_c}{M}
	\sum_{i,j = 1}^S
	{
		\nu_{ij}^{(N)}
	}
	\left(
		\rho 
		\expval{d_{i0} 
		d_{j0} }
		+ 
		\sqrt{1-\rho^2} 
		\expval{x_{i0}}
		\expval{d_{j0} }
	\right)
	\nonumber
	\\
	&\qquad =
	\overline R_0
	\sigma_e \sigma_c
	\frac{S}{M}\frac{1}{S}
	\sum_{i,j = 1}^S
	{
		\nu_{ij}^{(N)}
	}
	\left(
		\rho 
		\delta_{ij}
		+ 
		\sqrt{1-\rho^2} 
		\times 0 \times 0
	\right)
	=
	\rho \sigma_e\sigma_c \gamma^{-1}\nu\overline R_0,
\end{align}
where we have defined,
\begin{align}
	\nu
	\equiv
	\frac{1}{S}
	\sum_{i=1}^S
	{\nu_{ii}^{(N)}},
	\label{eq:nu-def}
\end{align}
which is the trace of the $\overline N_i \leftarrow m_j$ susceptibility matrix divided by $S$.
Discarding terms of order $O(M^{-1/2})$, we obtain,
\begin{align}
	0
	=
	\overline R_0
	\left(
		\kappa
		-
		\overline R_0
		-
		\frac{\sigma_e}{\sqrt{M}}
		\sum_{i=1}^S\overline N_{i \setminus 0}
		\left(
			\rho d_{i0}
			+
			\sqrt{1-\rho^2}x_{i0}
		\right)
		+
		\rho \sigma_e \sigma_c \gamma^{-1} \nu \overline R_0
		+
		\sigma_K \delta K_0
	\right)
	+
	O(M^{-1/2}).
	\label{eq:R0beforeconsist}
\end{align}
Now, observe the third and last terms are a sum of many independent random variables, meaning we can apply the central limit theorem and model the sum of these terms as a normal random variable.
The mean of these terms is,
\begin{align}
	&
	\expval{
		\sigma_K \delta K_0
		-
		\frac{\sigma_e}{\sqrt{M}}
		\sum_{i=1}^S
		\overline N_{i\setminus0}
		\left(
			\rho d_{i0} + \sqrt{1-\rho^2} x_{i0}
		\right)
	}
  =
	\sigma_K \expval{\delta K_0}
	-
	\frac{\sigma_e}{\sqrt{M}}
	\sum_{i=1}^S
	{\overline N_{i\setminus0}}
	\left(
		\rho \expval{d_{i0}} + \sqrt{1-\rho^2} \expval{x_{i0}}
	\right)
	\nonumber
	\\
	&\qquad=
	\sigma_K \times 0
	-
	\frac{\sigma_e}{\sqrt{M}}
	\sum_{i=1}^S
	{\overline N_{i\setminus0}}
	\left(
		\rho \times 0 + \sqrt{1-\rho^2} \times 0
	\right)
	=
	0.
\end{align}

The variance is,
\begin{align}
	\sigma_\kappa^2 
	&
	\equiv 
	\Var{
	\sigma_K \delta K_0
	-
	\frac{\sigma_e}{\sqrt{M}}
	\sum_{i=1}^S
	\overline N_{i\setminus0}
	\left(
		\rho d_{i0} + \sqrt{1-\rho^2} x_{i0}
	\right)
	}
	\\
	\nonumber
	&
	=
	\sigma_K^2 \Var{\delta K_0}
	+
	\frac{\sigma_e^2}{M}
	\sum_{i=1}^S
	\Var{\overline N_{i\setminus0}
	\left(
		\rho d_{i0} + \sqrt{1-\rho^2} x_{i0}
	\right)}
	\nonumber
	\\
	&
	=
	\sigma_K^2
	+
	\frac{\sigma_e^2}{M}
	\sum_{i=1}^S
	{\overline N_{i\setminus0}^2}
	\left(
		\rho^2 \Var{d_{i0}} + (1-\rho^2 )\Var{x_{i0}}
	\right)
	=
	\sigma_K^2
	+
	{\sigma_e^2}
	\frac{S}{M}
	\frac{1}{S}
	\sum_{i=1}^S
	{\overline N_{i\setminus0}^2}
	=
	\sigma_K^2 
	+
	\gamma^{-1}\sigma_e^2
	q_{N},
	\nonumber
\end{align}
where $
q_N \equiv
\frac{1}{S}
\sum_{i=1}^S {N_{i\setminus 0}^2}
$ which is approximately equal to $\expval*{\overline N_0^2}$.
The approximate steady-state condition for the added resource then becomes,
\begin{align}
	0 = \overline R_0
	\left(
		\kappa - \overline R_0 + \sigma_\kappa Z_R + \rho \sigma_e \sigma_c \gamma^{-1} \nu \overline R_0
	\right),
	\label{eq:R0selfconsist}
\end{align}
where $Z_R \sim \mathcal N (0,1)$ is a standard normal random variable.
Solving for $\overline R_0$ and discarding invadable solutions gives,
\begin{align}
	\boxed{\overline R_0 
	=
	\max\left\{
		0,
		\frac{\kappa+ \sigma_\kappa Z_R}{1- \rho \sigma_e \sigma_c \gamma^{-1} \nu}
	\right\}}
	.
	\label{eq:R0selfconsistsoln}
\end{align}

\subsection{ReLU function-transformed normal distributions\label{appendix:ReLUNormal}}

In these computations, we regularly work with normal distributions that are transformed by the `ReLU' function: $\mathrm{ReLU}(x) = \max\{0,x\} = x \Theta(x)$.
If $Z$ is a standard normal random variable, the PDF of $\mathrm{ReLU}(\sigma Z + \mu)$ is,
\begin{align}
	p_{\mathrm{ReLU}(\sigma Z + \mu)}(z)
	=
	\delta(z) \Phi(-\mu/\sigma)
	+
	\frac{1}{\sqrt{2\pi}\sigma} e^{-(z-\mu)^2 / 2\sigma^2}\Theta(z),
\end{align}
where,
\begin{align}
	\Phi(x)	= \frac{1}{\sqrt{2\pi}}\int_{-\infty}^x
	\mathrm{d}z\,
	e^{-z^2/2}
	=
	\frac{1}{2}\left(
		1+\mathrm{erf}(x/\sqrt{2})
	\right),
\end{align}
is the standard normal CDF.
The $j$th ($j\geq 1$) moment is then,
\begin{align}
	W_j(\mu,\sigma)
	&=
	\langle\mathrm{ReLU}(\sigma Z + \mu)^j\rangle
	=
	0+
	\frac{1}{\sqrt{2\pi}\sigma}
	\int_0^\infty 
	\mathrm{d}z\,
	z^j
	e^{-(z-\mu)^2/2\sigma^2}
	\\
	&
	=
	\sigma^j
	\int_{-\mu/\sigma}^\infty 
	\frac{\mathrm{d}z}{\sqrt{2\pi}}
	e^{-z^2/2}
	( z+\mu/\sigma)^j,
	\nonumber
	\\
	&=
	\frac{2^{-3/2}}{\sqrt{\pi}}(\sqrt{2} \sigma)^j
	\left[j\frac{\mu}{\sigma }  \Gamma \left(\frac{j}{2}\right) \, _1F_1\left(\frac{1-j}{2};\frac{3}{2};-\frac{\mu ^2}{2 \sigma ^2}\right)
	+
	\sqrt{2} \Gamma \left(\frac{j+1}{2}\right) \, _1F_1\left(-\frac{j}{2};\frac{1}{2};-\frac{\mu ^2}{2 \sigma ^2}\right)\right]
	\nonumber
	,
\end{align}
where $\,_1 F_1$ is the confluent hypergeometric function of the first kind, and $\Gamma$ is the gamma function.
Observe that $W_j(\mu/\alpha , \sigma/\alpha) = \alpha^{-j}W_j(\mu,\sigma)$.
Additionally,
\begin{align}
	W_0(x,1)
	&= 1,\\
	W_1(x,1)
	&=
	\frac{1}{\sqrt{2\pi}} e^{-x^2/2} + x \Phi(x),
	\\
	W_2(x,1)
	&=
	\frac{1}{\sqrt{2\pi}}x e^{-x^2/2} + (1+x^2) \Phi(x).
\end{align}
It follows from integration by parts,
\begin{align}
	W_2(x,1) = \Phi(x) + x W_1(x,1).
	\label{intByPartsID}
\end{align}
Additionally, for a random variable $\Theta(\sigma Z + \mu)$, the PDF is,
\begin{align}
	p_{\Theta(\sigma Z + \mu)} (z)
	=
	\frac{1}{2}
	\qty[
		1 + \mathrm{erf}\left(
			\frac{\mu}{\sigma\sqrt{2}}
		\right)
	]
	\delta(z-1)
	+
	\frac{1}{2}
	\mathrm{erfc}\left(
			\frac{\mu}{\sigma\sqrt{2}}
		\right)
	\delta(z),
\end{align}
so the $j$th moment ($j \geq 1$) is,
\begin{align}
	\langle\Theta(\sigma Z + \mu)^j\rangle
	=
	0
	+
	\frac{1}{2}
	\qty[
		1 + \mathrm{erf}\left(
			\frac{\mu}{\sigma\sqrt{2}}
		\right)
	] 1^j
	=
	\frac{1}{2}
	\qty[
		1 + \mathrm{erf}\left(
			\frac{\mu}{\sigma\sqrt{2}}
		\right)
	]
	=
	\Phi(\mu/\sigma).
\end{align}

\subsection{Final self-consistency equations}

Some essential quantities of interest are the expected fraction of surviving species $\phi_N$ and fraction of non-depleted resources $\phi_R$.
These quantities are computed using the moments calculated in section \ref{appendix:ReLUNormal} and Eqs.~\ref{eq:N0selfconsist},~\ref{eq:R0selfconsist}:
\begin{align}
	&
	\boxed{\phi_N =
	\expval{
		\Theta(\overline N_0)
	}
	=
	\Phi(\Delta_g)},
	\label{eq:phiNselfconsist}
	\\
	&
	\boxed{\phi_R = 
	\expval{
		\Theta(\overline R_0)
	}
	=
	\Phi(\Delta_\kappa)},
	\label{eq:phiRselfconsist}
\end{align}
where $\Delta_g = g/\sigma_g$ and $\Delta_\kappa = \kappa/\sigma_\kappa$ and $\Theta$ is the Heaviside step function with the convention $\Theta(0) = 0$.
Next, we can differentiate our expressions for $\overline N_0$ and $\overline R_0$ to find,
\begin{align}
	\nonumber
	\frac{\partial \overline N_0}{\partial m}
	&=
	\frac{\partial}{\partial m}
	\frac{g + \sigma_g Z_N}{\sigma_c \sigma_e \rho \chi}
	\Theta \left(\overline N_0\right)
	=
	-
	\frac{1}{\sigma_c \sigma_e \rho \chi}
	\Theta(\overline N_0)
	+
	\left[\text{$\overline N_0 \delta(\overline N_0)$-term}\right]
	\\
	\implies
	\expval{\frac{\partial \overline N_0}{\partial m}}
	&=
	\boxed{
		\nu
	=
	-\frac{\phi_N}{\sigma_c \sigma_e \rho \chi}}
	\label{eq:nuselfconsist}
	\\
	\nonumber
	\frac{\partial \overline R_0}{\partial K}
	&=
	\frac{\partial }{\partial K}
	\frac{\kappa + \sigma_\kappa Z_R}{1-\rho \sigma_e \sigma_c \gamma^{-1}\nu}
	\Theta(\overline R_0)
	=
	\frac{1}{1-\rho \sigma_e \sigma_c \gamma^{-1}\nu}
	\Theta(\overline R_0)
	+
	\left[\text{$\overline R_0 \delta(\overline R_0)$-term}\right]
	\\
	\implies
	\expval{
		\frac{\partial \overline R_0}{\partial K}
	}
	&=
	\boxed{\chi
	=
	\frac{\phi_R}{1-\rho \sigma_e \sigma_c \gamma^{-1}\nu}}
	\label{eq:chiselfconsist}
\end{align}
We can solve these two equations for $\chi,\nu$ to obtain the relations,
\begin{align}
	\nu = \frac{\gamma^{-1}\phi_N / \phi_R}{\rho \sigma_c \sigma_e \gamma^{-1}(\gamma^{-1}\phi_N / \phi_R - 1)}
	,
	\qquad
	\chi =
	\phi_R - \gamma^{-1}\phi_N.
	\label{eq:solvedSuscept}
\end{align}

Next, we use Eqs.~\ref{eq:N0selfconsist} and \ref{eq:R0selfconsist} and invoke our assumption that the system self-averages to find,
\begin{align}
	&
	\label{eq:Nselfconsist}
	\boxed{\expval{N}
	=
	\expval*{
		\overline N_0
	}
	=
	\frac{\sigma_g}{\sigma_c\sigma_e\rho\chi}
	W_1(\Delta_g,1)
	=
	\frac{\sigma_g}{\sigma_c\sigma_e\rho\chi}
	\left(
		\frac{e^{-\Delta_g^2/2}}{\sqrt{2\pi}}
		+
		\Delta_g \Phi(\Delta_g)
	\right)}
	,
	\\
	&
	\label{eq:Rselfconsist}
	\boxed{	\expval{R}
	=
	\expval*{\overline R_0}
	=
	\frac{\sigma_\kappa}{1-\rho \sigma_e \sigma_c \gamma^{-1}\nu}
	W_1(\Delta_\kappa,1)
	=
	\frac{\sigma_\kappa}{1-\rho \sigma_e \sigma_c \gamma^{-1}\nu}
	\left(
		\frac{e^{-\Delta_\kappa^2/2}}{\sqrt{2\pi}}
		+
		\Delta_\kappa \Phi(\Delta_\kappa)
	\right)}
	,
	\\
	&
	\label{eq:qNselfconsist}
	\boxed{q_N
	=
	\expval*{\overline N_0^2}
	=
	\left(
		\frac{\sigma_g}{\sigma_c\sigma_e\rho\chi}
	\right)^2
	W_2(\Delta_g,1)
	=
	\left(
		\frac{\sigma_g}{\sigma_c\sigma_e\rho\chi}
	\right)^2
	\left(
		\frac{\Delta_g e^{-\Delta_g^2/2}}{\sqrt{2\pi}}
		+
		(1+\Delta_g^2)
		\Phi(\Delta_g)
	\right)}
	,
	\\
	&
	\label{eq:qRselfconsist}
	\boxed{q_R
	=
	\expval*{\overline R_0^2}
	=
	\left(
		\frac{\sigma_\kappa}{1-\rho \sigma_e \sigma_c \gamma^{-1}\nu}
	\right)^2
	W_2(\Delta_\kappa,1)
	=
	\left(
		\frac{\sigma_\kappa}{1-\rho \sigma_e \sigma_c \gamma^{-1}\nu}
	\right)^2
	\left(
		\frac{\Delta_\kappa e^{-\Delta_\kappa^2/2}}{\sqrt{2\pi}}
		+
		(1+\Delta_\kappa^2)\Phi(\Delta_\kappa)
	\right)}
	.
\end{align}
The equations \ref{eq:phiNselfconsist}, \ref{eq:phiRselfconsist}, \ref{eq:nuselfconsist}, \ref{eq:chiselfconsist}, \ref{eq:Nselfconsist}, \ref{eq:Rselfconsist}, \ref{eq:qNselfconsist}, \ref{eq:qRselfconsist} constitute the cavity self-consistency equations for the aMCRM model.
They are eight independent nonlinear equations to solve for eight variables: $\phi_N, \phi_R, \chi, \nu, \expval{N}, \expval{R}, q_N, q_R$.
These equations are solved numerically using nonlinear least squares as discussed in SI section~\ref{appendix:solving-self-consist}.







\begin{figure}[t]
	\centering
	\includegraphics[width=0.85\linewidth]{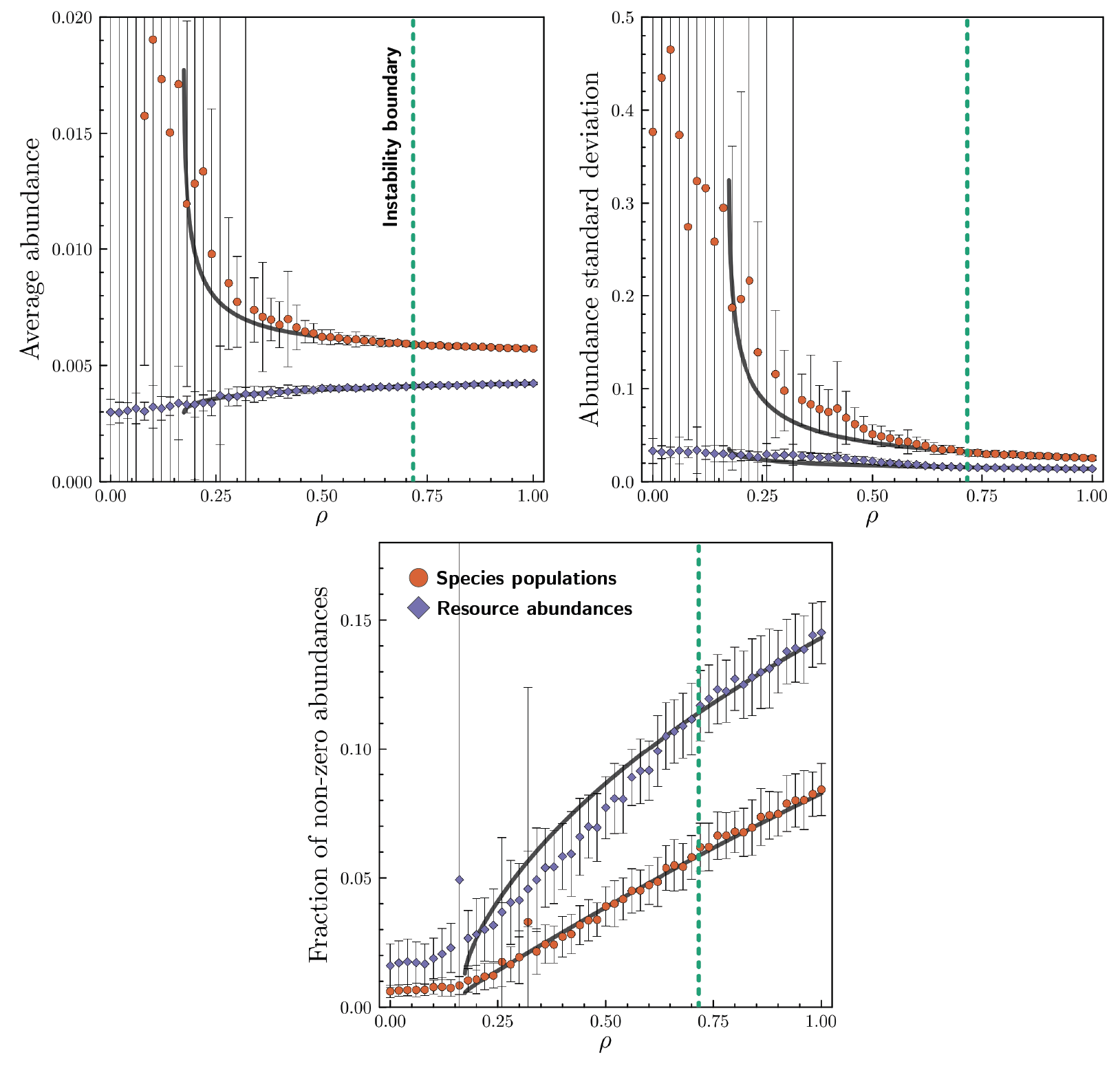}
	\caption{\label{fig:cavity-slice}
		Comparison of cavity results to numerical simulations for various $\rho$ and fixed $\sigma_c = \sigma_e = 3.5$, corresponding to the gray dashed line in Fig.~\ref{fig:phase-diagram}.
		(a) Average populations of species, $\langle N \rangle$, and abundances of resources, $\langle R \rangle$.
		(b) Standard deviations of species populations, $\sqrt{\langle N^2 \rangle - \langle N \rangle^2}$, and resource abundances, $\sqrt{\langle R^2 \rangle - \langle R \rangle^2}$.
		(c) Fractions of surviving species, $\phi_N$, and non-depleted resources, $\phi_R$.
		Points are means of numerical simulations, lines are cavity results.
		Error bars are standard deviations of numerical simulations; some error bars are smaller than the points.
		A dashed line is shown at $\rho^\star = 0.72$, the critical value of $\rho$ at which the system transitions between the stable and dynamic phases.
		For each value of $\rho$, $64$ simulations are performed with $M,S =512$ resources and species.
	}
\end{figure}

\begin{figure}[t]
	\centering
	\includegraphics[width=0.85\linewidth]{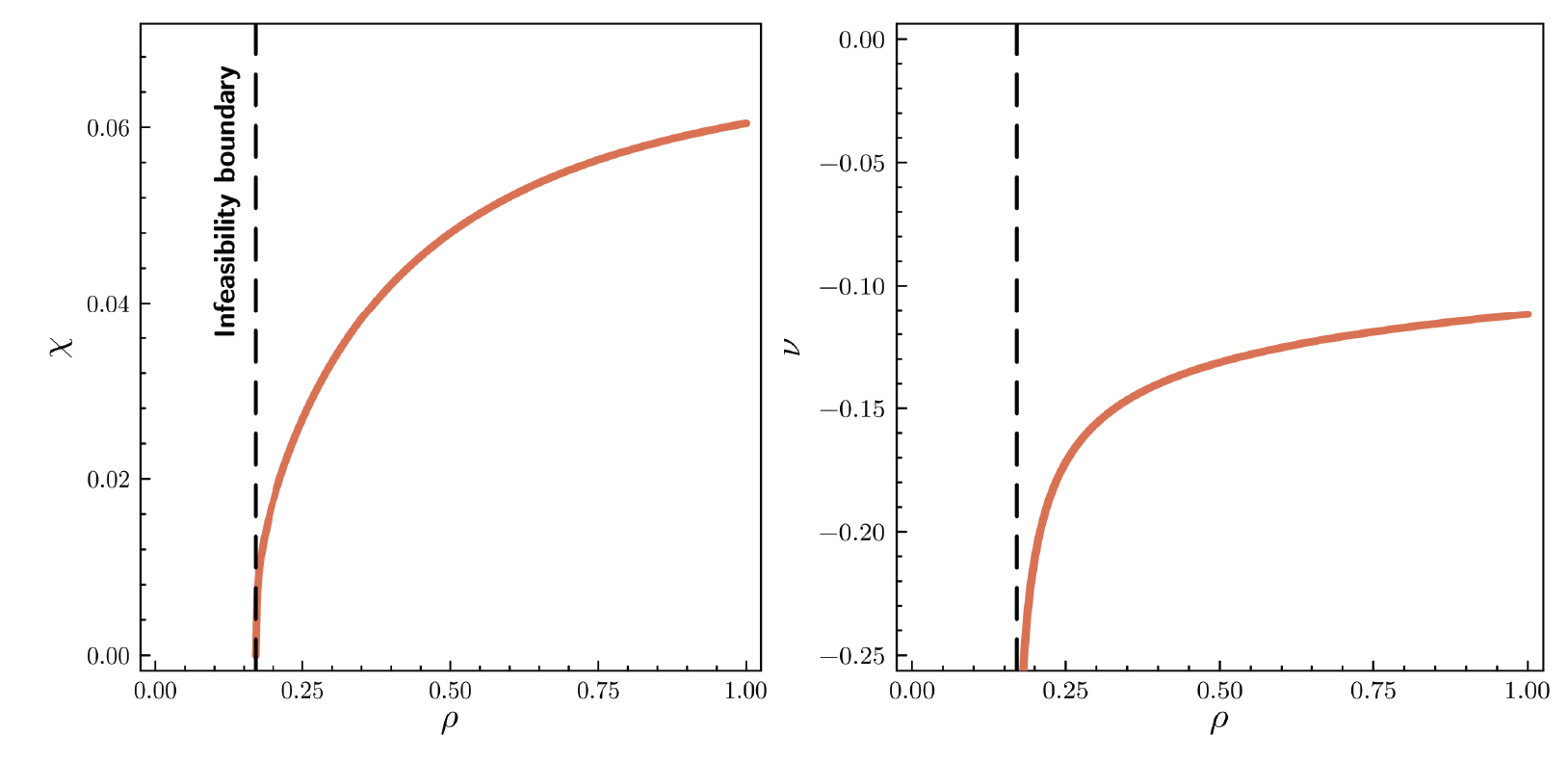}
	\caption{
		\label{fig:cavity-suscep-slice}
		Cavity susceptibilities for the same parameters as in Fig.~\ref{fig:cavity-slice}.
		The susceptibilities $\chi$, $\nu$ as defined in lines \ref{eq:chi-def}, \ref{eq:nu-def} are shown on the left and right, respectively.
	}
\end{figure}

\begin{figure}[t]
	\centering
	\includegraphics[width=0.495\linewidth]{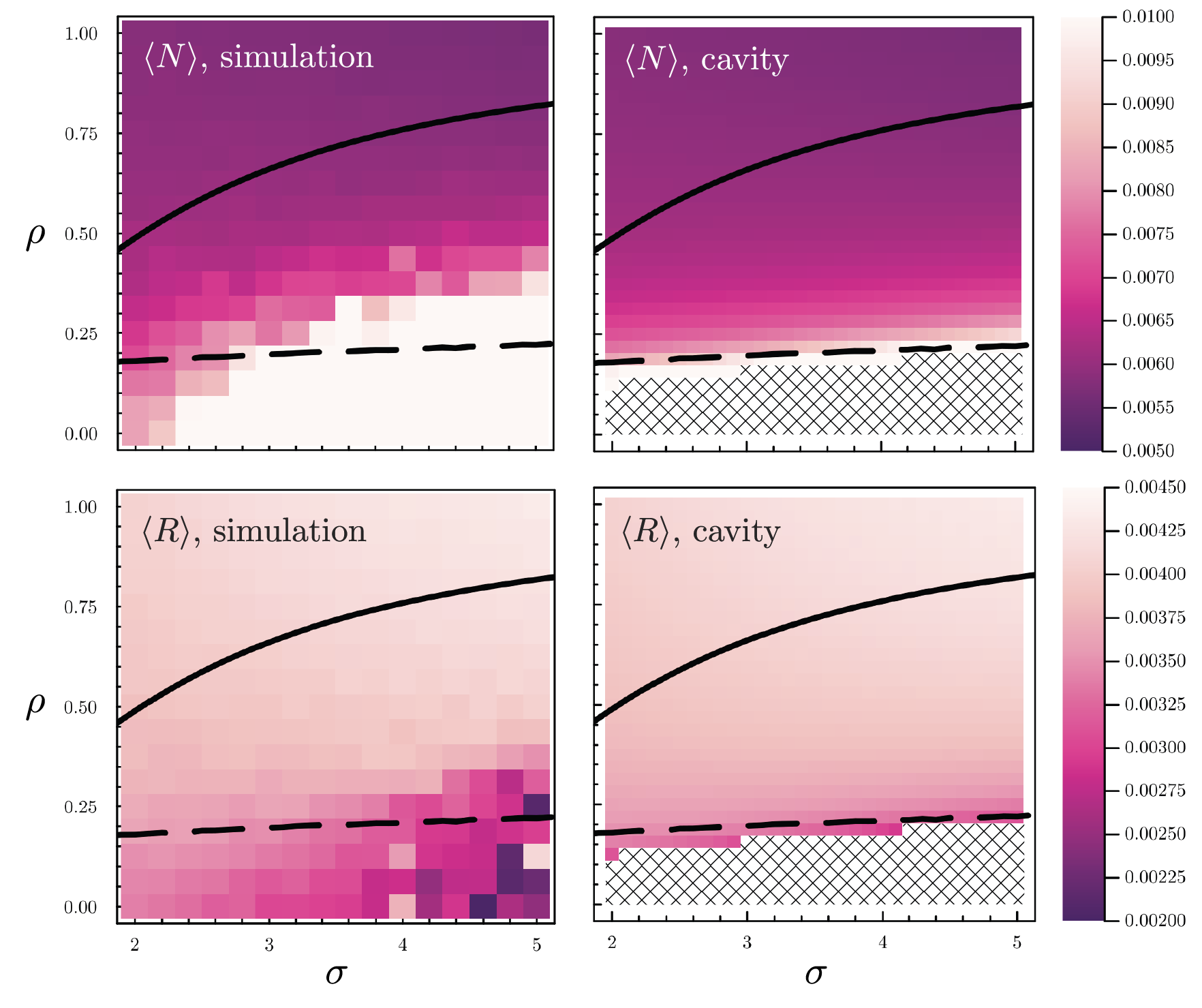}
	\includegraphics[width=0.495\linewidth]{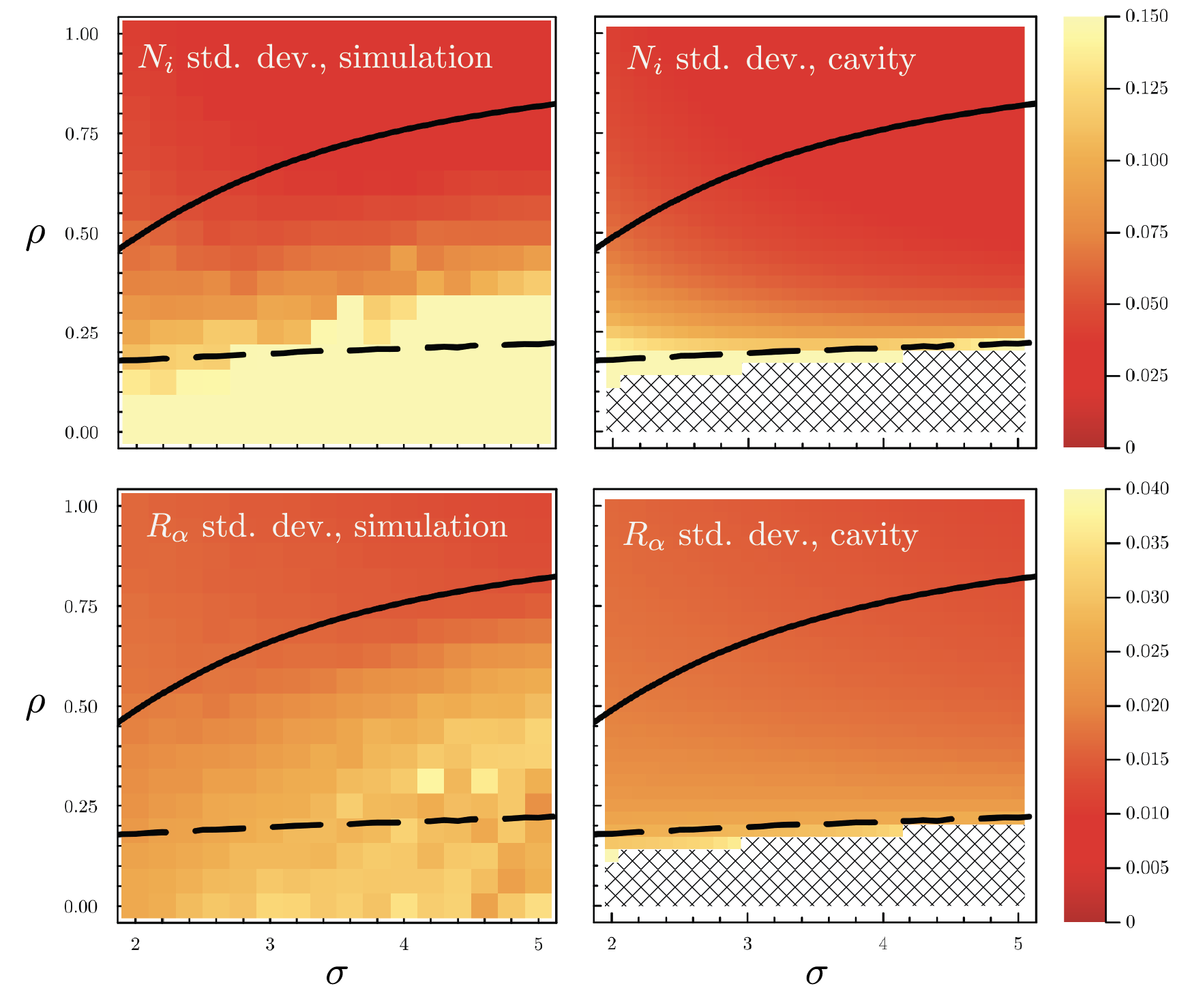}
	\includegraphics[width=0.495\linewidth]{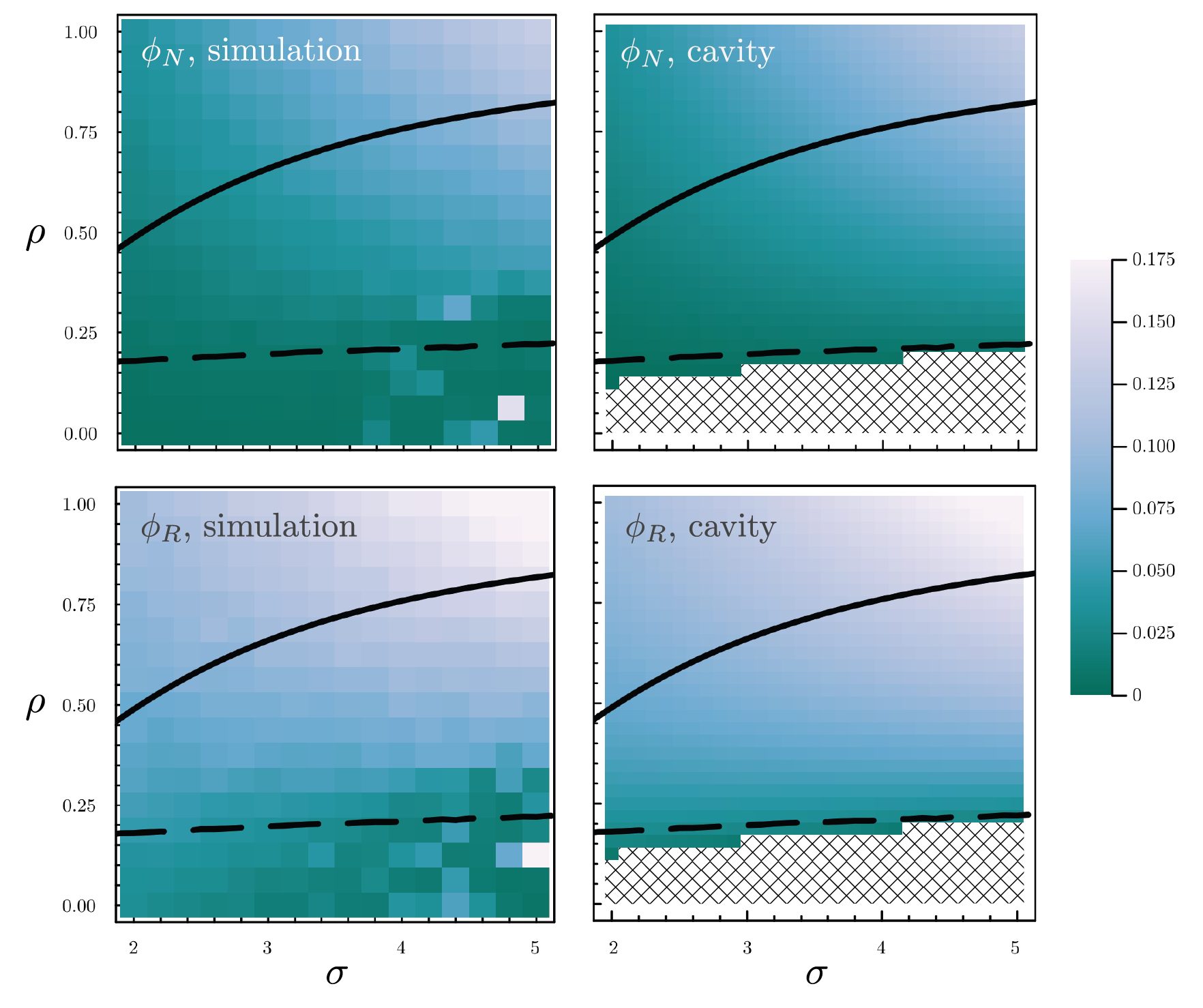}
	\caption{\label{fig:cavity-heatmaps}
	Heatmap comparison of cavity predictions and simulation results.
	Results from 32 simulations for various values of $\sigma$ and $\rho$ are displayed in the left column of each pane and the cavity predictions are displayed in the right column.
	The instability boundary is shown as a solid black line while the infeasibility boundary is shown as a dashed black line.
	For the cavity prediction plots, the parameter regions for which no cavity solution exists have a diamond pattern overlaid (c.f., Fig.~\ref{fig:LSQobjs} and SI section~\ref{appendix:solving-self-consist}).
	The parameters used are those used in Fig.~\ref{fig:phase-diagram}(a).
	}
\end{figure}

\begin{figure}[t]
	\centering
	\includegraphics[width=0.7\linewidth]{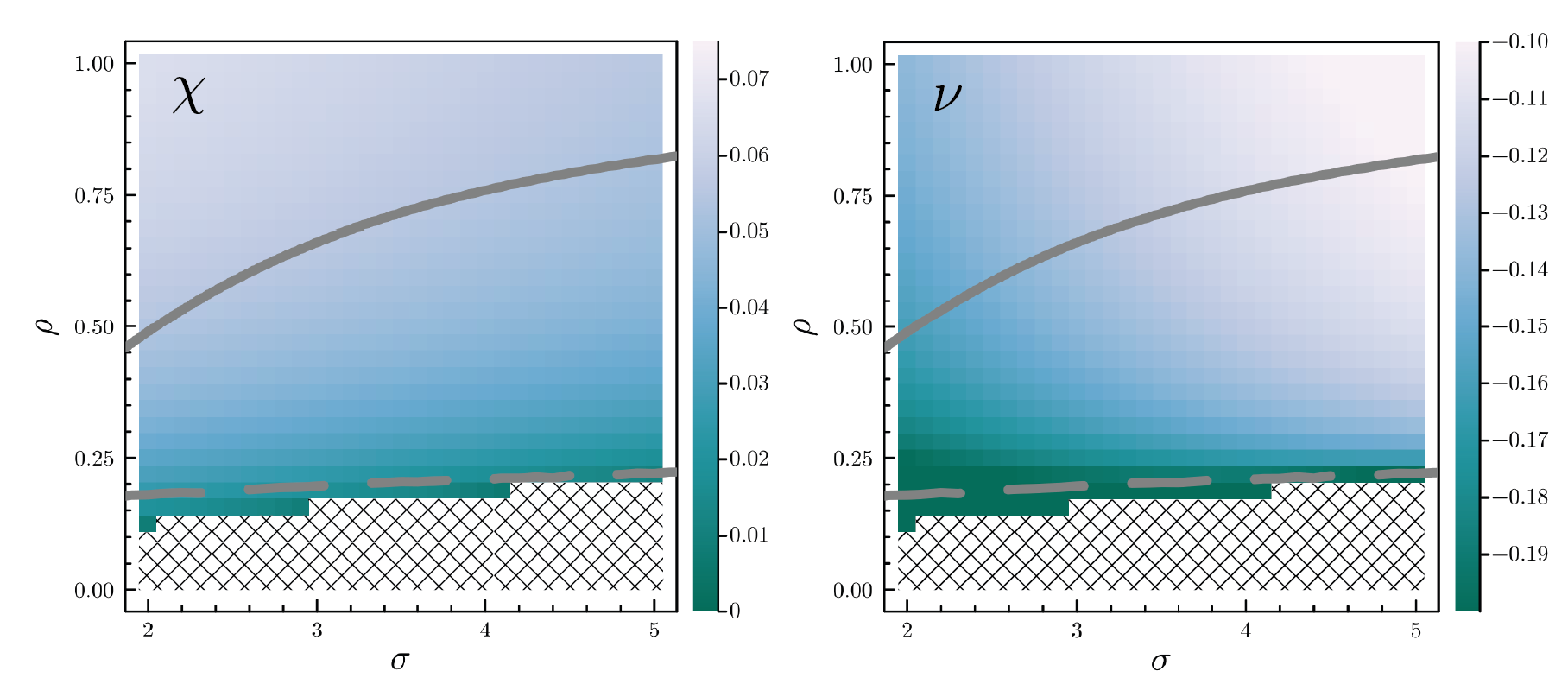}
	\caption{\label{fig:cavity-suscep-heatmap}
		Heatmap of cavity susceptibilities $\chi$, $\nu$ as defined in lines \ref{eq:chi-def}, \ref{eq:nu-def} for various values of $\sigma$ and $\rho$, as in Fig.~\ref{fig:cavity-heatmaps}.
	}
\end{figure}

\subsection{Infeasibility of the cavity solution\label{appendix:cavity-infeas}}

Next, we will investigate when there may exist a solution to the cavity self-consistency equations.
Observe that Eq.~\eqref{eq:nuselfconsist} implies that as $\chi \to 0$, $\nu$ diverges; Eqs.~\ref{eq:N0selfconsistsoln},~\ref{eq:R0selfconsistsoln} indicate that $\overline N_0$ and $\overline R_0$ become singular.
This is when, numerically, there fails to exist a solution to the cavity self-consistency equations.
From Eq.~\eqref{eq:chiselfconsist}, we see that $\chi \to 0$ when,
\begin{gather}
	\phi_R^{\text{CSI}}-\gamma^{-1} \phi_N^{\text{CSI}} = 0
	\Longleftrightarrow
	\frac{\phi_R^\text{CSI}}{\phi_N^\text{CSI}}
	\frac{M}{S}
	=
	1
	\\
	\nonumber
	\Downarrow
	\\
	\boxed{\text{cavity solution infeasibility boundary:}\qquad\text{\# of non-depleted resources} = \text{\# of surviving species}.}
	\label{eq:cavity-CSI}
\end{gather}
The cavity solution no longer has a solution when the ecosystem approaches the bound set by the principle of competitive exclusion.
According to this calculation, $\nu$ diverging here means that if a species becomes marginally more fit when all niches are packed, it will disrupt the ecosystem.
This relation defines a ninth equation for eight variables, so it determines a co-dimension-one boundary in the space of parameters.
This boundary can be calculated numerically using nonlinear least squares as discussed in SI section~\ref{appendix:solving-self-consist}.
When solving the cavity self-consistency equations numerically, we will see that there is a region of parameter space beyond this boundary where the least squares objectives are large compared to machine error (see Fig.~\ref{fig:LSQobjs} in SI section~\ref{appendix:solving-self-consist}).

This boundary corresponds to the transition to the ``unbounded growth'' phase in the Lotka--Volterra literature (see Ref.~\cite{Bunin2017}) because $\ev{N} \to \infty$ as $\chi\to0$ as can be seen in Eq.~\eqref{eq:Nselfconsist}.
There is no unbounded growth in this model due to the negative feedback structure of consumer-resource models.
In the cavity solution, this can be seen through $\ev{R} \to 0$ as $\nu \to -\infty$ in Eq.~\eqref{eq:Rselfconsist}.
That is, species abundances are predicted to diverge while abundances of all resources become zero.
Additionally, as $\chi \to 0$, $q_R \to 0$, so $\sigma_g = \sigma_m$ and $g = -m$, meaning $\phi_N^\text{CSI} = 1 - \Phi(m/\sigma_m)$; therefore $\phi_N$ and $\phi_R$ approach nonzero values at this boundary.

\subsection{Comparison of cavity and simulation results\label{appendix:cavity-sim-compare}}

For higher reciprocity levels $\rho$ when the system is in the stable phase, the analytic solution produced from the cavity method matches the simulation results remarkably well.
In Fig.~\ref{fig:cavity-slice}, we compare the cavity results to numerical simulations for various $0 \leq \rho \leq 1$ and fixed $\sigma_c = \sigma_e = 3.5$, corresponding to the gray dashed line in Fig.~\ref{fig:phase-diagram}(a).
For all predicted quantities, the chi-squared statistic is near the number of degrees of freedom, indicating that the cavity solution is a good fit to the simulation results.
The cavity susceptibilities $\chi, \nu$ for this slice are shown in Fig.~\ref{fig:cavity-suscep-slice}. In this figure, we see that the infeasibility boundary occurs when $\chi \to 0$ and $\nu \to -\infty$.

In Fig.~\ref{fig:cavity-heatmaps}, we compare the cavity results to numerical simulations for the various values $\rho$ and $\sigma_c = \sigma_e \equiv \sigma$ shown on the grid in Fig.~\ref{fig:phase-diagram}(a).
Again, the analytic solutions produced from the cavity method match the simulation results remarkably well in the stable phase.
Cavity susceptibilities $\chi, \nu$ for this grid are shown in Fig.~\ref{fig:cavity-suscep-heatmap}.

\subsection{Cavity susceptibilities}

The cavity susceptibilities,
\begin{equation}
\begin{gathered}
	\chi
	\equiv
	\frac{1}{M}
	\sum_{\alpha=1}^M
	\pdv{R_\alpha}{K_\alpha},
	\qquad
	\nu
	\equiv
	\frac{1}{S}
	\sum_{i=1}^S
	\pdv{N_i}{m_i},
\end{gathered}
\end{equation}
have physical significance: $\chi$ and $\nu$ are the average linear-order response of a resource's abundance to a small change in its carrying capacity and a species' population to a small change in its natural mortality rate, respectively.
These susceptibilities could be measured numerically by introducing small, nonzero-mean perturbations to $K_\alpha$ and measuring $\sum_{\alpha=1}^M \Delta R_\alpha/\sum_{\alpha=1}^M \Delta K_\alpha$ when the system is self-averaging.

At steady state, the full susceptibility matrices can be computed using just the knowledge of which species and resources.
Consider that the surviving species and resources satisfy the following matrix equation:
\begin{equation}
\begin{aligned}
	\begin{bmatrix}
		0 & c^\ast
		\\
		(e^\ast)^{\operatorname{T}} & 1
	\end{bmatrix}
	\begin{bmatrix}
		N^\ast
		\\
		R^\ast
	\end{bmatrix}
	=
	\begin{bmatrix}
		m^\ast\\K^\ast
	\end{bmatrix},
\end{aligned}
\end{equation}
where $X^\ast$ represents a matrix or vector with rows and/or columns corresponding to species and resource which are not surviving removed.
This block-matrix equation can be solved to find,
\begin{equation}
	\begin{bmatrix}
		N^\ast
		\\
		R^\ast
	\end{bmatrix}
	=
	\begin{bmatrix}
		-\qty( c^\ast (e^\ast)^{\operatorname{T}} )^{-1}
		&
		\qty( c^\ast (e^\ast)^{\operatorname{T}} )^{-1} c^\ast
		\\
		(e^\ast)^{\operatorname{T}}\qty( c^\ast (e^\ast)^{\operatorname{T}} )^{-1}
		&
		1 - (e^\ast)^{\operatorname{T}}( c^\ast (e^\ast)^{\operatorname{T}} )^{-1}c^\ast
	\end{bmatrix}
	\begin{bmatrix}
		m^\ast \\ K^\ast
	\end{bmatrix}.
\end{equation}
Given that perturbations are sufficiently small so that no species or resources go extinct and no new species or resources can invade, the susceptibility matrices are,
\begin{equation}
	\begin{aligned}
		&
		[\nu_{ij}^{(N)}]=
		-\qty( c^\ast (e^\ast)^{\operatorname{T}} )^{-1}
		,\qquad
		&&
		[\chi_{i\beta}^{(N)}]=
		\qty( c^\ast (e^\ast)^{\operatorname{T}} )^{-1} c^\ast,
		\\
		&
		[\nu_{\alpha j}^{(R)}]=
		(e^\ast)^{\operatorname{T}}\qty( c^\ast (e^\ast)^{\operatorname{T}} )^{-1}
		,\qquad
		&&
		[\chi_{\alpha\beta}^{(R)}]=
		1 - (e^\ast)^{\operatorname{T}}( c^\ast (e^\ast)^{\operatorname{T}} )^{-1}c^\ast.
	\end{aligned}
\end{equation}
Therefore,
\begin{equation}
	\nu
	=
	-
	\frac{1}{S}
	\tr \qty(c^\ast (e^\ast)^{\operatorname{T}})^{-1},
	\qquad
	\chi
	=
	\frac{1}{M}
	\tr
	\qty[1
	-
	\qty(e^\ast)^{\operatorname{T}}( c^\ast (e^\ast)^{\operatorname{T}} )^{-1}c^\ast].
\end{equation}
This means that when a steady state exists, the cavity susceptibilities can be computed exactly with knowledge of which species and resources are surviving.
Notice that $\qty(e^\ast)^{\operatorname{T}}( c^\ast (e^\ast)^{\operatorname{T}} )^{-1}c^\ast$ is an oblique projector because it is idempotent.
The trace of this matrix is the dimension of the range of the projector.
Therefore, the infeasibility boundary in the cavity solution occurs (or equivalently, the niches are fully packed) when the projector is full-rank.

\section{Stability phase transition in the thermodynamic limit\label{appendix:stability-phase-transition}}

In nonlinear dynamics, a common approach to assess the appearance of instability is to begin with an assumption of stability and see what breaks down when the system is perturbed.
In the case of the aMCRM, when there is stability, the results from the cavity solution will be valid, so we will use these results to determine the critical threshold for instability.

We consider perturbing all surviving species and resources, $\overline N_i^+$, $\overline R_\alpha^+$, by small amounts, $\varepsilon \eta_i^{(N)}$, $\varepsilon \eta_\alpha^{(R)}$, respectively, where $\eta_i^{(N)}$ and $\eta_\alpha^{(R)}$ are independent random variables all with mean zero and variance one~\cite{Bunin2017}.
By considering perturbations on surviving species and resources and using cavity results (c.f.,~Eqs.~\ref{eq:R0selfconsistsoln},~\ref{eq:N0selfconsistsoln}), we are assuming that we are working with an {\em uninvadable} steady state.
From Eqs.~\ref{eq:N0beforeconsist},~\ref{eq:R0beforeconsist}, for surviving species and non-depleted resources,
\begin{gather}
	\overline N_0^+
	= 
	\frac{1}{\sigma_c \sigma_e \rho \chi }
	\qty[	g 
	+
	\frac{\sigma_c}{\sqrt{M}}
	\sum_{\alpha :\overline R_\alpha>0} d_{0\alpha} \overline R_{\alpha \setminus 0}^+
	-
	\sigma_m \delta m_0
	],
	\\
	\overline R_0^+
	=
	\frac{1}{1
	-
	\rho \sigma_e \sigma_c \gamma^{-1} \nu }
  \qty[
	\kappa
	-
	\frac{\sigma_e}{\sqrt{M}}
	\sum_{i:\overline N_i>0}
	\overline N_{i \setminus 0}^+
	\left(
		\rho d_{i0}
		+
		\sqrt{1-\rho^2}x_{i0}
	\right)
	+
	\sigma_K \delta K_0
	].
\end{gather}
Applying the perturbation gives,
\begin{align}
	\overline N_0^+
	&= 
	\frac{1}{\sigma_c \sigma_e \rho \chi }
	\qty[
	g 
	+
	\frac{\sigma_c}{\sqrt{M}}
	\sum_{\alpha :\overline R_\alpha>0} d_{0\alpha} \left(
		\overline R_{\alpha \setminus 0}^+
		+
		\varepsilon \eta_{\alpha}^{(R)}
	\right)
	-
	\sigma_m \delta m_0
	],
	\\
	\overline R_0^+
	&=
	\frac{1}{1
	-
	\rho \sigma_e \sigma_c \gamma^{-1} \nu }
	\qty[
	\kappa
	-
	\frac{\sigma_e}{\sqrt{M}}
	\sum_{i:\overline N_i>0}
	\left(
		\overline N_{i \setminus 0}^+
		+
		\varepsilon \eta_i^{(N)}
	\right)
	\left(
		\rho d_{i0}
		+
		\sqrt{1-\rho^2}x_{i0}
	\right)
	+
	\sigma_K \delta K_0
	].
\end{align}
Differentiating with respect to $\varepsilon$ yields,
\begin{align}
	\dv{\overline N_0^+}{\varepsilon}
	&= 
	\frac{1}{ \sigma_e \rho \chi \sqrt{M}}
	\sum_{\alpha :\overline R_\alpha>0} d_{0\alpha} \left(
		\dv{\overline R_{\alpha \setminus 0}^+}{\varepsilon}
		+
		\eta_{\alpha}^{(R)}
	\right),
	\\
	\dv{\overline R_0^+}{\varepsilon}
	&=
	-\frac{{\sigma_e}/{\sqrt{M}}}{1
	-
	\rho \sigma_e \sigma_c \gamma^{-1} \nu }
	\sum_{i:\overline N_i>0}
	\left(
		\dv{\overline N_{i \setminus 0}^+}{\varepsilon}
		+
		\eta_i^{(N)}
	\right)
	\left(
		\rho d_{i0}
		+
		\sqrt{1-\rho^2}x_{i0}
	\right).
\end{align}
Because $\overline R_{\alpha \setminus 0}^+$, $d_{0\alpha}$, $\eta_\alpha^{(R)}$ are all independent, $\langle d \overline R_0^+/ \mathrm d\varepsilon\rangle = 0$; similarly, $\langle d \overline N_0^+/ \mathrm d\varepsilon\rangle = 0$.
The first moment of the response to the perturbation does not give useful information about the stability; therefore, we turn to the second moment.
First, we square the above quantities:
\begin{align}
	\left[\dv{\overline N_0^+}{\varepsilon}\right]^2
	&= 
	\frac{1/M}{ (\sigma_e \rho \chi)^2 }
	\sum_{\alpha,\beta :\overline R_\alpha>0,\overline R_\beta>0}
	d_{0\alpha} 
	d_{0\beta} 
	\left(
		\dv{\overline R_{\alpha \setminus 0}^+}{\varepsilon}
		+
		\eta_{\alpha}^{(R)}
	\right)
	\left(
		\dv{\overline R_{\beta \setminus 0}^+}{\varepsilon}
		+
		\eta_{\beta}^{(R)}
	\right)
	,
	\\
	\left[
		\dv{\overline R_0^+}{\varepsilon}
	\right]^2
	&=
	\frac{\sigma_e^2/M}{
		(1
	-
	\rho \sigma_e \sigma_c \gamma^{-1} \nu)^2 }
	\sum_{i,j:\overline N_i>0,\overline N_j>0}
	\left(
		\dv{\overline N_{i \setminus 0}^+}{\varepsilon}
		+
		\eta_i^{(N)}
	\right)
	\left(
		\dv{\overline N_{j \setminus 0}^+}{\varepsilon}
		+
		\eta_j^{(N)}
	\right)
	\\
	\nonumber
	&\hspace{5cm}
	\times
	\left(
		\rho d_{i0}
		+
		\sqrt{1-\rho^2}x_{i0}
	\right)
	\left(
		\rho d_{j0}
		+
		\sqrt{1-\rho^2}x_{j0}
	\right)
	.
\end{align}
Averaging over all sources of randomness and using
$
\expval{\left[\dv{N_{0}}{\varepsilon}\right]^2}=S^{-1}\sum_{i=1}^S \left[\dv{N_{i\setminus 0}}{\varepsilon}\right]^2
$
and
$
\expval{\left[\dv{R_{0}}{\varepsilon}\right]^2}=M^{-1}\sum_{\alpha=1}^M \left[\dv{R_{\alpha\setminus 0}}{\varepsilon}\right]^2
$,
which follows from the assumption that the system self averages, we obtain:
\begin{align}
	\expval{
	\left[\dv{\overline N_0^+}{\varepsilon}\right]^2
	}
	&= 
	\frac{1/M}{ (\sigma_e \rho \chi)^2 }
	\sum_{\alpha,\beta :\overline R_\alpha>0,\overline R_\beta>0}
	\expval{
	d_{0\alpha} 
	d_{0\beta}
	}
	\left(
			\dv{\overline R_{\alpha \setminus 0}^+}{\varepsilon}
		\dv{\overline R_{\beta \setminus 0}^+}{\varepsilon}
		+
		\dv{\overline R_{\alpha \setminus 0}^+}{\varepsilon}
		\expval{\eta_{\beta}^{(R)}}
		\right.
		\\
		\nonumber
		&
		\hspace{6cm}
		\left.
		+
		\expval{\eta_{\alpha}^{(R)}}
		\dv{\overline R_{\beta \setminus 0}^+}{\varepsilon}
		+
		\expval{
		\eta_{\alpha}^{(R)}
		\eta_{\beta}^{(R)}}
	\right)
	\\
	&=
	\frac{1/M}{ (\sigma_e \rho \chi)^2 }
	\sum_{\alpha,\beta:\overline R_\alpha,\overline R_\beta>0}
	\delta_{\alpha\beta}
	\left(
				\dv{\overline R_\alpha\setminus}{\varepsilon}
				\dv{\overline R_{\beta\setminus0}}{\varepsilon}
		+
		\delta_{\alpha\beta}
	\right)
	\\
	&=
	\frac{\phi_R}{ (\sigma_e \rho \chi)^2 }
	\left(
		\expval{
			\left[
				\dv{\overline R_{0}^+}{\varepsilon}
			\right]^2
		}
		+
		1
	\right)
	,
\end{align}
\begin{align}
	\expval{
	\left[
		\dv{\overline R_0^+}{\varepsilon}
	\right]^2
	}
	&=
	\frac{\sigma_e^2/M}{
		(1
	-
	\rho \sigma_e \sigma_c \gamma^{-1} \nu)^2 }
	\sum_{i,j:\overline N_i>0,\overline N_j>0}
	\left(
		\dv{\overline N_{i \setminus 0}^+}{\varepsilon}
		\dv{\overline N_{j \setminus 0}^+}{\varepsilon}
		+
		\dv{\overline N_{i \setminus 0}^+}{\varepsilon}
		\expval{\eta_j^{(N)}}
	\right.
		\\
		\nonumber
		&\hspace{0.4cm}
	\left.
		+
		\expval{\eta_i^{(N)}}
		\dv{\overline N_{j \setminus 0}^+}{\varepsilon}
		+
		\expval{
		\eta_i^{(N)}
		\eta_j^{(N)}
		}
	\right)
	\\
	\nonumber
	&
	\qquad
	\times 
	\left(
		\rho^2 
		\expval{
		d_{i0}
		d_{j0}}
		+
		\rho
		\sqrt{1-\rho^2}
		\left(
		\expval{d_{i0}}
		\expval{x_{j0}}
		+
		\expval{x_{i0}}
		\expval{d_{j0}}
		\right)
		+
		(1-\rho^2)
		x_{i0}
		x_{j0}
	\right)
	\\
	&=
	\frac{\sigma_e^2(S/M)/S}{(1-\rho\sigma_e\sigma_c\gamma^{-1}\nu)^2}
	\sum_{i,j:\overline N_i>0,\overline N_j>0}
	\left(
			\dv{\overline N_{i\setminus0}^+}{\varepsilon}
			\dv{\overline N_{j\setminus0}^+}{\varepsilon}
		+
		\delta_{ij}
	\right)
	\left(
		\rho^2 \delta_{ij}
		+
		(1-\rho^2)\delta_{ij}
	\right)
	\\
	&=
	\frac{\sigma_e^2\gamma^{-1} \phi_N}{(1-\rho\sigma_e\sigma_c\gamma^{-1}\nu)^2}
	\left(
		\expval{
			\left[
				\dv{\overline N_0^+}{\varepsilon}
			\right]^2
		}
		+
		1
	\right)
	.
\end{align}

Solving this system of equations yields:
\begin{align}
	\label{eq:dN0deps-sol}
	&
	\boxed{\expval{
		\left[
			\dv{\overline N_0^+}{\varepsilon}
		\right]^2
	}
	=
	\frac{
		\phi_R \left(
			(1-\nu\rho\sigma_c\sigma_e
			\gamma^{-1})^2
			\sigma_e^{-2}
			+
			\gamma^{-1}\phi_N
		\right)
	}{
		\left[
			\rho\chi\left(
				1-\nu \rho \sigma_c \sigma_e\gamma^{-1}
			\right)
		\right]^2
		-\gamma^{-1}\phi_N\phi_R
	}},
	\\
	\label{eq:dR0deps-sol}
	&
	\boxed{
	\expval{
		\left[
			\dv{\overline R_0^+}{\varepsilon}
		\right]^2
	}
	=
	\frac{
		\gamma^{-1}\phi_N\left(
			\phi_R+(\rho\sigma_e\chi)^2
		\right)
	}{
		\left[
			\rho\chi\left(
				1-\nu \rho \sigma_c \sigma_e\gamma^{-1}
			\right)
		\right]^2
		-\gamma^{-1}\phi_N\phi_R
	}}.
\end{align}
\begin{figure}[t]
	\centering
	\includegraphics[width=0.7\linewidth]{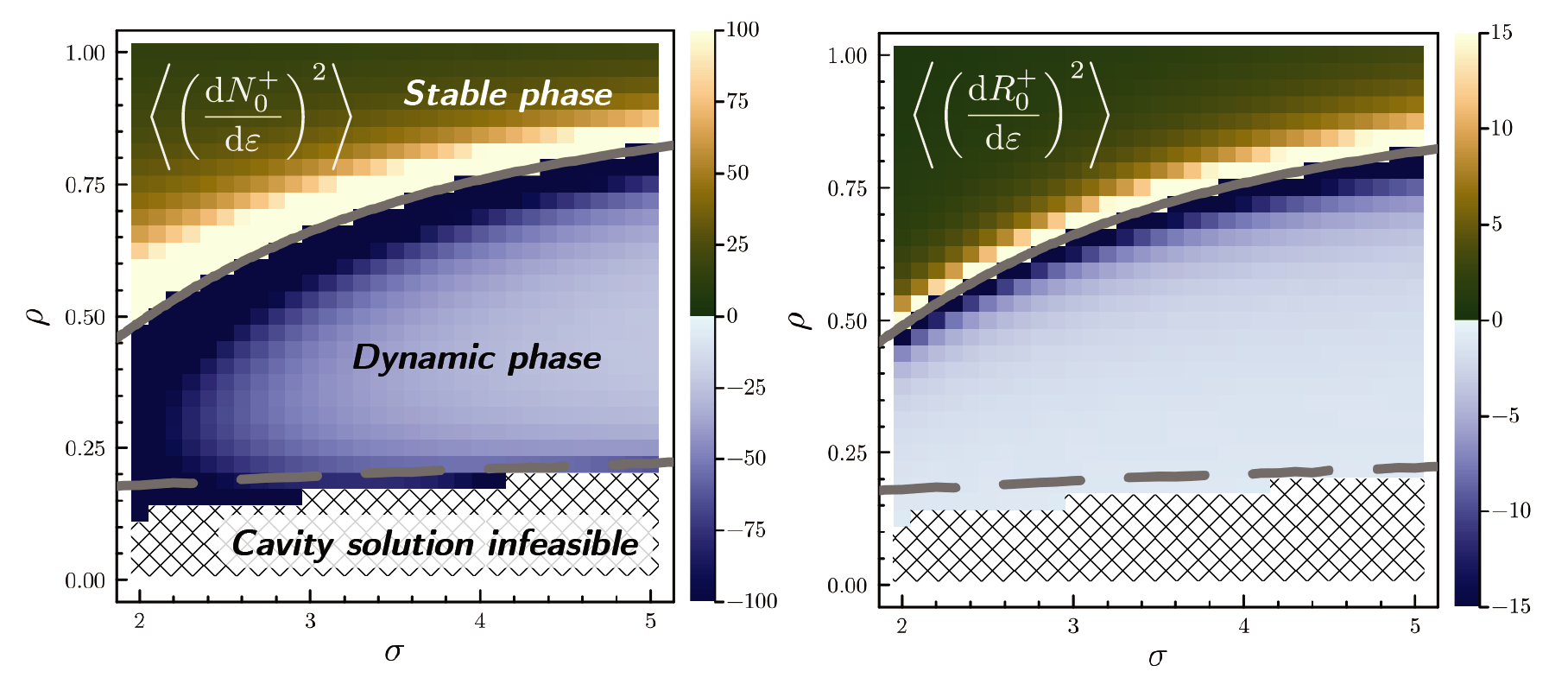}
	\caption{\label{fig:suscep-heatmaps}
		Heatmaps of the variances of the response of a surviving species and resources to a random perturbation (Eqs.~\ref{eq:dN0deps-sol},~\ref{eq:dR0deps-sol}) as a function of the parameters $\sigma$ and $\rho$ with other parameters matching those in Fig.~\ref{fig:phase-diagram}(a).
		The instability boundary is shown as a gray solid line, and the infeasibility boundary is shown as a gray dashed line.
	}
\end{figure}

\begin{figure}[ht]
	\centering
	\includegraphics[width=250px]{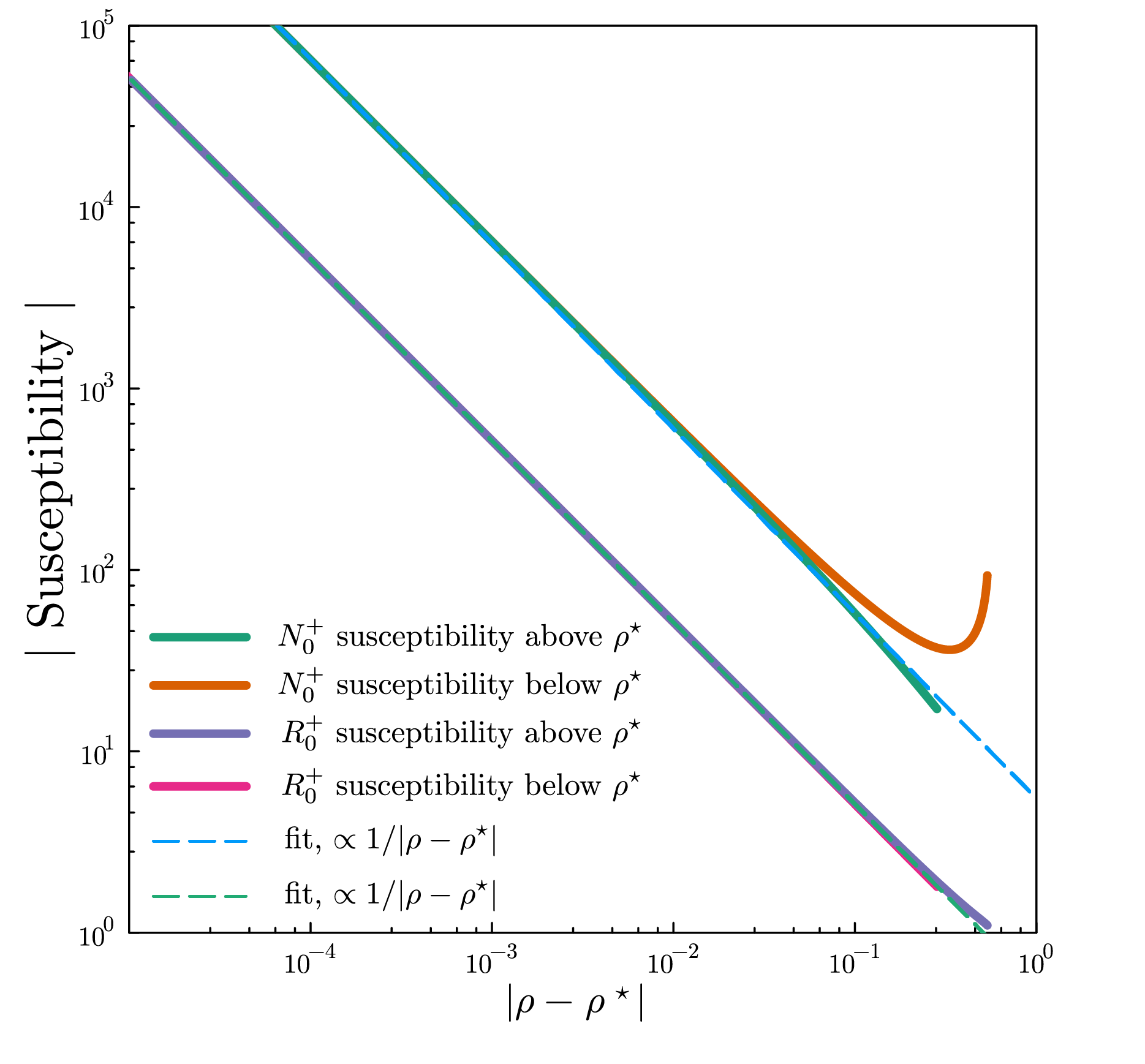}
	\caption{
		\label{fig:susceptibility-log-log}
		Log-log plot of the absolute value of the variances of the susceptibilities $ \langle \left(d N_0^+/\mathrm d\varepsilon \right)^2 \rangle$, $ \langle \left(d R_0^+/\mathrm d\varepsilon \right)^2 \rangle$ plotted as a function of the distance from the instability boundary, $|\rho-\rho^\star|$.
		The data is exactly that which is used in Fig.~\ref{fig:phase-diagram}(b).
		A fit $\propto 1/|\rho-\rho^\star|$ is shown for each susceptibility with dashed lines.
	}
\end{figure}
These susceptibilities are the variance of the response of a surviving species or resource when the system is subject to a random perturbation.
The divergence of these variances represent the breakdown of the mean-field approximation.
Further, when these susceptibilities diverge, the uninvadable fixed point predicted by the mean-field approximation becomes dynamically unstable.
Therefore, when these susceptibilities are divergent, whenever invasion is attempted, the system will exhibit dynamical fluctuations.
We call the boundary in parameter space at which these susceptibilities diverge the instability boundary.
These susceptibilities diverge when 
$
		0=\left[
		\rho\chi\left(
			1-\nu \rho \sigma_c \sigma_e\gamma^{-1}
		\right)
	\right]^2
	-\gamma^{-1}\phi_N\phi_R
	.
$
This condition along with the cavity self-consistency equations (Eqs.~\ref{eq:phiNselfconsist},~\ref{eq:phiRselfconsist},~\ref{eq:nuselfconsist},~\ref{eq:chiselfconsist},~\ref{eq:Nselfconsist},~\ref{eq:Rselfconsist},~\ref{eq:qNselfconsist},~\ref{eq:qRselfconsist}) determines a co-dimension-one boundary in the space of model parameters.
By using the relations in line~\ref{eq:solvedSuscept}, we obtain the following relation for the boundary at which the system becomes unstable:
\begin{gather}
	\phi_R^\star 
	\left(
		(\rho^\star)^2 \phi_R^\star  - \gamma^{-1}\phi_N^\star
	\right)
	=
	0
	\implies
	\left(
		\rho^\star
	\right)^2
	=
	\gamma^{-1} \frac{\phi_N^\star}{\phi_R^\star}
	\\
	\nonumber
	\Downarrow
	\\
	\boxed{\text{instability boundary occurs when:}
	\quad
	\left(\rho^\star\right)^2 = 
		\frac{(\text{\# of surviving species})^\star}{(\text{\# of non-depleted resources})^\star},
	}
	\label{eq:instability-boundary}
\end{gather}
where $X^\star$ denotes the value of a quantity $X$ at the instability boundary.
We are able to determine the critical level of asymmetry, $\rho^\star$, at which the ecosystem becomes unstable by solving a nonlinear least squares problem (see SI section~\ref{appendix:solving-self-consist}).
The variances of the susceptibilities $ \langle \left(\mathrm d N_0^+/\mathrm d\varepsilon \right)^2 \rangle$, $ \langle \left(\mathrm d R_0^+/\mathrm d\varepsilon \right)^2 \rangle$ are shown for various values of $\sigma = \sigma_c = \sigma_e$ and $\rho$ in Fig.~\ref{fig:suscep-heatmaps} with the instability and infeasibility boundaries overlaid.
These variances of susceptibilities additionally have an asymptotic power law dependence on the distance from the instability boundary, $|\rho-\rho^\star|$, as shown in Fig.~\ref{fig:susceptibility-log-log}.

The transition to the dynamic phase coincides in method to the transition to the ``multiple attractors'' phase discussed in the random generalized Lotka--Volterra model literature. This was first described in Ref.~\cite{Bunin2017}.

\section{Chaotic nature of dynamics\label{appendix:chaos}}


In the dynamic phase of the aMCRM, when an ecosystem is sufficiently large, the dynamics are chaotic.
In this section, we provide numerical evidence that the dynamics of the aMCRM are chaotic in the classical sense (i.e., sensitive dependence on initial conditions) and in the sense of unpredictability (i.e., no clear patterns in the dynamics).
As shown in Fig.~\ref{fig:diverging-trajecs-maxlyap_dotplot}(b) in the main text, the trajectories of the abundances of the surviving species and resources diverge from one another, indicating that the dynamics are chaotic in the classical sense.
Evidence of chaos and unpredictability in the dynamic phase is present in other numerical experiments, as well.

In Fig.~\ref{fig:dynamics-projection}, we show that the dynamics of the aMCRM in the dynamic phase are unpredictable in the sense that a trajectory does not display any clear patterns in the projection onto the first three principal components of the correlation matrix of the time series of the abundances of the surviving species and resources.
Such a projection is necessary to visualize the dynamics of the system, as they are very high-dimensional.

\begin{figure}[ht]
	\centering
	\includegraphics[width=0.7\linewidth]{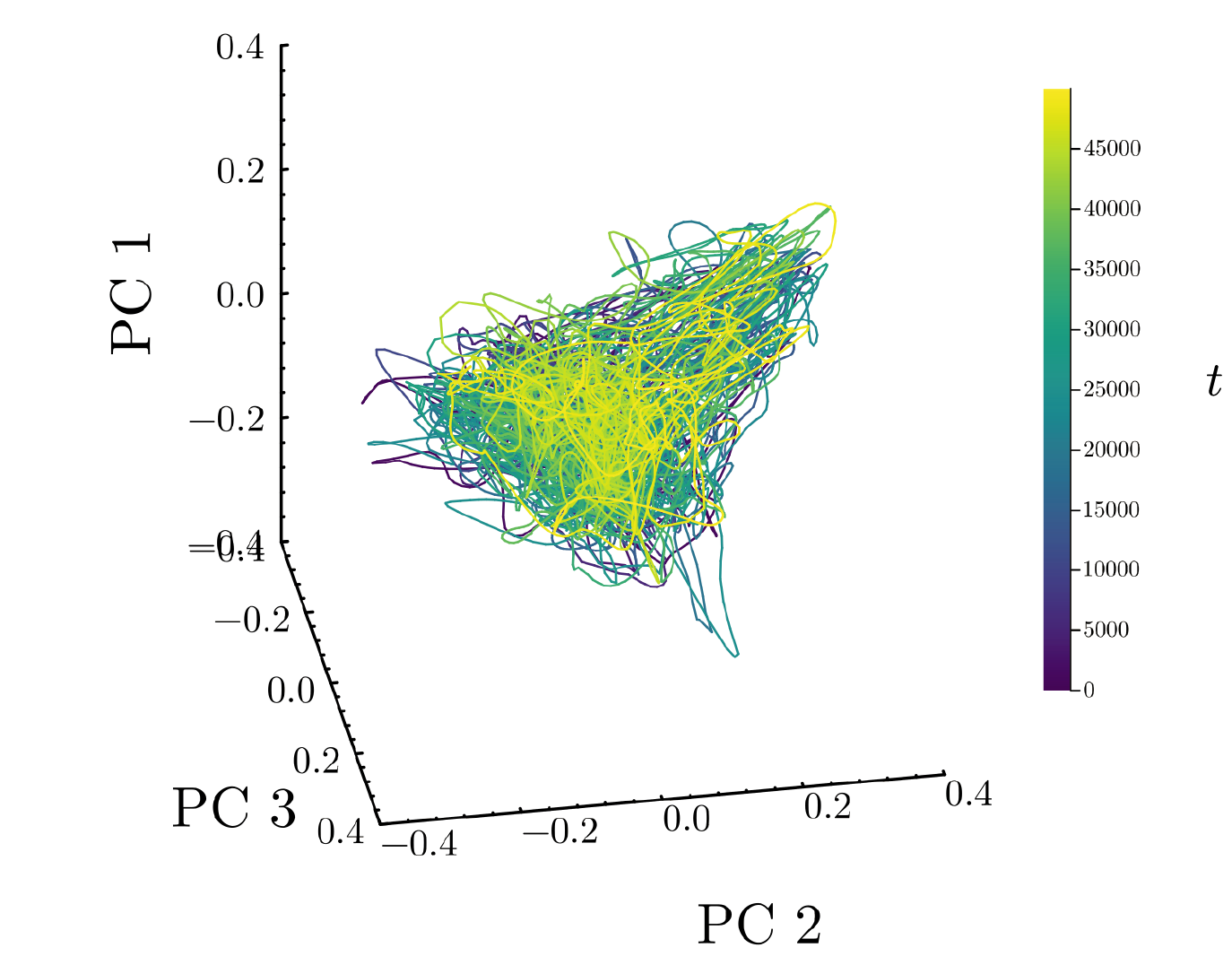}
	\caption{\label{fig:dynamics-projection}
	Projection of $M+S = 256+256 = 512$-dimensional dynamics onto the 3 highest-ranked eigenvectors of the correlation matrix (first 3 principal components) of the time series of the abundances.
	The lack of clustering of the trajectories in the projection indicates that the dynamics are chaotic and unpredictable.
	The parameters correspond to the gray star in Fig.~\ref{fig:phase-diagram}(a).
	Before plotting, the simulation was run for $10^4$ time units to eliminate any potential transients.
	}
\end{figure}


\subsection{Analysis of the Lyapunov exponents\label{appendix:lyapunov-exponent}}


In Fig.~\ref{fig:diverging-trajecs-maxlyap_dotplot}(a), we show the maximal Lyapunov exponent for simulations classified by whether they reach a steady state for various values of $\rho$.
To determine the Lyapunov spectrum, we use the \texttt{lyapunovspectrum} function in the \texttt{ChaosTools.jl} \textit{Julia} package which employs the `H2' method of Geist, originally stated in Benettin et al.~\cite{Datseris2018,Datseris2022,Benettin1980,Geist1990}.
This algorithm is described in detail in SI section~A of Ref.~\cite{Datseris2022}, and its applications are described in Chapter~3.2.
Conceptually, in order to compute the $k$th largest Lyapunov exponent, the algorithm evolves $n \geq k$ deviation vectors in the tangent space of the system and computes how the shape of the $n$-dimensional parallelepiped spanned by the deviation vectors evolve; the eigenvalues of the matrix describing the evolution of the parallelepiped are asymptotically related to the Lyapunov exponents.
The Jacobian of the system when the dynamical variable is considered as $(N_1,\dots,N_S,R_1,\dots,R_M)$ is:
\begin{align}
	\mathbf J (N_1,\dots,N_S,R_1,\dots,R_M) = 
	\left[
		\begin{array}{c|c}
			\frac{\partial \dot N_i}{ \partial N_j} & \frac{\partial \dot N_i}{ \partial R_\beta }\\
			\hline
			\frac{\partial \dot R_\alpha}{ \partial N_j }& \frac{\partial \dot R_\alpha}{ \partial R_\beta}
		\end{array}
	\right]
	=
	\left[
		\begin{array}{c|c}
			\delta_{ij}(\sum_{i=1}^S c_{i\alpha} R_\alpha - m_i) & N_ic_{i\beta} \\
			\hline
			-e_{j\alpha} R_\alpha  & \delta_{\alpha\beta}(K_\alpha - 2R_\alpha - \sum_{i=1}^S c_{i\alpha} N_i)
		\end{array}
	\right].
\end{align}
The Lyapunov spectrum is computed with $M=S=512$ species and resources, and the initial deviation vectors are chosen to be $8$ unit vectors in the direction of randomly chosen species and resources.
The parameters passed to the \texttt{lyapunovspectrum} function are the number of steps, \texttt{N = 1000}, the time step, \texttt{$\Delta$t = 5}, and the time to wait before starting to record the Lyapunov spectrum, \texttt{Ttr = 1000}.
The initial conditions are $N_i(0) = 1/S$ and $R_\alpha(0) = 1/M$.

As seen in Fig.~\ref{fig:diverging-trajecs-maxlyap_dotplot}(a), the maximal Lyapunov exponent for a simulation that does not reach a steady state is positive while the maximal Lyapunov exponent for a simulation that does reach a steady state is negative.
A simulation is categorized as reaching a steady state if the mean absolute value of all derivatives of the abundances at the end of the simulation is less than $10^{-6}$.
These results indicates that the dynamics are chaotic in the classical sense for the simulations which have persistent fluctuations.
The impact of system size on the probability of reaching a steady state is discussed in SI section~\ref{appendix:finite-size-scaling}.



\subsection{Analysis of the generalized alignment index (GALI)\label{appendix:gali}}

\begin{figure}[ht]
	\centering
	\includegraphics[width=0.7\linewidth]{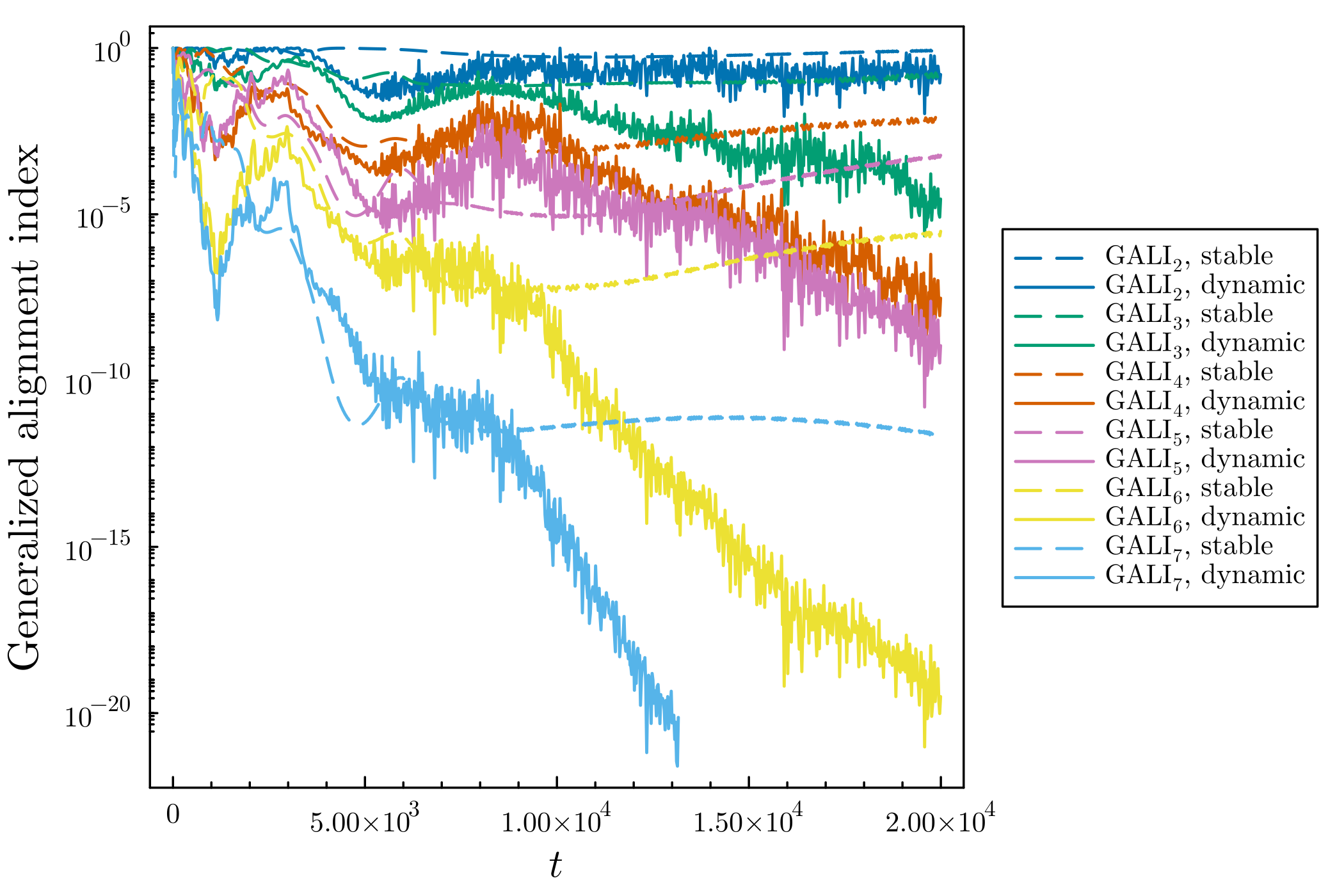}
	\caption{\label{fig:gali}
	Generalized alignment indices (GALIs) for simulations in the dynamic and stable phases with $M=S=512$ species and resources.
	The simulations all have the same sampled parameters; only $\rho$ and order of the GALI are varied.
	For the simulations in the dynamic phase (solid lines), the GALI asymptotically decays exponentially with time, indicating chaos, while for the simulations in the stable phase (dashed lines), the GALI asymptotically approaches a nonzero value, indicating the dynamics achieve a steady state.
	Simulation parameters for the GALI in the stable and dynamic phases correspond to the gray dot and star in Fig.~\ref{fig:phase-diagram}(a), respectively.
	}
\end{figure}

An alternative method to determine whether a system is chaotic is to use the generalized alignment index~(GALI)~\cite{galiSkokos2007,Datseris2018,Datseris2022}.
The GALI is a measure of how vectors in the tangent space of a trajectory align with each other.
Let $\hat w_1(0),\dots,\hat w_k(0) \in \mathbb R^{M+S}$ be linearly-independent unit deviation vectors that evolve according to,
\begin{align}
	\dv{ t} \hat w_i(t) = \mathbf J(x(t)) \hat w_i(t),
\end{align}
where $x(t) =( N_1(t),\dots,N_S(t),R_1(t),\dots,R_M(t))$ and $\mathbf J(x(t))$ is the Jacobian matrix of the system in state $x(t)$.
These deviation vectors are normalized to have unit length at regular (small) time intervals.
The order-$k$ GALI is defined as,
\begin{align}
	\mathrm{GALI}_k (t)
	= 
	\left\|
	\hat w_1(t) \wedge \hat  w_2(t) \wedge \cdots \wedge \hat w_k(t)
	\right\|
	=
	|\det
	(\hat w_1(t), \hat w_2(t), \dots, \hat w_k(t))
	|,
\end{align}
which is the volume of the $k$-dimensional parallelepiped spanned by the time-evolved deviation vectors.
If the system is chaotic, the GALI will decay exponentially with time, indicating that the deviation vectors are becoming more aligned with each other due to the exponential divergence of nearby trajectories.
If the system is not chaotic and reaches a steady state, the GALI will asymptotically approach a nonzero value, indicating that the deviation vectors form a nonzero volume in the tangent space because no single direction dominates the dynamics.
Alternatively, if the system is asymptotically periodic or quasi-periodic and does not reach a steady state, the GALI will decay as a power law with time.

In Fig.~\ref{fig:gali}, we show the GALI for simulations in the dynamic and stable phases for $k = 2,\dots, 7$ with $M=S=512$ species and resources.
We use the \texttt{gali} function from the \texttt{ChaosTools.jl} \textit{Julia} package~\cite{Datseris2018}.
The simulations all have the same sampled parameters; only $\rho$ and the order of the GALI are varied.
For the simulations in the dynamic phase (solid lines), the GALI asymptotically decays exponentially with time, indicating chaos, while for the simulations in the stable phase (dashed lines), the GALI asymptotically approaches a nonzero value, indicating the dynamics achieve a steady state.
The initial deviation vectors are chosen to be $k$ unit vectors in the direction of randomly chosen species and resources.
The arguments passed to the \texttt{gali} function are the run time, \texttt{T = 2e4}, the time step, \texttt{$\Delta$t = 5.}, the threshold at which to stop the simulation, \texttt{threshold = 1e-22}, and the order of the GALI, \texttt{k}.

\subsection{Analysis of effective dimension of dynamics\label{appendix:effective-dimension}}

Chaotic systems can often be characterized by their effective dimension, which is the number of degrees of freedom that are relevant to the dynamics.
One measure of the effective dimension is the Kaplan--Yorke (KY) dimension~\cite{kaplan1979dimension,frederickson1983dimension,datseris2023FractalDimensions}, which is the number of Lyapunov exponents that are positive.
The KY dimension is the linearly interpolated point at which the cumulative sum of the Lyapunov exponents crosses zero:
\begin{equation}
	D_\text{KY}
	=
	k +
	\frac{\sum_{i=1}^k \lambda_i}{\abs{\lambda_{k+1}}},
	\qquad
	k = \max_j \qty(
		\sum_{i=1}^j \lambda_i > 0
	)
\end{equation}
We estimate the KY dimension of the aMCRM using the Lyapunov spectrum measured using the methods described in SI section~\ref{appendix:lyapunov-exponent} and the \texttt{kaplanyorke\_dim} function from the \texttt{FractalDimensions.jl} \textit{Julia} package~\cite{datseris2023FractalDimensions}.

We find that the KY dimension of the aMCRM in the dynamic phase for the parameters corresponding to the gray star in Fig.~\ref{fig:phase-diagram}(a) is $D_\text{KY} = 15 \pm 7$, which is an average over 64 simulations.
The number of surviving species and resources in these simulations at the end of the simulation is $S^\star + M^\star = 74 \pm 17$.
The ratio of the KY dimension to the number of surviving species and resources is $D_\text{KY}/(S^\star + M^\star) = 0.21 \pm 0.09$.
The number of species and resources that are transitioning between high- and low-abundance states is $M_t + S_t = 38 \pm 19$, and the ratio of the KY dimension to the number of species and resources that are transitioning between high- and low-abundance states is $D_\text{KY}/(M_t + S_t) = 2.6 \pm 1.2 \approx O(1)$.
From this, we hypothesize that the effective dimension of the dynamics is approximately this number of species and resources `jumping' between high- and low-abundance states.
The intuition behind this phenomenon is that the hypothesis that dynamics of the aMCRM in the dynamic phase are dominated by the species and resources that are transitioning between high- and low-abundance states~\cite{depirey2023manyspecies}.
We hope to explore this connection further in future work.


\clearpage
\section{Simulation and numerical methods\label{appendix:simulation-and-numerical-methods}}

\subsection{Numerical integration\label{appendix:numerical-integration}}

Numerical integration of the differential equations is performed using the \texttt{Tsit5} solver from the \texttt{DifferentialEquations.jl} package~\cite{rackauckas2017differentialequations} in the \textit{Julia} programming language~\cite{julia2017}.
\texttt{Tsit5} is the Tsitouras 5/4 Runge--Kutta method, a 5th order Runge--Kutta method with an embedded 4th order method for error estimation and step size control~\cite{tsitouras2011runge}.
The solver is configured to use a relative tolerance of $10^{-14}$ and an absolute tolerance of $10^{-14}$.
The aMCRM differential equations are also effectively integrated using the \texttt{VCABM} solver which an adaptive order adaptive time Adam--Moulton method~\cite{VCABM1993,rackauckas2017differentialequations}.
When using \texttt{VCABM}, the solver is configured to use a relative tolerance of $10^{-11}$ and an absolute tolerance of $10^{-11}$.
The \texttt{VCABM} is often faster than \texttt{Tsit5} for the aMCRM differential equations, but is less stable for some parameter values, particularly in the dynamic phase with low $\rho$ and high $\sigma$.
Therefore, in nearly all cases, especially those where the parameter space is explored, \texttt{Tsit5} is used.


\subsection{Surviving/depletion cutoffs, numerical instability, and small immigration\label{appendix:immigration-and-cutoffs}}

Simulating ecological dynamics often involves including very small immigration rates, $\lambda_N$, $\lambda_R$ for species and resources, respectively, to numerically regularize dynamics and ensure an uninvadable steady state is approached in simulations.
Furthermore, the inclusion of small immigration ensures that chaotic fluctuations in the dynamics do not drive species or resources to extinction, killing chaos due to finite size effects.
The equations used for numerical simulations are,
\begin{align}
	\dv{N_i}{t}
	&=
	N_i
	\left(
		\sum_{\alpha=1}^M
		c_{i\alpha} R_\alpha - m_i
	\right)
	+
	\lambda_N,
	\\
	\dv{ R_\alpha}{t}
	&=
	R_\alpha
	\left(
		K_\alpha - R_\alpha
	\right)
	-
	\sum_{i=1}^S N_i e_{i\alpha} R_\alpha
	+
	\lambda_R.
\end{align}
In all our simulations, we take $\lambda_N, \lambda_R = 10^{-10}$.
Including this small level of immigration regularizes the dynamics in that the simulated steady-state abundances of species and resources now have a clear gap between the abundances of those that are surviving and those that are extinct, as shown in Fig.~\ref{fig:cutoff-log-hist}.
This gap is used to define a cutoff to identify fractions of surviving species and resources.
Including this small level of immigration additionally stabilizes the numerical integrator, leading to fewer simulations that report errors due to numerical instability.
The effects of immigration on dynamics in the random Generalized Lotka--Volterra model is discussed in detail in~\cite{depirey2023manyspecies}; we expect that many of the intuitions presented there apply to the aMCRM as well.

\begin{figure}[ht]
	\centering
	\includegraphics[width=0.6\linewidth]{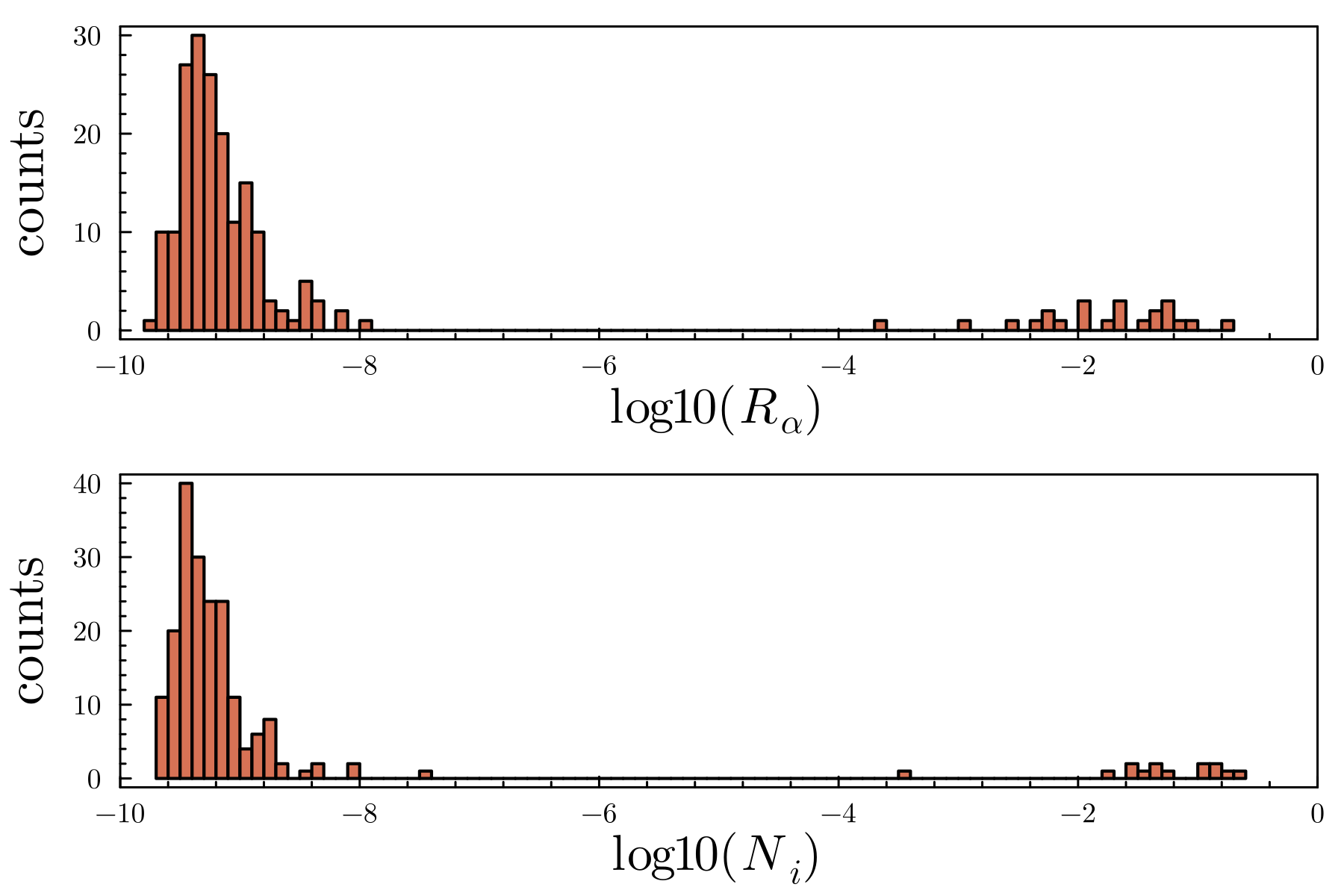}
	\caption{\label{fig:cutoff-log-hist}
		Log histogram of the species and resource abundances at steady state for a system with $M,S=256$ and $\rho = 0.9$.
		There is a clear gap in abundance between species/resources that are surviving and those that are extinct.
		The immigration rate used is $\lambda = 10^{-10}$.
	}
\end{figure}


\subsection{Numerically solving the cavity self-consistency equations and calculating phase boundaries\label{appendix:solving-self-consist}}

\begin{figure}[ht]
	\centering
	\includegraphics[width=0.5\linewidth]{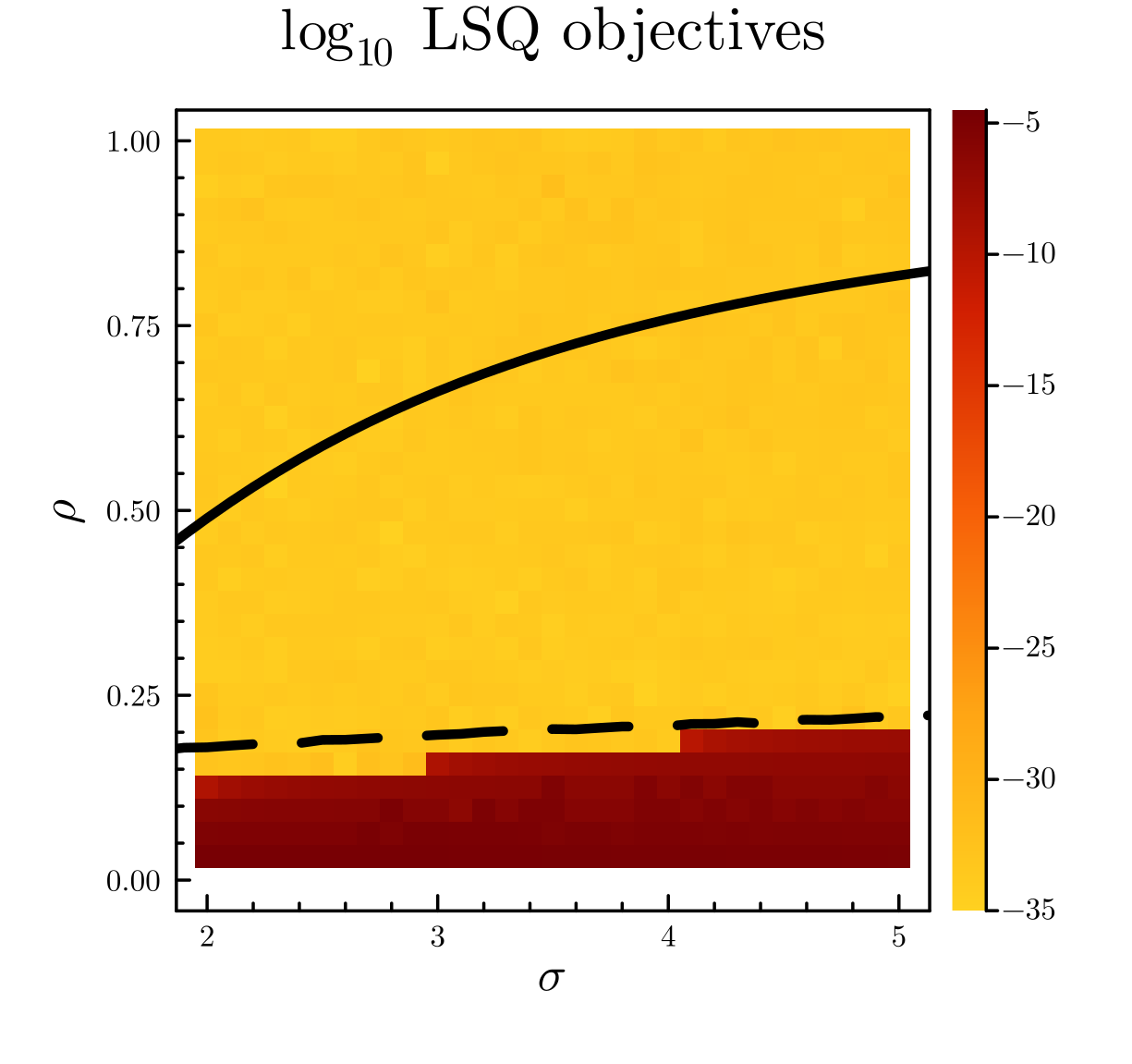}
	\caption{
		\label{fig:LSQobjs}
		Least-squares objectives for solving the cavity self-consistency equations.
		The solid curve is the instability boundary, and the dashed curve is the infeasibility boundary.
		The region in which the objective function is very small is the region in which the cavity self-consistency equations are solved; the region in which the objective function is large represents invalid solutions which are not included in plots.
		The parameters used are those used in Fig.~\ref{fig:phase-diagram}(a).
		}
\end{figure}


In order to solve the self-consistency equations for the cavity method, we the use nonlinear least squares method.
We define the objective function to be the sum of the squares of the differences between the left and right-hand sides of the self-consistency equations (\ref{eq:phiNselfconsist}, \ref{eq:phiRselfconsist}, \ref{eq:Nselfconsist}, \ref{eq:Rselfconsist}, \ref{eq:qNselfconsist}, \ref{eq:qRselfconsist}, \ref{eq:chiselfconsist}, \ref{eq:nuselfconsist}):
\begin{align}
	L(\phi_R, \phi_N, \langle N \rangle, \langle R \rangle, q_N, q_R, \chi, \nu)
	=
	(\text{LHS $-$ RHS of Eq.~\eqref{eq:phiNselfconsist}})^2
	+
	\cdots
	+
	(\text{LHS $-$ RHS of Eq.~\eqref{eq:nuselfconsist}})^2.
\end{align}
When the self-consistency equations are solved, the objective function is zero.
The objective is minimized using adaptive differential evolution solver, implemented as\\ \texttt{BBO\_adaptive\_de\_rand\_1\_bin\_radiuslimited} in the \textit{Julia} package \texttt{BlackBoxOptim.jl}~\cite{JuliaBBO}.

There are choices of parameters $\mu_c, \mu_e, \sigma_c, \sigma_e, K, \sigma_K, m, \sigma_m, \rho$ for which the objective function does not reach zero.
In Fig.~\ref{fig:LSQobjs}, we plot the objective function values for various choices of $\rho$ and $\sigma = \sigma_e = \sigma_c$; the other parameters for this figure are those used in Fig.~\ref{fig:phase-diagram}.
We can see that the objective function values are very small for $\rho$ near one and suddenly jumps to a much larger value for $\rho$ near zero.
The cause of this increase in the objective function is the term (LHS $-$ RHS of Eq.~\eqref{eq:nuselfconsist}) in the objective function.
Observe that as $\chi \to 0$, $-\nu \to \infty$, which leads to the objective function value blowing up.
In order to find the boundary at which the objective function blows up, we simply add a term to the objective function that is zero when $\chi = 0$ and allow one model parameter (here, $\rho$) to vary:
\begin{align}
	\label{eq:LSQobj-CSI}
	L^\text{CSI}(\phi_R, \phi_N, \langle N \rangle, \langle R \rangle, q_N, q_R, \chi, \nu,\rho^\text{CSI})
	=&
	L(\phi_R, \phi_N, \langle N \rangle, \langle R \rangle, q_N, q_R, \chi, \nu,\rho^\text{CSI})
	\\
	&
	\qquad\qquad\qquad
	\nonumber
	+
	(\chi - 0)^2.
\end{align}
Similarly, to find the instability boundary, we add a term to the objective function corresponding to the instability condition (Eq.~\eqref{eq:instability-boundary}):
\begin{align}
	L^\star(\phi_R, \phi_N, \langle N \rangle, \langle R \rangle, q_N, q_R, \chi, \nu,\rho^\star)
	=&
	L(\phi_R, \phi_N, \langle N \rangle, \langle R \rangle, q_N, q_R, \chi, \nu,\rho^\star)
	\\
	\nonumber
	&
	\qquad\quad
	+
	(\text{LHS $-$ \text{RHS} of Eq.~\eqref{eq:instability-boundary}})^2.
\end{align}
Minimizing these objective functions while allowing one additional model parameter to vary gives the infeasibility and instability boundaries shown in all figures in this paper.

\clearpage
\section{Finite size scaling analysis\label{appendix:finite-size-scaling}}

In order to understand whether the aMCRM achieves a steady state in the thermodynamic limit, we must understand how finite system size and a finite simulation time impact simulation results.
To assess the existence and dynamical nature of a steady state in a simulation, we numerically calculate a time to steady state (TtSS).
After running a simulation, we iterate through the time series and find the time at which the mean absolute value of the derivatives last dips below a threshold, $10^{-10}$.
The tolerance of the solver (\texttt{Tsit5}~\cite{tsitouras2011runge,rackauckas2017differentialequations}) is $10^{-14}$, so this threshold is well above the numerical error of the simulation.
Because we simulate to a finite time $T_\text{max} =2^{16} + 10^4 =75536$, it is not possible to distinguish between a system that will reach a steady state at some time beyond $T_\text{max}$ and a system that will never reach a steady state.
Furthermore, when performing these simulations, especially with large system sizes, errors occasionally occur in the solver, and the simulation is aborted; while we do not present finite-size scaling results in regions where the simulations are frequently aborted, a robust analysis should account for this.
Here, we develop a custom maximum likelihood estimation (MLE) technique to fit the distributions TtSS's at different system sizes, taking into account these two situations.
We then examine the scaling of the resulting fit parameters to determine their behavior in the thermodynamic limit.

\subsection{Modeling simulation outcomes}

To construct a MLE model, we first must model the statistical process which determines each possible outcome of a simulation.
We observe that simulation outcomes fall into three categories: 
\begin{enumerate}[(i)]
	\item The simulation reaches steady state within the maximum simulation time $T_{\max}$. 
	\item The simulation does not reach steady state within $T_{\max}$.
	\item The solver encounters an error and the simulation aborts.
\end{enumerate}

First, we represent the probability of a simulation encountering an error as $p_\text{err}$.
Next, we consider whether a simulation could possibly reach a steady state if given enough simulation time. We represent the probability of a simulation reach steady state in the infinite time limit as $\phi_\text{SS}$.
Finally, we define the probability density function $p_\text{SS}(t)$ of TtSS's for those simulations that can reach steady state when given enough time.

Using these definitions, we can now construct the probabilities for each of the cases defined above.
Let $T$ be a random variable representing the observed outcome of a simulation with possible values ${T\in [0, T_{\max}]\cup \{\texttt{noSS}, \texttt{error}\}}$,
where $T$ takes a finite value in the interval $[0, T_{\max}]$ for case (i),
or the special values \texttt{noSS} or \texttt{error} in cases (ii) and (iii), respectively.

To model case (i), we note that it results in an outcome represented by a continuous numerical value, while cases (ii) and (iii) represent categorical outcomes.
For ease of derivation, we also convert $T$ to a categorical outcome in the prior case by artificially breaking our simulation interval into small discrete time steps of size $\Delta t$.
Later, we will take the limit $\Delta t \rightarrow 0$, removing the dependency of our model on this step size.
The probability of a simulation completing successfully and reaching a steady state in the time interval $t < T < t+\Delta t$ with $T< T_{\max}$ is approximately
\begin{equation}
	\P(t < T \leq t + \Delta t \mid T < T_{\max}) \approx (1-p_\text{err}) \phi_\text{SS}  \Delta t \, p_\text{SS}(t).
\end{equation}

For case (ii), simulations do not encounter an error, but do not reach steady state within $T_{\max}$. We must consider two separate possibilities: a simulation may never reach steady state, even if $T_{\max}\rightarrow \infty$, or would reach steady-state after further simulation, $T > T_{\max}$.
In the first case, the probability is simply $(1-p_\text{err}) (1 - \phi_\text{SS})$. If a simulation would eventually have reached steady-state if we had continued simulating, we must use our probability density function $p_\text{SS}(t)$. However, we do not know exactly when the simulation would have finished, so we must sum across all discrete time intervals greater than $T_{\max}$. The total probability for case (ii) is then
\begin{equation}
\begin{aligned}
	\P(\texttt{noSS}) &\approx (1-p_\text{err}) (1 - \phi_\text{SS}) + (1-p_\text{err})\phi_\text{SS}\sum_{n=0}^\infty \Delta t \, p_\text{SS}(T_{\max} + n\Delta t)\\
	&\approx (1-p_\text{err})\qty(1 - \phi_\text{SS} F_\text{SS}(T_{\max})),
\end{aligned}
\end{equation}
where we have taken the limit $\Delta t \rightarrow 0$ and simplified, defining $F_\text{SS}(T) = \int_0^T dt p_\text{SS}(t)$ as the cumulative distribution function of $p_\text{SS}(t)$.

Finally, for case (iii), the probability is simply
\begin{equation}
	\P(\texttt{error}) = p_\text{err}.
\end{equation}

\subsection{Likelihood function}

We now use the outcome probabilities in the previous section to construct a likelihood function to fit the data.
We are given $N$ (independent) simulations resulting in observed TtSSs $\{T_1,\cdots,T_N\}$ which may take on values ${T_i\in [0, T_{\max}]\cup \{\texttt{noSS}, \texttt{error}\}}$.
Our fit parameters $\theta$ include $p_\text{err}$, $\phi_\text{SS}$, along with the parameters that define the probability density function $p_\text{SS}(t)$.
Defining the likelihood of the data given the parameters $\theta$ as $\P(\{T_i\}_{i=1}^N | \theta)$, the MLE estimate of the parameters is found by minimizing the negative log-likelihood,
\begin{equation}
	\theta^\star = \argmax\limits_\theta \P(\{T_i\}_{i=1}^N | \theta)
=
- \argmin\limits_\theta \frac{1}{N}\log \P(\{T_i\}_{i=1}^N | \theta).
\end{equation}
Using our outcome probabilities, the negative log-likelihood is then
\begin{equation}
\begin{aligned}
\mathcal{L}(\{T_i\}_{i=1}^N ; \theta) =&  -\frac{1}{N}\log \P(\{T_i\}_{i=1}^N | \theta) = -\frac{1}{N}\log \prod_{i=1}^N \P(T_i|\theta) = -\frac{1}{N}\sum_{i=1}^N\log \P(T_i|\theta)\\
=& -\frac{1}{N}\sum_{i=1 : T_i < T_{\max}}^N \log \qty[(1-p_\text{err}) \phi_\text{SS}  \Delta t \, p_\text{SS}(T_i)]\\ 
& - \frac{1}{N}\sum_{i=1 : T_i = \texttt{noSS}}^N \log \qty[(1-p_\text{err})\qty(1 - \phi_\text{SS} F_\text{SS}(T_{\max}))] -\frac{1}{N}\sum_{i=1 : T_i = \texttt{noSS}}^N \log p_\text{err}.
\end{aligned}
\end{equation}
Next, we define $f_\text{err}$ as the fraction of simulations that encounter errors, and the $f_\text{SS}$ as the fraction of simulations that did not encounter errors and reached a steady state within time $T_{\max}$.
Using these definitions, we get
\begin{equation}
\begin{aligned}
\mathcal{L}(\{T_i\}_{i=1}^N ; \theta) =& -\frac{1}{N}\sum_{i=1 : T_i < T_{\max}}^N \log p_\text{SS}(T_i) - f_\text{SS}\log \qty[(1-p_\text{err}) \phi_\text{SS}  \Delta t ]\\ 
& - (1-f_\text{err} - f_\text{SS}) \log \qty[(1-p_\text{err})\qty(1 - \phi_\text{SS} F_\text{SS}(T_{\max}))] - f_\text{err} \log p_\text{err}\\
=& -\frac{1}{N}\sum_{i=1 : T_i < T_{\max}}^N \log p_\text{SS}(T_i) - (1-f_\text{err} - f_\text{SS}) \log \qty(1 - \phi_\text{SS} F_\text{SS}(T_{\max}))\\
& - f_\text{SS}\log \phi_\text{SS}  - (1-f_\text{err})\log  (1-p_\text{err}) - f_\text{err} \log p_\text{err} - f_\text{SS}\log \Delta t.
\end{aligned}
\end{equation}
We note that the last term is a constant, so we may choose to ignore it and take the limit $\Delta t \rightarrow 0$ as mentioned previously, giving us the final form for the likelihood,
\begin{equation}
\begin{aligned}
	\mathcal{L}(\{T_i\}_{i=1}^N ; \theta) =& -\frac{1}{N}\sum_{i=1 : T_i < T_{\max}}^N \log p_\text{SS}(T_i) - (1-f_\text{err} - f_\text{SS}) \log \qty(1 - \phi_\text{SS} F_\text{SS}(T_{\max}))\\
& - f_\text{SS}\log \phi_\text{SS}  - (1-f_\text{err})\log  (1-p_\text{err}) - f_\text{err} \log p_\text{err} \label{eq:likelihood}
\end{aligned}
\end{equation}

\subsection{Fitting the model}

To find the maximum likelihood estimates of the parameters, we minimize Eq.~\eqref{eq:likelihood} using the BFGS algorithm implemented in the \textit{Julia} package $\texttt{Optimization.jl}$~\cite{optimizationRackauckas2023}.
We found empirically that the Fr\'echet distribution is a good choice for $p_\text{SS}(t)$.
The Fr\'echet distribution has the following forms for the cumulative distribution function and probability density function:
\begin{align}
  F_\text{SS}(t)
  =
  \begin{cases}
    e^{
      -(t/\tau)^{-\alpha}
    }
    , & t > 0,\\
    0, & t \le 0,
  \end{cases}
  \qquad
  p_\text{SS}(t)
  =
  \begin{cases}
    \frac{\alpha}{\tau}
    \left(
      \frac{t}{\tau}
    \right)^{-\alpha-1}
    e^{
      -(t/\tau)^{-\alpha}
    }
    , & t > 0,\\
    0, & t \le 0,
  \end{cases}
\end{align}
where $\tau$ is the timescale parameter and $\alpha$ is the shape parameter.
The errors in the fitted parameters $\phi_\text{SS}, p_\text{err}, \tau,\alpha$ are computed by bootstrapping the data.
That is, we randomly sample $N$ data points from the simulation data set with replacement and fit the model to the sampled data repeatedly, providing us with an empirical distribution for each fit parameter.

\subsection{Scaling of parameters with system size and discussion}

The simulated data for different system sizes in the stable and dynamic phases can be visualized by analyzing the empirical cumulative distribution functions (CDFs) of the observed times to steady state called $T_\text{SS}$ above.
In Fig.~\ref{fig:scaling-cdfs}, we show the CDFs for the stable and dynamic phases for $4 \leq M,S \leq 1024$ along with the CDFs for the fitted distributions.
A clear difference is apparent between the stable and dynamic phases, and the fitted distributions match the simulated data well.
A scaling collapse of the CDF based on the fitted distributions is shown in Fig.~\ref{fig:scaling-collapses}.

The fit parameters for the stable and dynamic phases are shown in Fig.~\ref{fig:finite-size-scaling}.
In the stable phase, the estimated probability of reaching steady state, $\phi_\text{SS}$ remains approximately equal to $1$ for all system sizes, and the estimated timescale $\tau$ asymptotically increases with system size as a power law.
In the dynamic phase, $\phi_\text{SS}$ decreases with system size towards $0$, and $\tau$ asymptotically increases with system size also as a power law, but with a larger exponent.
In both phases, $\alpha$ approaches $1$ asymptotically from above.

We provide fits of $\phi_\text{SS}$ as a function of system size $M$ using curves of the form $\phi_\text{SS}(M) = \phi_\text{max} - \Delta (1 - \exp(-(M/\xi)^\kappa))$.
For the dynamic phase $\phi_\text{max} = 1.0$, $\Delta = 1.0$, $\xi = 278.5$, and $\kappa = 1.5$; for the stable phase, $\phi_\text{max} = 1.0, \Delta = 0.0$,  $\kappa$ remains approximately equal to the value at which it is initialized in the optimization algorithm, and $\xi$ approaches the maximum value set in the optimization algorithm.

We also provide fits of the timescale $\tau$ as a function of system size $M$ using power laws of the form  $\tau(M) = b M^a $. Curves are fit for data $M,S \geq 2^{9.5}$.
In the stable phase, $a = 0.4$ and $b = 580$;
in the dynamic phase, $a = 0.9$ and $b = 70$.



This finite size scaling is run in the range $512 \leq M,S \leq 1024$ but is not shown in Fig.~\ref{fig:finite-size-scaling} because solver errors are occasionally present.
More significantly, as very few simulations reach steady state in the dynamic phase when the system size is sufficiently large, the MLE procedure cannot clearly distinguish between cases where $\tau$ is very large or $\phi_\text{SS}$ is very small; this ambiguity can be seen more clearly in Fig.~\ref{fig:scaling-cdfs}.
With the prior knowledge of the finite size scaling as shown in Fig.~\ref{fig:finite-size-scaling} where the MLE procedure succeeds and there are no solver errors, we conclude that at these larger system sizes, it is both the case that $\tau$ grows large (in polynomial order of system size) and $\phi_\text{SS}$ approaches either zero or one. That fact both the timescale $\tau$ diverges and $\phi_\text{SS}$ approaches either zero at large system size leads us to conclude that in the thermodynamic limit, all systems in the stable phase eventually reach steady state for long enough times, while in the dynamic phase, no systems reach steady state.

\begin{figure}[t]
	\centering
	\includegraphics[width=0.8\linewidth]{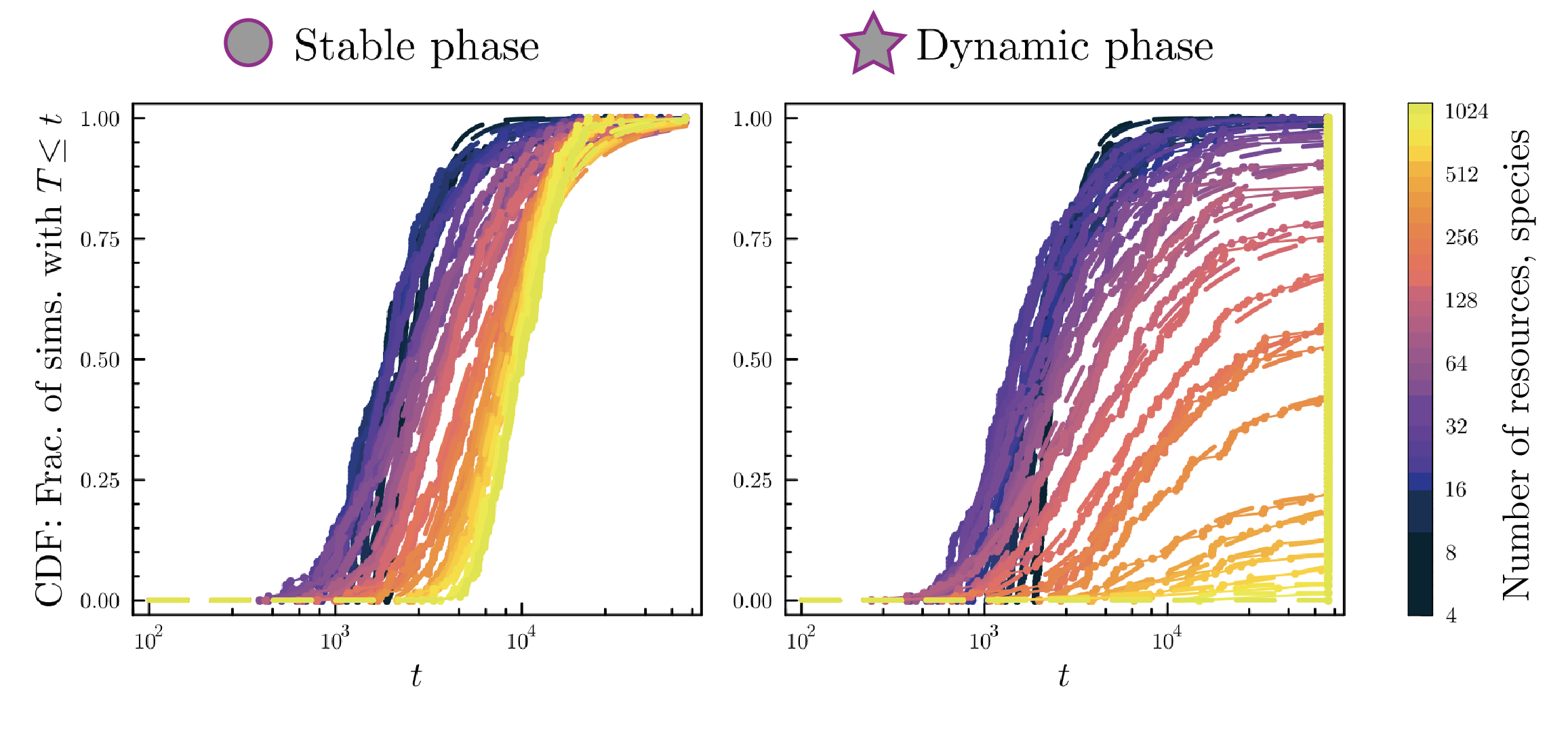}
	\caption{\label{fig:scaling-cdfs}
		Empirical cumulative distribution functions of $T$, the time to steady state, for simulations of different system sizes in the stable phase (left) and the dynamic phase (right).
		The maximum likelihood estimate distributions as described in SI section~\ref{appendix:finite-size-scaling} for the time to steady state for the various system sizes are shown with dashed lines.
		The maximum value shown on the horizontal axis is the maximum time simulated, $T_\text{max} = 2^{16} + 10^4 = 75536$.
		}
  \end{figure}

  \begin{figure}[t]
	\centering
	\includegraphics[width=0.8\linewidth]{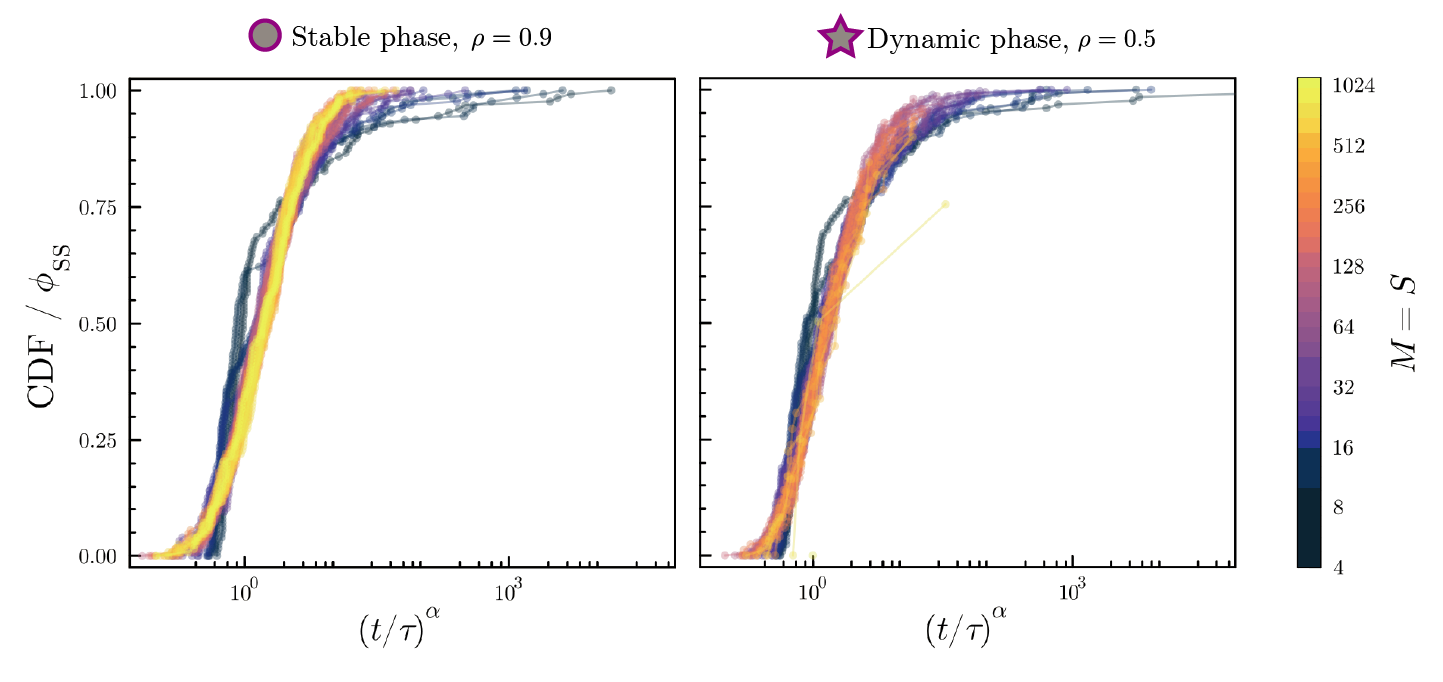}
	\caption{\label{fig:scaling-collapses}
		Scaling collapse of TtSS CDFs.
		CDFs of observed TtSSs are scaled by the MLE-fitted value of $\phi_\text{SS}$ where the horizontal axis is scaled by the MLE-fitted values of $\tau$, $\alpha$.
		Note that the curves collapse onto each other in both phases for system sizes that are moderately large $M,S\gtrapprox 2^{4.5}$.
		This indicates that we have successfully identified the asymptotic scaling of the form of the TtSS distribution with system size.
	}
  \end{figure}

\begin{figure}[t]
	\centering
	\includegraphics[width=0.85\linewidth]{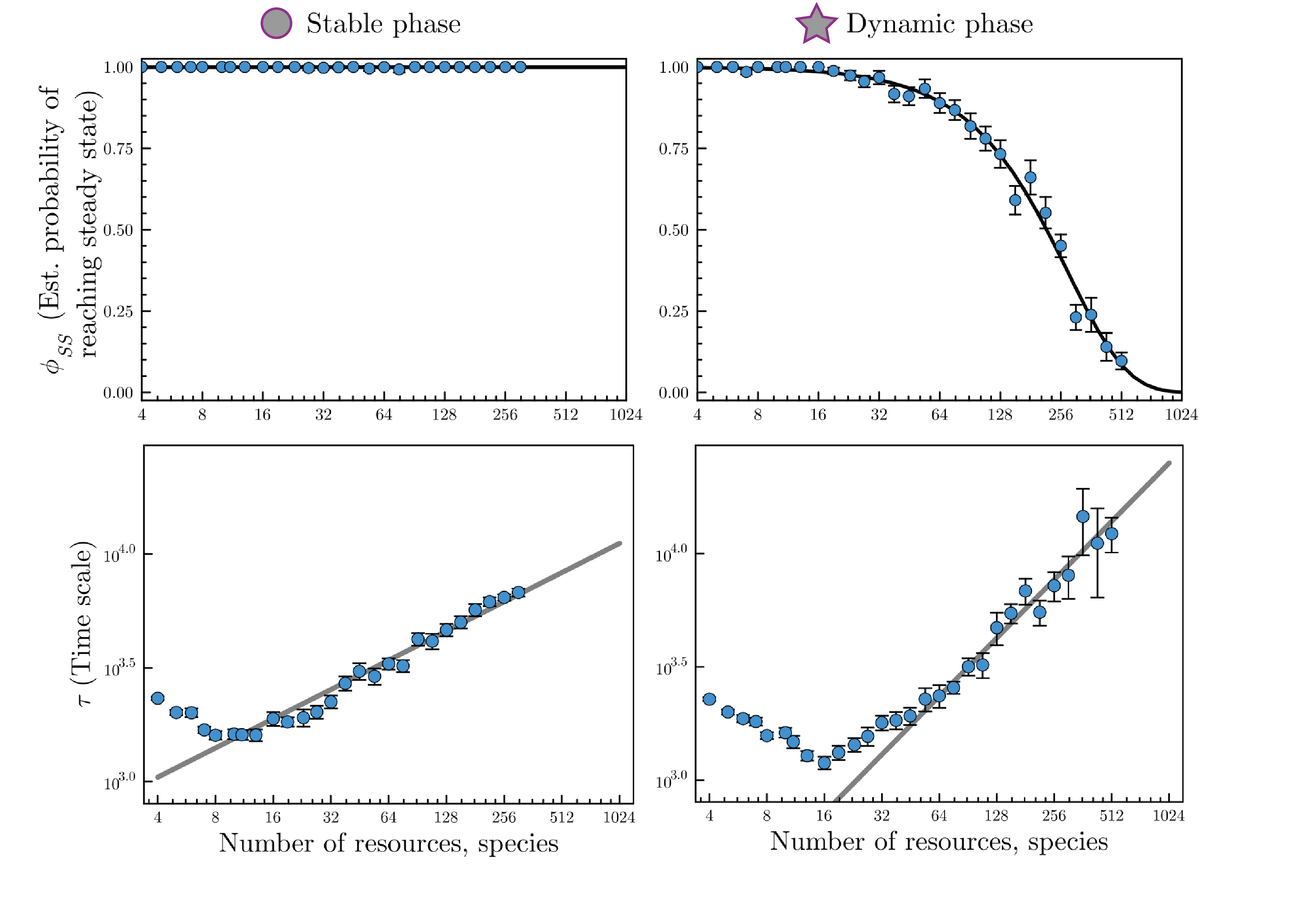}
	\caption{
		\label{fig:finite-size-scaling}
	Impact of finite size effects on the dynamics of the aMCRM.
	Rows correspond to different parameters describing dynamics: the estimated probability a simulation of a given system size will reach steady state (top) and the timescale of the time to steady state of simulations of a given system size that do reach steady state (bottom).
	Columns correspond to simulation data from different phases: the stable phase (left) and the dynamic phase (right); the parameters corresponding these simulations are marked in the phase diagram in Fig.~\ref{fig:phase-diagram} with a circle and star, respectively.
	The probability of reaching steady state in the stable phase remains at $1$ for all system sizes, while the probability of reaching steady state in the unstable phase decreases to zero with increasing system size, indicating that stable dynamics are not observed within the dynamic phase in the thermodynamic limit.
	The timescale of the time to steady state increases with system size in both the stable and dynamic phases.
	}
  \end{figure}

\clearpage

\section{Additional simulations}

\begin{figure}[h]
	\centering
	\includegraphics[width=0.49\linewidth]{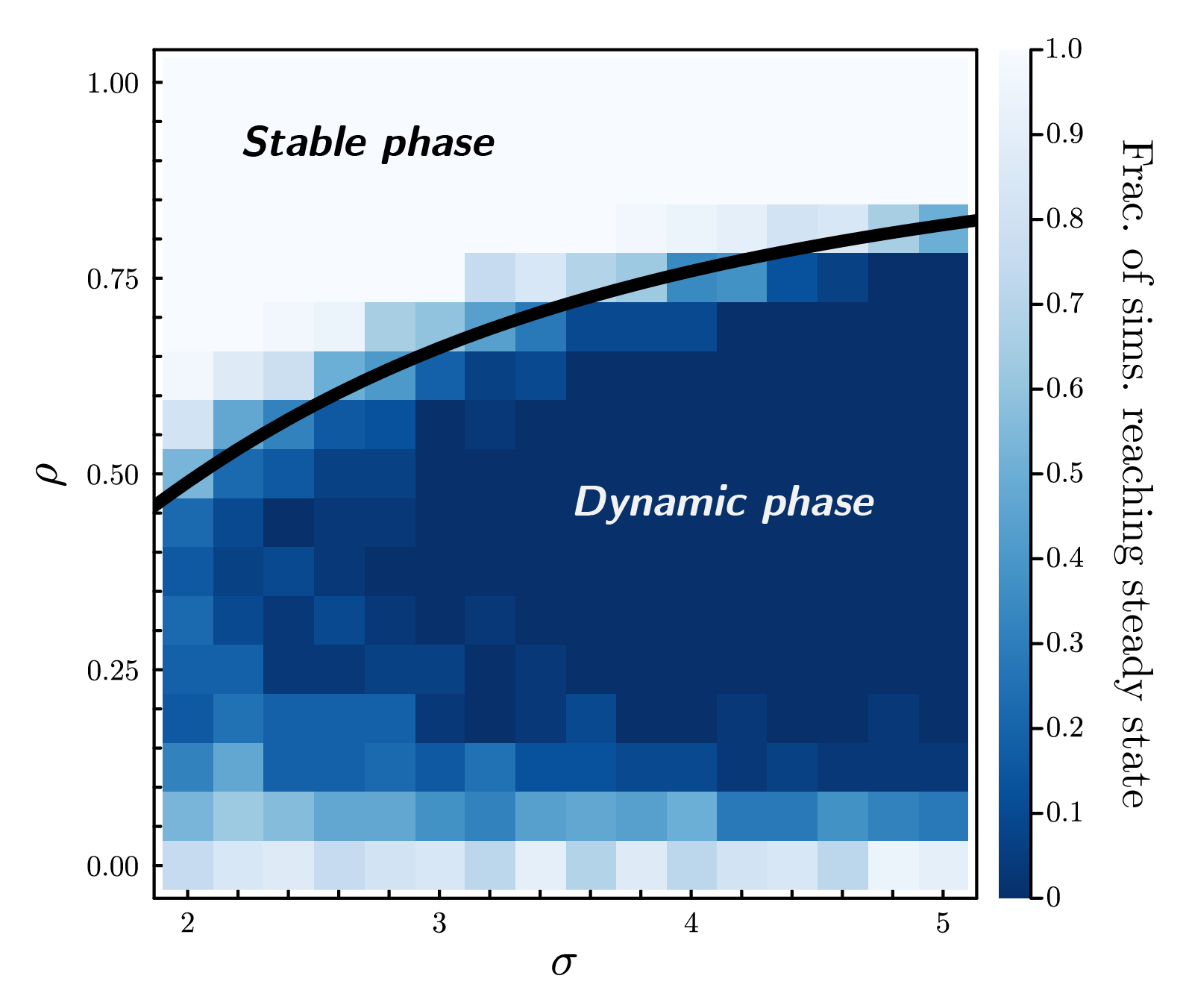}
	\caption{\label{fig:uniform-heatmap}
		Heatmap of the fraction of simulations at steady state when the sampling distributions are uniform.
		Parameters are the same as in Fig.~\ref{fig:phase-diagram}(a).
		Because the calculations are performed in the thermodynamic limit, the phase boundary is agnostic to the choice of sampling distribution.
		Further details about sampling distribution and parameters are in appendix \ref{appendix:fig-details}.
	}
\end{figure}

\begin{figure}[h]
	\centering
	\includegraphics[width=0.82\linewidth]{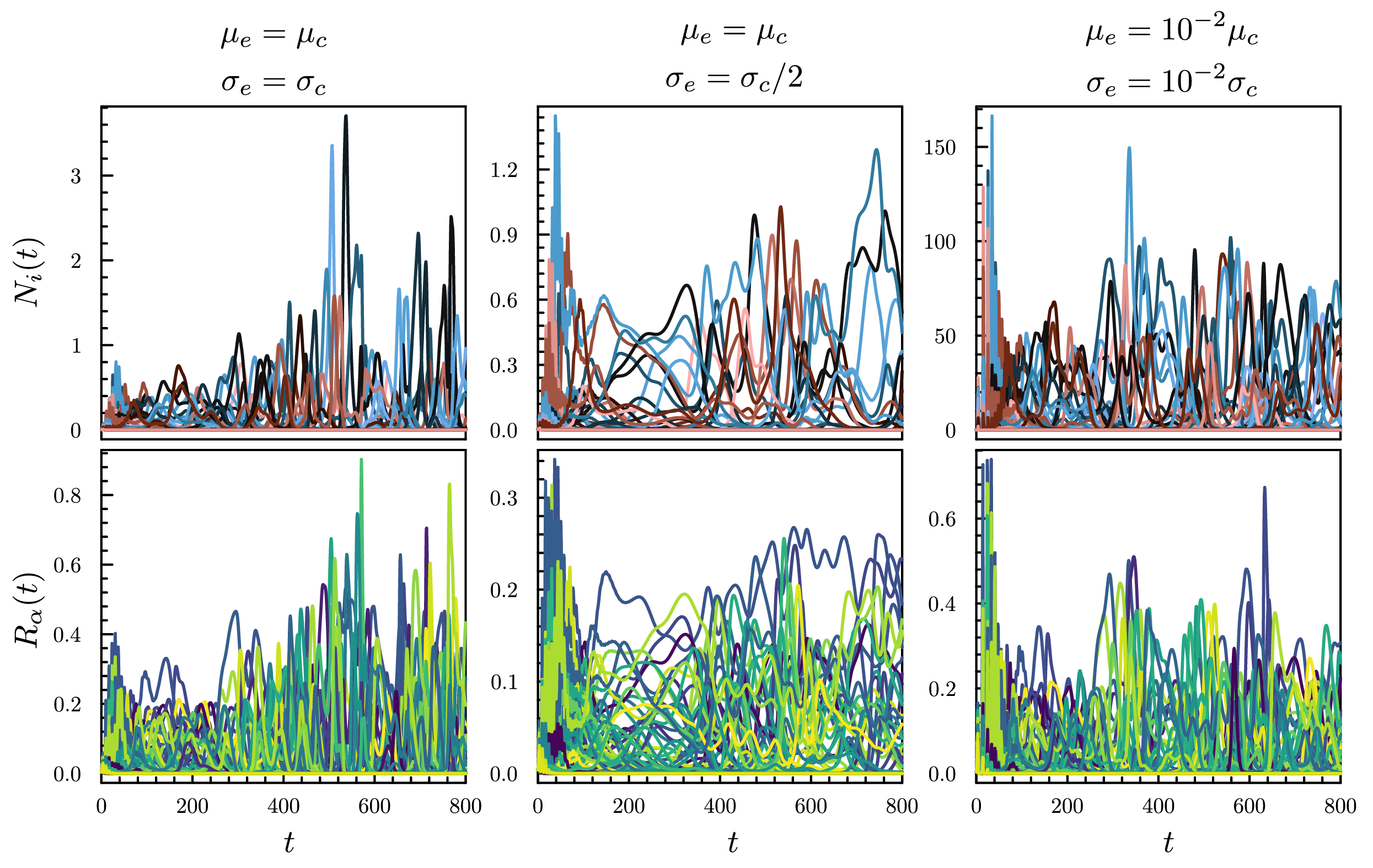}
	\caption{
		Example simulations of the aMCRM with (left column) $\mu_e = \mu_c$, $\sigma_e = \sigma_c$; (center column) $\mu_e = \mu_c$, $\sigma_e = \sigma_c /2$; and (right column) $\mu_e = 10^{-2}\mu_c$, $\sigma_e = 10^{-2}\sigma_c$.
		The sampled random matrices and parameters are the same in each of the cases, as is the coloring of the species and resources.
		The chosen values of $\mu_c, \sigma_c$, and $\rho$ correspond to the gray star in Fig.~\ref{fig:phase-diagram}(a).
	}
\end{figure}

\newpage


\section{Parameters in figures\label{appendix:fig-details}}

\subsubsection*{Figure~\ref{fig:aMCRM-schematic} (schematic) details}
Species drawings are traces of images generated using Adobe Firefly Generative AI. Prompts: ``simple 2D flat unshaded vector graphics of microbes,'' ``simple 2D flat digital art sketches of bacteria and microbes in icon-style with no shading.''

\subsubsection*{Figure~\ref{fig:stable-dynamic-dyn} (example dynamics) simulation parameters}
Both simulations have random variables $d_{i\alpha}, x_{i\alpha}, K_\alpha, m_i$ drawn from standard normal distributions.
The parameters are,
\begin{align}
	K = 1, \quad
	\sigma_K = 0.1, \quad
	m = 1, \quad
	\sigma_m = 0.1, \quad
	\mu_c = 200, \quad
	\sigma_c = 3.5, \quad
	\mu_e = 200, \quad
	\sigma_e = 3.5, \quad
	\lambda_N = \lambda_R = 10^{-10},
\end{align}
with $M = S = 256$ resources and species.
For dynamics in the stable phase, $\rho = 0.9$, and for the dynamics in the unstable phase, $\rho = 0.5$.
Between the two simulations, the model parameters ($\delta K_\alpha, \delta m_i, d_{i\alpha}, x_{i\alpha}$) are sampled once and only the level of nonreciprocity ($\rho$) is changed.
For these choices of parameters, $\rho^\star = 0.72$.
The initial conditions are $N_i(0) = 1/S$ and $R_\alpha(0) = 1/M$.
Numerical integration is performed using a `fourth-order, five-stage explicit Runge-Kutta method with embedded error estimator of Tsitouras' implemented as \texttt{Tsit5} in the \textit{Julia} \texttt{DifferentialEquations.jl} package~\cite{rackauckas2017differentialequations,tsitouras2011runge}

\subsubsection*{Figure~\ref{fig:phase-diagram} (phase diagram) simulation parameters}
Each point in the heatmap represents an average over 32 simulations where the parameters were independently sampled from normal distributions: $d_{i\alpha}, x_{i\alpha}, K_\alpha, m_i \sim \mathcal{N}(0,1)$.
The distribution parameters are,
\begin{align}
	K = 1, \quad
	\sigma_K = 0.1, \quad
	m = 1, \quad
	\sigma_m = 0.1, \quad
	\mu_c = 200, \quad
	\mu_e = 200, \quad
	\lambda_N = \lambda_R = 10^{-10},
\end{align}
with $M = S = 256$ resources and species.
The parameters $\sigma_c$ and $\sigma_e$ are taken to be the same: $\sigma \equiv \sigma_c = \sigma_e$.
The dashed gray line represents a slice through the parameter space with the above parameters fixed and $\sigma = 3.5$ where $\rho$ varies from $0$ to $1$.
The gray star represents the parameters $\sigma = 3.5$ and $\rho = 0.5$; the gray circle represents the parameters $\sigma = 3.5$ and $\rho = 0.9$.
The values of parameters $\mu_c,\mu_e,\sigma_c,\sigma_e$ are chosen so that when $M=S=256$, $\langle c_{i\alpha} \rangle ,\langle e_{i\alpha}\rangle = O(1)$ and $\Var{c_{i\alpha}}, \Var{e_{i\alpha}} = O(1)^2$; this abates some issues involving solver errors and instabilities.
The initial conditions are $N_i(0) = 1/S$ and $R_\alpha(0) = 1/M$.
Numerical integration is performed using a `fourth-order, five-stage explicit Runge-Kutta method with embedded error estimator of Tsitouras' implemented as \texttt{Tsit5} in the \textit{Julia} \texttt{DifferentialEquations.jl} package~\cite{rackauckas2017differentialequations,tsitouras2011runge}.
The simulations are run until $5 \times 10^4$ time units, and a simulation is considered to have reached steady state if the average absolute value of all derivatives is less than $10^{-7}$ within the last $0.05 \times 10^4$ time units.
\\
The instability phase boundary (solid black) is found by solving the cavity self-consistency equations (see SI section~\ref{appendix:cavity-calc} and Eq.~\eqref{eq:instability-boundary} simultaneously using nonlinear least squares, as described in SI section~\ref{appendix:solving-self-consist}).

\subsubsection*{Figure~\ref{fig:diverging-trajecs-maxlyap_dotplot} (Lyapunov dot plot and diverging trajectories) simulations details}
\textbf{(a)}
Details on the calculation and analysis of Lyapunov exponents are discussed in appendix \ref{appendix:lyapunov-exponent}.
Parameters correspond to a slice of the phase diagram shown in Fig.~\ref{fig:phase-diagram}(a) along the dashed gray line.
\\ \indent
\textbf{(b)}
Simulation parameters correspond to the gray star ($\sigma = 3.5$, $\rho = 0.5$) in Fig.~\ref{fig:phase-diagram}(a).
The initial conditions for the trajectory plotted in (solid) red is the state given after evolving the system for $5 \times 10^3$ time units from initial conditions, $N_i(0) = 1/S$, $R_\alpha(0) = 1/M$.
The initial conditions for the trajectory plotted in (dotted) blue are generated by selecting 8 species with abundance greater than $10^{-3}$ at random and perturbing their abundances by a random amount drawn from a uniform distribution $U([-10^{-4}, 10^{-4}])$; the highlighted species is not one of these 8 species.
The Lyapunov exponent is computed using the `H2' method of Geist~\cite{Benettin1980,Geist1990,Datseris2022,Datseris2018} for the trajectory plotted in red; see SI section~\ref{appendix:lyapunov-exponent} for details.

\subsubsection*{Figure~\ref{fig:cavity-eCDFs} (cavity distribution comparison) simulation parameters}
\begin{gather}
	K = 1, \quad 
	\sigma_K = 0.1, \quad
	m = 1, \quad
	\sigma_m = 0.1, \quad
	\mu_c = 8, \quad
	\sigma_c = 1, \quad
	\mu_e = 8, \quad
	\sigma_e = 1, \quad
	\rho=0.9
	\\
	\phi_N = 0.318
	,\;
	\phi_R = 0.729
	,\;
	\nu = -0.858
	,\;
	\chi = 0.412
	,\;
	\langle N \rangle = 0.105
	,\;
	\langle R \rangle = 0.114
	,\;
	\langle N^2 \rangle = 0.0575
	,\;
	\langle R^2\rangle = 0.0258.
\end{gather}
The uninvadable steady state is found using an iterative constrained optimization method described in~\cite{marsland2020mepp}.
Cavity parameters are found by solving the self-consistency equations using non-linear least squares as described in appendix \ref{appendix:solving-self-consist}.

\subsubsection*{Figure~\ref{fig:uniform-heatmap} (phase diagram with uniform sampling) simulation parameters}
All parameters are identical to those in Fig.~\ref{fig:phase-diagram}(a) except that the random matrices and vectors are sampled as:
\begin{align}
	d_{i\alpha},x_{i\alpha},\delta K_\alpha, \delta m_i \sim \text{Uniform}([-\sqrt{3},\sqrt{3}]).
\end{align}
This choice of distributions ensures that $d_{i\alpha},x_{i\alpha},\delta K_\alpha, \delta m_i$ are still standard random variables and appropriate thermodynamic scaling are used to obtain cavity results that are identical to those in Fig.~\ref{fig:phase-diagram}(a).

\end{widetext}


\end{widetext}

\end{document}